\documentclass[useAMS,usenatbib]{mn2e}
\usepackage{times}
\usepackage{graphicx}
\usepackage{longtable} 
\usepackage{tablefootnote} 
\usepackage{changes}  
\usepackage{xcolor}
\usepackage{cancel}
\usepackage{ulem}
\usepackage{textcomp}
\newcommand{\mch}{$\mathrm{M_{Ch}}$}

\newcommand{\msun}{$\mathrm{M_{\odot}}$\ }
\newcommand{\msune}{$\mathrm{M_{\odot}}$}
\newcommand{\mdot}{$\mathrm{\dot{M}}$\ }
\newcommand{\mdote}{$\mathrm{\dot{M}}$}
\newcommand{\acef}{$\mathrm{\eta_{acc}}$\ }
\newcommand{\acefe}{$\mathrm{\eta_{acc}}$}

\def\sna{SN~Ia}
\def\sne{SNe~Ia}

\def\cwd{CO~WD}
\def\mwd{$M_{\rm WD}$}

\def\apgt{\ {\raise-.5ex\hbox{$\buildrel>\over\sim$}}\ }
\def\aplt{\ {\raise-.5ex\hbox{$\buildrel<\over\sim$}}\ }
\newcommand{\ms}{\mbox {$\mathrm M_{\odot}$}}
\newcommand{\pyr}{\mbox {{\rm yr$^{-1}$}}}
\def\apgt{\ {\raise-.5ex\hbox{$\buildrel>\over\sim$}}\ }
\def\aplt{\ {\raise-.5ex\hbox{$\buildrel<\over\sim$}}\ }
\newcommand{\rsun}{\mbox {$\mathrm{R_{\odot}}$}}
\newcommand{\ls}{\mbox {$L_{\odot}$}}

\newcommand{\myr}{\mbox {~${\rm M_{\odot}~yr^{-1}}$}}
\newcommand{\porb}{\mbox {$P_{\mathrm orb}$}}

\title[He-accreting WDs]
      {He-Accreting WDs: accretion regimes and final outcomes}
\author[L. Piersanti et al.]{L. Piersanti$^{1,4}$\thanks{E-mail:
piersanti@oa-teramo.inaf.it (LP); tornambe@oa-teramo.inaf.it (AT); lry@inasan.ru (LY)} 
and A. Tornamb\'e$^{2}$ and L.R. Yungelson$^{3}$\\
$^{1}$INAF-Osservatorio Astronomico di Collurania Teramo
              via Mentore Maggini, snc, 64100, Teramo, IT\\
$^{2}$INAF-Osservatorio Astronomico di di Roma
              via di Frascati, 33, 00040, Monte Porzio Catone, IT\\
$^{3}$Institute of Astronomy, Pyatnitskaya 48, 119017 Moscow, Russia\\
$^{4}$INFN-sezione di Napoli, 80126 Napoli, Italy}
\begin{document}

\date{}

\pagerange{\pageref{firstpage}--\pageref{lastpage}} \pubyear{2013}

\maketitle

\label{firstpage}

\begin{abstract}
The behaviour of carbon-oxygen white dwarfs (WDs) subject to direct helium 
accretion is extensively studied. We aim to analyze the thermal 
response of the accreting WD to mass deposition at different  
time scales. The analysis has been performed for 
initial WDs masses and accretion rates in the range (0.60 - 1.02)\,\ms\ 
and $(10^{-9} - 10^{-5})$\,\myr, respectively. 
Thermal regimes in the parameters space $\mathrm{M_{WD} - 
\dot{M}_{He}}$, leading to formation of red-giant-like structure, steady 
burning of He, mild, strong and dynamical flashes have been identified 
and the transition between those regimes has been studied in detail.
In particular, the physical properties of WDs experiencing the He-flash 
accretion regime have been investigated in order to determine the mass 
retention efficiency as a function of the accretor total mass and accretion 
rate. We also discuss to what extent the building-up of a He-rich layer via 
H-burning could be described according to the behaviour of models accreting 
He-rich matter directly. Polynomial fits to the obtained results are 
provided for use in binary population synthesis computations. Several 
applications for close binary systems with He-rich donors and CO WD 
accretors are considered and the relevance of the results for the 
interpretation of He-novae is discussed. 
\end{abstract}

\begin{keywords}
Binaries: general, Supernovae:general, White Dwarfs, Accretion
\end{keywords}

\section{Introduction}
\label{s:intro}

Accretion of helium onto carbon-oxygen white dwarfs (CO~WDs) plays an 
important role in several astrophysical processes. Most significantly, 
it may have relevance to the problem of the Supernovae Ia (\sne) 
progenitors. Among hypothetical evolutionary paths to \sne\ are 
semidetached close binary stars in which CO WDs grow in mass up to the 
Chandrasekhar mass limit (\mch) by accretion of matter directly from 
non-degenerate or degenerate helium-rich companions \citep[e.g.,][]
{ty96,yl03,sy05,wang09}, as well as explosions of sub-Chandrasekhar 
mass CO~WD via ``edge-lit'' or ``double-detonation'' mechanism, in 
which detonation in the He-layer at the surface of an accreting WD triggers 
detonation of CO-accretor via shock waves that compress the latter 
\citep[e.g.,][]{nomoto80,nom82b,livne1990,lg91,lt91,ww94,la95,
garciasenz99,fink07,sim2010,fink2010,woosley_kasen_dd11,schwab2012,
townsley12,moll_woosley_dd13,shen_bildsten_dd13,moore_dd13}.   

Recent modification of the classical ``double-degenerate'' scenario 
\citep{web84,iben1984} envisions ``violent'' or ``prompt'' merger so 
that detonation is initiated at the interface of two merging WD in 
C-O or He-C-O mixture and determines the final complete burning of 
both WDs \citep[e.g.,][]{pakmor2010,pakmor2011,pakmor_violent12,
pakmor_violent_13,kromer_violent13,moll_prompt13}. It is worth noting 
that some authors questioned whether carbon detonation is really prompt 
\citep[e.g.,][]{raskin_remnants12}. Moreover it has been shown that 
the location of initial explosion does depend on the numerical 
resolution as well as on the initial configuration adopted in the 
computations \citep{dan2012,dan2013}.
Double-detonation and violent merger scenarios are currently considered 
as alternative or complementary to the ``classical'' single-degenerate 
and double-degenerate scenarios for \sne. For a review of the \sne\ 
progenitors and the observational constraints to theoretical models see, 
e.g., \citet{hillen13,hoeflich_snia_2013,mmn_snia_rev13,postnov14}, 
while for the 
recent results of binary population synthesis (BPS) for \sne\ from 
different channels see \citet{ruiter09,wang09,mennekens10,ruiter2011,
toonen12,nele2012,ruiter_13_violent,wang13}.

For the double-detonation scenario, the onset of the initial He-detonation 
depends on the accretion rate and the mass of He retained at the WD 
surface\footnote{We do not consider here He-detonation initiated by 
instability in the accretion flow \citep{Guillochon_10}.}. For both 
scenarios the percentage of the He-rich matter transferred from the donor 
and effectively retained by the accreting WD and nuclearly processed into C/O 
rich mixture 
up to either the He-detonation or the merger may be important since 
simulations show that physical conditions adequate to reproduce normal \sne\ 
are more favourable in massive accretors -- $M_{\rm WD} \apgt 0.8$\,\ms\ 
for double detonation \citep{lg91,fink2010,kromer2010,piro2013} and 
$M_{\rm WD} \apgt 0.9$\,\ms\ for violent mergers \citep{pakmor_violent_13,
kromer_violent13}. This means that the transfer of He-rich matter onto 
new-born WD and its conversion into C and O may be necessary, as CO~WDs 
typically form with lower masses. 

In the ``classical'' single-degenerate scenario for Chandrasekhar mass 
\sne\ retention efficiency of He is a crucial parameter, since the most 
important physical process involved is the nuclear burning of H into He 
and then into C-O-Ne mixture \citep[see, e.g.,][for recent discussion]
{bours_retention_13}. Retention efficiency of helium and He-burning 
regimes are, for instance, also important for formation and evolution 
of AM CVn stars \citep{nyp01a}, the origin of faint and fast transients 
possibly associated with single He-layer detonation at the surface of 
accreting WDs, like still hypothetical SNe~.Ia \citep[see, e.g.,][]
{bild2007,shen_.ia10,wald2011,woosley_kasen_dd11,sim2012,
raskin_remnants12,shen_bildsten_dd13,kasliwal_bridge12} and, possibly, 
some subluminous \sna\ \citep{wang13}. Other problems include, e.g., 
the interpretation of events suggested to be ``Helium Novae'' 
\citep{ashok03b,rosenbush_he_nov08}. 

The large majority of studies devoted to the thermal response of CO 
WDs accreting He-rich matter, focused on defining physical conditions 
for the onset of either a very strong He-flash or a He-detonation. 
{\citep{sufu1978,nks1979,nns1980,taam1980a,taam1980b,
nomoto1982a,nom82b,fu_su1982,wtw1986,nh1987,iben1991,lt91,
ww94,pier2001,bilds2006,shen2007,shen2009,woosley_kasen_dd11}.
As a result, various accretion regimes were roughly defined, depending 
both on the accretion rate and on the CO WD mass. In particular, 
\cite{nomoto1982a} suggested that for a high value of the accretion rate 
\mdote, but still below the Eddington limit, an extended He-rich layer is 
piled-up and WD expands to giant dimensions. The minimum value of \mdot 
for this is defined by the rate at which He burns into CO-rich matter 
at the base of the He-envelope. Basing on the computations of massive 
pure-He stars by \citet{uus1970}, Nomoto derived an analytical formula 
for such a limiting value:
\begin{equation}
\mathrm{
\left({\frac{dM}{dt}}\right)_{\rm RHe}=7.2\times 10^{-6}\left({M_{\rm CO}}-0.60\right)\ \ \ \ \ \ \
(in\ M_\odot yr^{-1})},
\end{equation}
where $\mathrm{M_{CO}}$ is mass of the CO core in \ms, varying in the 
range (0.75 -- 1.38)\,\msune. 
For \mdot lower than $\sim 10^{-6}$ \myr, but larger than 
$\sim 10^{-7}$ \myr, He-burning is stable and the CO core increases 
in mass steadily \citep{iben1989}. For lower values of the accretion rate, 
but larger than $\sim 3\times 10^{-8}$\myr, He-burning proceeds 
through recurrent flashes, whose strength increases with decreasing \mdote, 
for a fixed mass of the CO core. In this case the He-flashes are too weak 
to develop any dynamical effects, even if the large energy release can 
trigger the expansion of the accreting structure and, hence, the interaction 
with its binary companion \citep{taam1980a,fu_su1982}. Finally, for \mdot
$\le 3\times 10^{-8}$\myr\ a dynamical He-flash occurs 
when a critical amount of He-rich matter has been accreted. 
\citet{woosley_kasen_dd11} investigated in great details the physical 
properties of He-accreting WDs at low accretion rates and define very accurately 
the transition from novalike He-flashes to He-detonation (see their Figure 19). 
They demonstrated that the He-layer above the CO core could detonate only if 
the density at the ignition point is larger than a critical value, around 
${\rm 10^6\,g\,cm^{-3}}$.

\citet{kato2004} (hereinafter KH04) calculated the retention efficiency 
\acef of He-accreting WD for CO cores in the range (0.6 -- 1.3)\,\ms\ and 
$3\cdot 10^{-8}\le \dot{M}\le 1.5\cdot 10^{-6}$\,\myr. 
\acef is defined as the ratio of the mass effectively accumulated onto 
the CO core after one full He-flash driven cycle and the amount of matter 
transferred from the donor during the same cycle. The computations by 
KH04 were based on the ``optically thick wind theory'', which presumably 
describes the continuum-radiation driven wind operating very deeply 
inside the photosphere \citep{kato1994}. Even if 
a huge efforts have been applied to investigate the thermal response of 
CO WDs accreting He-rich matter, an overall picture is still missing.
In particular, in the regime characterised by recurrent He-flashes the 
possibility that accreting objects could overfill their Roche lobe has 
not been explored so far, despite this issue is of a paramount importance 
to determine the long-term evolution of binaries harbouring He-donors. 
The present work is aimed to the systematic analysis of the parameter 
space $M_{\rm WD} - \mathrm{\dot{M}}$, considering fully evolutionary models 
of CO WDs with initial mass in the range (0.6 -- 1.05)\,\msune\footnote{1.02\,
\ms\ is the maximum mass of the new-born CO core at the beginning of the 
TP-AGB phase in our evolutionary code. If the core exceeds this limit, 
C-burning will be ignited at the centre or off-centre, depending on the 
total mass of the core. By assuming that during the AGB phase the CO 
core almost does not grow in mass, 1.05\,\ms\ represents the maximum 
mass for a CO WD.} and accretion rates $10^{-9}$ -- $10^{-5}$\myr. 
Basing on our results, we intend to define the limits of different 
accretion regimes and the possible outcomes of the accretion process. 
Moreover, at variance with previous studies, we investigate the effects 
of the previous evolution, namely of the accretion history, on the actual 
thermal response of accreting WDs. 
\begin{table*}  
\caption{Physical properties of the initial CO WDs ({\sl Cool Models}) 
         and after the first mass transfer episode ({\sl Heated Models}). 
         We list the total mass of the model $\mathrm{M_{tot}}$, the 
         mass fraction abundance ratio of carbon over oxygen at the center 
         C/O, the mass extention of the He-deprived core $\mathrm{M_{CO}}$, 
         the mass extention of the more external layer where the helium 
         abundance by mass fraction is larger than 0.05 $\mathrm{\Delta M_{He}}$,
         the temperature $\mathrm{T_c}$ in K and the density $\mathrm{\rho_c}$ in 
         $\mathrm{g\,cm^{-3}}$ at the center, the surface luminosity $\mathrm{L}$, 
         the effective temperature $\mathrm{T_{eff}}$ in K and the 
         surface radius $\mathrm{R}$.
         For more details see text.}
\label{t:initial} 
\centering 
 \begin{minipage}{160mm}
  \begin{tabular}{l r r r r r | r r r r r }
   \hline 
   {\sl }  & \multicolumn{5}{c} {\sl Cool Models} & 
             \multicolumn{5}{c} {\sl Heated Models} \\ \hline                    
   LABEL                     & \multicolumn{1}{c}{M060}  & \multicolumn{1}{c}{M070}  & 
   \multicolumn{1}{c}{M081}  & \multicolumn{1}{c}{M092}  & \multicolumn{1}{c}{M102}  &    
   \multicolumn{1}{c}{M060}  & \multicolumn{1}{c}{M070}  & \multicolumn{1}{c}{M081}  &    
   \multicolumn{1}{c}{M092}  & \multicolumn{1}{c}{M102}  \\
   $\mathrm{M_{tot}}$ (in \msune)& 0.60000 & 0.70000 & 0.80833 & 0.91962 & 1.02061 & 0.59678 &  0.70185 &  0.81033 &  0.91897 &  1.02048 \\
   $\mathrm{C/O}$            & 0.3839  & 0.4308  & 0.5396  & 0.5371  & 0.4870  & 0.3839  &  0.4308  &  0.5396  &  0.5371  &  0.4870  \\
   $\mathrm{M_{CO}}$ (in \msune)& 0.5158  & 0.6580  & 0.7921  & 0.9114  & 1.0157  & 0.5173  &  0.6580  &  0.7921  &  0.9116  &  1.0158  \\
   $\mathrm{\Delta M_{He}}$ (in $10^{-2}$\msune) & 3.37  & 1.40  & 0.73  & 0.39  & 0.18  & 1.62  &  0.67  &  0.30  &  0.13 &  0.06 \\
   $\mathrm{\log(T_c)}$      & 7.8647  & 7.8644  & 7.8759  & 7.9915  & 8.0798  & 7.8318  &  7.8495  &  7.8593  &  7.9012  &  7.9528  \\
   $\mathrm{\log(\rho_c)}$   & 6.4981  & 6.7518  & 7.0219  & 7.2792  & 7.5403  & 6.4552  &  6.7387  &  7.0187  &  7.2995  &  7.5779  \\
   $\mathrm{\log(L/L_\odot)}$& 0.5358  & 0.5637  & 0.4881  & 1.3760  & 2.7042  & 3.3358  &  3.7567  &  4.0224  &  4.2488  &  4.4`87  \\
   $\mathrm{\log(T_{eff})}$  & 4.7996  & 4.8424  & 4.8606  & 5.0989  & 5.4154  & 5.2821  &  5.4382  &  5.5655  &  5.6850  &  5.7789  \\
   $\mathrm{\log(R/R_\odot)}$&-1.8083  &-1.8799  &-1.9542  &-1.9868  & -1.9557 &-1.3745  & -1.4752  & -1.5969  & -1.7235  & -1.8254  \\
   \hline                  
  \end{tabular}
\end{minipage}
\end{table*}
\begin{figure*}   
 \centering
  \includegraphics[width=\textwidth]{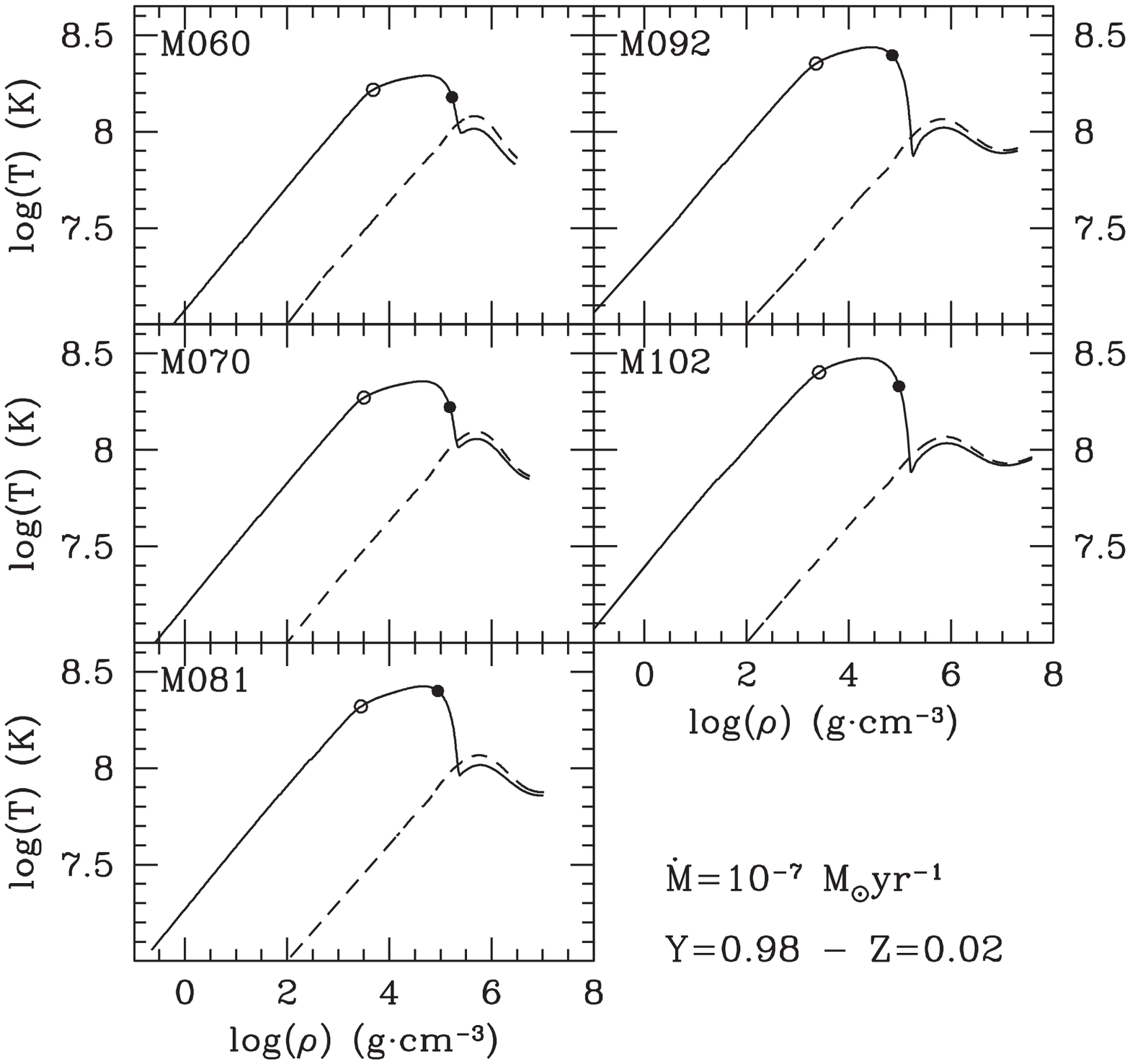}
  \caption{Profiles in the $\rho - T$ plane for the {\sl Cool} (dashed lines) 
           and {\sl Heated Models} (solid line). Each panel refers to one CO 
           WD as labelled. In the profiles of {\sl Heated Models} we mark 
           with a filled circle the He/CO interface and with and open circle 
           the He-burning shell.} 
  \label{f:map}
\end{figure*}

In \S~\ref{s:code} we describe our evolutionary code and the input physics. 
We also present the main properties of the initial CO WD models. In 
\S~\ref{s:regimes} we present our results and define various accretion 
regimes. In \S~\ref{s:c_igni} we discuss under what conditions central 
C-ignition could occur in a Chandrasekhar-mass WD, thus triggering an 
explosion. \S~\ref{s:h_eff} is devoted to the accumulation efficiency in 
models accreting directly He-rich matter and as a by-product of H-burning 
in H-accreting WDs. In \S~\ref{s:applic} we discuss the formation of He-rich 
donors in close binaries and we analyze the applications of our results to 
several types of systems in which stable accretion of pure He onto a CO WD 
may occur. Our final considerations are reported in \S~\ref{s:final}.

\section{Input physics and numerical methods}
\label{s:code}

All the models presented in this study have been computed with an updated 
version of the F.RA.N.E.C code, the original version being described in 
\cite{chieffi1989}. The setup of the code is the same as in \cite{pier2007}; 
in particular, we consider solar chemical composition models adopting 
the solar mixture GS98 derived in \cite{pier2007}. Tables of radiative 
opacities for temperature lower than $5\times 10^8$ K have been derived 
through the web facility provided by the OPAL group 
\citep[http://opalopacity.llnl.gov/new.html\ see][]{igles1996}, while at 
higher temperatures we adopt the tables derived from the Los Alamos 
Opacity library \citep{hueb1977}. The contribution of electron conduction 
to the total opacity has been included according to the prescription by 
\cite{pote1999}. We adopt the equation of state computed by \cite{stra1988} 
and successive upgrades \citep{prada2002}. Matter is accreted by assuming 
that it has the same specific entropy as the external layers of the CO WD; 
this is equivalent to the assumption that the energy excess is radiated 
away \citep[][and\ references\ therein]{pier2000}. We fix the chemical 
composition of the accreted matter to X=0, Y=0.98, Z=0.02, where X, Y 
and Z are the mass fraction abundance of hydrogen, helium and metals, 
respectively. Since the He-donor has formed via nuclear burning of H-rich 
matter, we assume that all the initial CNO elements have been converted 
into ${^{14}\mathrm{N}}$ (i.e., 
$\mathrm{Y^{i}_{^{12}C}+Y^{i}_{^{13}C}+Y^{i}_{^{14}N}+Y^{i}_{^{15}N}+
         Y^{i}_{^{16}O}+Y^{i}_{^{17}O}+Y^{i}_{^{18}O}=Y^{f}_{^{14}N}}$ 
where $\mathrm{Y_{j}}$ is the abundance by number of the $j$-isotope 
and the subscripts $i$ and $f$ refer to the initial MS star and the 
final He-donor star, respectively). For all the elements heavier than 
oxygen we assume a scaled-solar chemical composition.

The adopted nuclear network includes elements from H to Fe, linked by 
$\alpha$-, $p$- and $n$-capture reactions as well as $\beta^{\pm}$ 
decays. We do not consider the NCO chain ($\mathrm{^{14}N(e^-,\nu)^{14}C
(\alpha,\gamma)^{18}O}$) because it has been demonstrated that its 
contribution to the energy budget does not have a sizable effects 
on the physical properties of accreting WDs \citep{pier2001}.

The convective mixing is modelled by means of the time dependent 
algorithm introduced by \citet{sparks1980} and successively modified 
by \citet{stran2006} \citep[for a recent review see also][]{stran2014}.
In this case the degree of convective mixing between two points inside 
the convective zone depends on the corresponding turnover timescale.

In order to obtain the initial CO WD models we evolved intermediate mass 
stars from the pre-main sequence to the cooling sequence, by assuming 
that they are components of interacting binary systems. For models M060 
and M070 we assumed that their progenitors experienced a common envelope 
episode during the red giant branch phase, while for progenitors of models 
M081, M092 and M102 we assumed a Roche lobe overflow (RLOF) in the 
hydrogen-shell burning stage while they possessed radiative envelopes. 
Some physical properties of the initial WD models are summarised in 
Table~\ref{t:initial} ({\sl Cool Models}).

The deposition onto the {\sl Cool Models} of He-rich matter at a relatively 
high rate (let say, \mdot$>(5-10)\times10^{-8}$ \myr) determines 
the heating of the physical base of the He-rich layer, because the adopted 
initial models are compact and cool (their luminosity is definitively 
lower than that of a post-AGB star of the same mass). Hence, the compressional 
heating timescale at the surface of the accretor is definitively lower than 
the inward thermal diffusion timescale. As a consequence the temperature 
at the base of the He-rich layer increases and He-burning is ignited when the 
local temperature attains $\mathrm{\sim 10^{8}}$\,K. However, due to the 
partial degeneracy at the ignition point\footnote{We define the ignition 
point as the mass coordinate where the nuclear production via 3$\alpha$-reactions 
is at a maximum at the epoch when the energy per unit of time delivered 
by He-burning is 100 times the surface luminosity. This implies that at 
the ignition point $\mathrm{\varepsilon_{nuc} > \varepsilon_{\nu}}$.}, 
a thermonuclear runaway occurs, even if the resulting flash does not 
become dynamical. Accreting WD reacts to the injection of a huge amount 
of nuclear energy by expanding to giant dimensions ($\mathrm {R_{WD}>
100\,R_\odot}$). When considering that these objects are components of 
binary systems, it turns out that they have to experience a RLOF, losing 
part of the matter previously accreted. 

We simulated this first flash episode by adopting \mdot$=10^{-7}$\myr.
Moreover, for all the considered models we fix the radius of the Roche 
lobe to $\mathrm{R_{Roche}=10 R_\odot}$. It is worth noticing that, even 
if such an assumption is completely arbitrary, the physical properties 
of the final structure do not depend on the exact value of $\rm R_{Roche}$, 
since the expansion to giant dimensions occurs on a very short timescale, 
smaller that the nuclear timescale of the He-burning shell\footnote{The 
He-burning shell is defined as the mass coordinate where the energy 
production via 3$\alpha$ reactions is at a maximum.} and of the inward 
thermal diffusion timescale. 
\begin{figure}  
 \centering
  \includegraphics[width=\columnwidth]{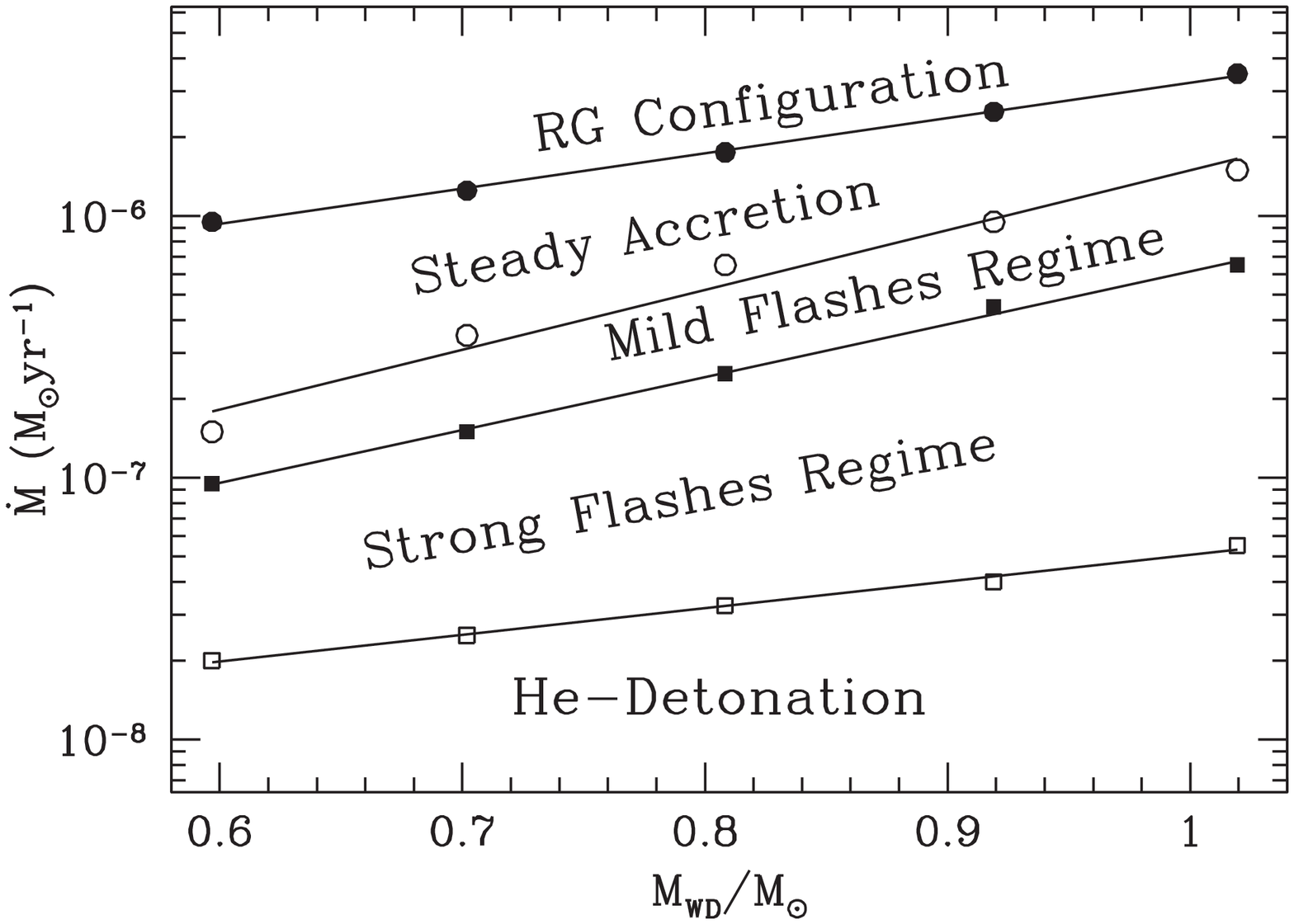}
  \caption{Possible accretion regimes as a function of the WD initial mass 
           and accretion rate. Solid lines represent the transition 
           between different accretion regime as obtained by linearly 
           interpolating the results of our computations (open and filled 
           symbols). Interpolation formulae are presented in Appendix 
           \ref{app1}.}
  \label{f:regime}
\end{figure}

We find that during this first mass transfer episode, 
the amount of material effectively deposited onto the WDs is practically 
zero (in some cases the pre-existing He-rich zone is also eroded). The only 
effect of the first He-flash is just to modify the physical properties of 
the He-rich layer (mainly temperature and density profiles), so that the He-shell 
attains the conditions for steady burning. This is clearly depicted in Fig. 
\ref{f:map} where we show, for each considered model, the profile in the 
$\rho-T$ plane for WDs at the beginning of the first mass transfer episode 
(dashed lines) and after the RLOF episode, when the WDs has attained its maximum 
effective temperature at the beginning of the cooling sequence (solid lines). 
Henceforth, we 
address these models with modified thermal structure of the envelope as 
{\sl Heated Models}. In Table~\ref{t:initial} we report also some relevant 
physical quantities referring to each model after the first He-flash episode.
The CO core is defined as the portion of the star where the helium abundance is 
lower than $10^{-20}$ by mass fraction, while the He-burning shell is defined 
as the mass coordinate where He-burning is at maximum. The physical and chemical 
conditions for $\rm \varepsilon_{nuc}$ is a maximum are far from the He/CO 
interface, but close to the mass coordinate where He abundance is $\sim$ 0.05 
by mass fraction. The He-burning shell does not correspond to the maximum 
temperature because, after the RLOF episode, it moves outward, where temperature 
is lower, while the new-synthesized CO layer contracts and heats up.

\section{Accretion Regimes}\label{s:regimes}

We accrete He-rich matter directly onto {\sl heated} CO WDs (see 
previous section) by adopting values for the accretion rate in the 
range $\mathrm{10^{-9}\le \dot{M} \le 10^{-5} M_\odot yr^{-1}}$. The 
starting point for all the computations is defined along the high 
luminosity branch, during the blueward evolution of the 
post-first-flash steady-state structure.

Our results are summarised in Fig.~\ref{f:regime} where we show the possible 
accretion regimes as a function of the WD initial mass and accretion rate. 
The lines marking the transition from one accretion regime to another have 
been obtained by considering the thermal response of {\sl Heated Models} to 
the mass transfer. For example, for the model M081 we fix the transition 
from the Steady Accretion to the Mild Flashes at \mdote=$\mathrm{6.5\times 
10^{-7}}$\myr, since for \mdote=$\mathrm{7\times 10^{-7}}$\myr\ the model 
is in a steady state while for \mdote=$\mathrm{6\times 10^{-7}}$\myr\ it 
experiences recurrent mild flashes (see below for the definitions of 
different regimes).

\subsection{Dynamical He-Flashes regime }
\label{s:detain}

At low accretion rates, the physical base of the He-rich mantle cools down 
and becomes degenerate. Later on, due to the continuous deposition of matter, 
it heats-up and He-burning is ignited. Owing to the degeneracy at the 
ignition point, the nuclear energy delivered by the $3\alpha$-reactions 
is  stored locally producing the increase of temperature and causing a 
thermonuclear runaway (He-flash). However, the degeneracy level at the 
ignition point is only a necessary condition to trigger a dynamical 
burning event. Indeed, the He-flash triggers the formation of a 
convective shell which extends outward as the temperature increases at 
the base of the He-rich zone. Convective shell injects fresh He in the 
burning zone fuelling the thermonuclear runaway and, hence, the 
continuous increase of the local temperature. As a matter of fact, if 
the mass of the He-rich layer (and, hence, the total He abundance by mass) 
is too small, the propelling effect of convective mixing rapidly 
exhausts, the flash quenches and the accreting WD expands (for a more 
detailed discussion see \S~\ref{s:strongfl}).
                                   
For accretion rate lower than $\sim 5\times 10^{-8}$\myr\ the whole 
structure becomes isothermal very rapidly, independently of the initial 
temperature profile, since the inward thermal diffusion timescale is 
very small as compared to the accretion timescale ($\mathrm{\tau_{diff}
\sim 10^5}$ yr, while $\mathrm{\tau_{acc} > 10^7}$ yr). This also implies 
that the energy delivered by the mass deposited on the surface of the WD 
can not produce a local increase of temperature. As a consequence, the 
nature of the final He-flash (explosion or not) depends only on the 
interplay between the neutrino cooling of the physical base of the He-shell 
and the heating driven by the compression of the whole structure. Since 
the latter quantity depends on the mass growth timescale, it turns out 
that, for a fixed total mass of the initial WD, the nature of the final 
He-flash depends only on the accretion rate: if it 
is smaller than a critical value $\mathrm{\dot{M}(He)_{expl}}$, the 
He-shell becomes strongly degenerate and the resulting flash turns into an 
explosion.
\begin{table} 
\caption{Models experiencing dynamical He-flashes. For each model we list as a function of 
         the accretion rate $\rm \dot{M}$ (in $10^{-9}$\myr), the final mass 
         $\rm M_{fin}$  in \msune\, the accreted mass $\rm M_{acc}$ in \msune, 
         the accretion time $\rm T_{acc}$  in $10^6$ yr, the mass coordinate 
         where He-burning is ignited $\rm M_{ig}$  in \msune, the temperature 
         $\rm T_{ig}$  in $10^7$ K and density $\rm \rho_{ig}$ in $10^{6}\,
         \mathrm{g\,cm^{-3}}$ when He-burning is ignited. The last column 
         gives the mass of the zone where helium abundance by mass fraction is 
         larger than 0.01 $\rm \Delta M_{He}^{pk}$ in 
         \msune. A more detailed table with small increments in $\rm \dot{M}$ 
         is available online.} 
\label{t:detona}
\begin{tabular}{r r r r r c c c }
\hline
 $\rm \dot{M}$ & $\rm M_{fin}$   & $\rm M_{acc}$ & $\rm T_{acc}$ & $\rm M_{ig}$ & 
 $\rm T_{ig}$  & $\rm \rho_{ig}$ & $\rm \Delta M_{He}^{pk}$ \\
\hline
\multicolumn{8} c {M060} \\
  1 & 1.255 & 0.658 & 658.4 & 0.587 & 5.246 & 72.061 & 0.737 \\
  5 & 1.092 & 0.495 &  99.1 & 0.601 & 7.579 & 13.452 & 0.571 \\
 15 & 0.971 & 0.374 &  24.9 & 0.608 & 8.619 &  4.995 & 0.448 \\
\multicolumn{8} c {M070} \\
  1 & 1.277 & 0.575 & 698.0 & 0.698 & 5.124 & 75.664 & 0.616 \\
  5 & 1.123 & 0.421 &  84.2 & 0.715 & 7.648 & 12.802 & 0.457 \\
 20 & 0.913 & 0.211 &  10.6 & 0.735 & 9.442 &  1.839 & 0.252 \\
\multicolumn{8} c {M081} \\
 1.5 & 1.263 & 0.453 & 301.7 & 0.818 & 6.071 & 45.067 & 0.461 \\
 6   & 1.144 & 0.333 &  55.6 & 0.825 & 7.851 & 10.913 & 0.342 \\
25   & 0.912 & 0.102 &  4.1 & 0.826 & 9.947 & 0.929  & 0.111 \\
\multicolumn{8} c {M092} \\
 1.5 & 1.301 & 0.383 & 255.2 & 0.918 &  5,799 & 59.903 & 0.392 \\
 7   & 1.180 & 0.262 &  37.4 & 0.930 &  7,844 & 10.624 & 0.271 \\
30   & 0.963 & 0.045 & 1.491 & 0.922 & 10.534 & 0,590  & 0.053 \\
\multicolumn{8} c {M102} \\
 1.5 & 1.318 & 0.299 & 199.7 & 1.019 &  5.883 & 92.234 & 0.305 \\
 8   & 1.204 & 0.185 &  23.1 & 1.028 &  8.075 &  8.791 & 0.190 \\
50   & 1.039 & 0.020 &   0.4 & 1.021 & 11.148 &  0.427 & 0.025 \\
\hline
\end{tabular}
\end{table}

In our computations, following \cite{bild2007} and \cite{shen2009}, 
we check the onset of a dynamical flash by comparing $\rm \tau_{heat}$, 
the heating timescale due to release of nuclear energy, and 
$\rm \tau_{dyn}$, the dynamical timescale, at the He-burning shell. 
If $\rm \tau_{heat}$ becomes smaller than $\rm \tau_{dyn}$, the 
He-rich zone above the He-burning shell can not readjust to a new 
equilibrium configuration and, hence, their evolution decouples, 
driving to the formation at the interface of an overpressure which 
triggers the explosion. These two characteristic timescales have 
been computed according to the following relations:
\begin{eqnarray}
\mathrm{\tau_{heat}={\frac{C_P T}{\varepsilon_{nuc}}},}\nonumber\\
\mathrm{\tau_{dyn}={\frac{H_P}{v_{sound}}},}
\end{eqnarray}
where $\rm C_P$ is the specific heat at constant pressure, T --- the 
temperature, $\rm \varepsilon_{nuc}$ --- the rate of energy production 
via nuclear burning, $\rm H_P$ --- the pressure scale height and $\rm 
v_{sound}$ --- the local value of the sound velocity. 
\begin{figure}  
 \centering
  \includegraphics[width=\columnwidth]{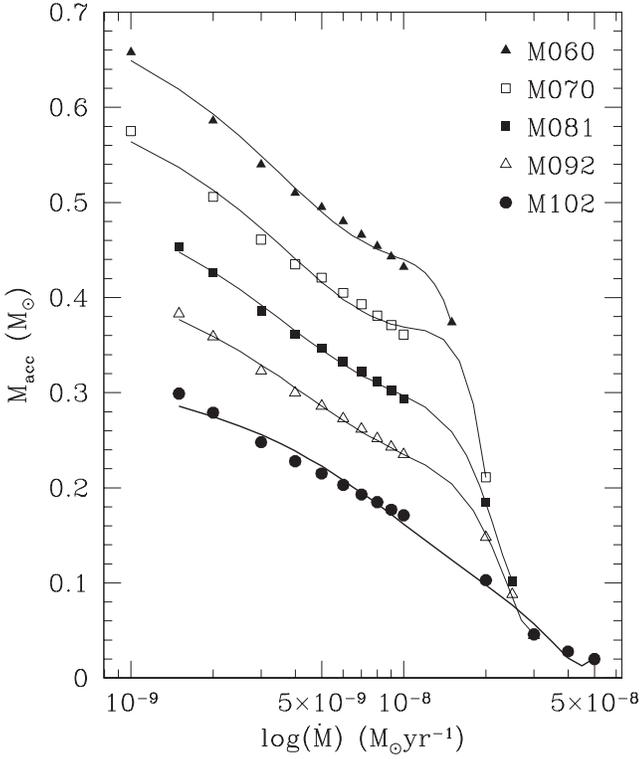}
  \caption{Accreted mass as a function of the accretion rate for models 
           experiencing a dynamical He-flash. Each set of symbols refers 
           to a different initial model, as labelled in the figure. 
           Solid lines represent polynomial fits to the data 
           (see Appendix \ref{app2}).}
  \label{f:detmas}
\end{figure}

Our results are summarised in Table~\ref{t:detona}, where we report, 
as a function of the accretion rate, the total final mass ($\rm M_{fin}$), 
the accreted mass ($\rm M_{acc}$) and the corresponding accretion time 
($\rm T_{acc}$), the mass coordinate of the point where He-flash is ignited 
($\rm M_{ig}$) and the value of temperature ($\rm T_{ig}$) and density 
($\rm \rho_{ig}$) at the ignition point for some representative computed 
models. In the last column we report $\rm \Delta M_{He}^{pk}$, defined as 
the total mass of the layer where He abundance is larger than 0.01 by 
mass fraction. Note that the latter is slightly 
larger than the accreted mass since the initial WD models are capped by 
a He-rich layer, determined by the previous evolution, which is much 
smaller than the mass to be accreted 
for igniting a dynamical He-flash (see \S~\ref{s:code}). 
Polynomial fits of $\mathrm{M_{acc}}$ as a function of the accretion rate
for all the models displayed in Fig. \ref{f:detmas} are provided in  
Appendix \ref{app2}. 
Where comparable, for M$_{\rm WD}$=(0.6 -- 1.0)\,\ms, minimum accreted 
masses necessary for a dynamical He-flash are in very good agreement, within factor 
of less than 2, with the estimates obtained by \citet{shen_.ia10}. As 
well, in agreement with the latter study we find that the ignition mass 
depends on \mdote. In Fig. \ref{f:detmas} we present the accreted mass 
as a function of the accretion rate for each initial WD model. 
At variance with \citet{nomoto1982a} and in agreement with \citet{shen_.ia10}, 
we find that the maximum value of the accretion rate still producing a 
dynamical He-flash slightly depends on the WD mass (see Fig.~\ref{f:regime}). 
By interpolating the data reported in Table~\ref{t:detona} we obtained 
(in \myr )\footnote{Equation \ref{eq:detonation} gives the 
\textit{highest} value of $\rm \dot{M}_{He}$ for which in our computations 
He definitely detonates, while the limits shown in Fig.~\ref{f:regime} 
and the corresponding formulae in Appendix \ref{app1} represent results of 
interpolation between models experiencing different burning regimes, 
because of finite resolution of the adopted accretion rates grid.}
\begin{equation}
\mathrm{
\log(\dot{M}(He)_{expl})\approx (1.15\pm 0.11) {\frac{M_{WD}}{M_\odot}}-(8.52\pm 0.07). 
}
\label{eq:detonation}
\end{equation}

\begin{table}  
\caption{The same as in Table~\ref{t:detona}, but for model M092 accreting 
         He-rich matter at $\rm\dot{M}=(2.75\cdot e^{-{\frac{t}{\tau}}}+0.15)\dot 10^{-8}$
         \myr, with different values of the characteristic timescale 
         $\tau$, as listed in the first column in $\mathrm{10^{6}}$ yr. 
         See text for more details.}
\centering
\label{t:tau}
\begin{tabular}{r r r r r c c c }
\hline\hline
 $\rm \tau$ & $\rm M_{fin}$ & $\rm M_{acc}$& $\rm T_{acc}$ & $\rm M_{ig}$ & 
 $\rm T_{ig}$ & $\rm \rho_{ig}$ & $\rm \Delta M_{He}^{pk}$ \\
\hline
 4 & 1.300 & 0.382 & 181.239 & 0.918 & 5.823 & 58.871 & 0.393 \\
 6 & 1.298 & 0.380 & 143.235 & 0.918 & 5.870 & 56.824 & 0.391 \\
 8 & 1.290 & 0.372 & 101.557 & 0.918 & 6.040 & 49.957 & 0.383 \\
 9 & 1.259 & 0.341 &  62.223 & 0.918 & 6.606 & 30.342 & 0.351 \\
10 & 1.140 & 0.222 &  13.316 & 0.928 & 8.259 &  6.808 & 0.233 \\
15 & 0.995 & 0.077 &   2.903 & 0.928 & 9.912 &  1.053 & 0.085 \\
20 & 0.984 & 0.066 &   2.403 & 0.926 & 0.096 &  0.882 & 0.074 \\
\hline
\end{tabular}
\end{table}

In order to investigate the dependence of our results on the history 
of the accretion rate we computed an additional set of models, by 
adopting as initial CO WD the M092 model and a time-dependent accretion 
rate, given by: 
\begin{equation}
\mathrm{
\dot{M}=(2.75\cdot e^{-{\frac{t}{\tau}}}+0.15)\times 10^{-8} M_\odot yr^{-1},
}
\end{equation}
where $\tau$ is a characteristic timescale varying in the range (4 - 20) 
Myr. The results are summarised in Table~\ref{t:tau}, where we list the 
same quantities as in Table~\ref{t:detona}. 
\begin{figure}  
 \centering
  \includegraphics[width=\columnwidth]{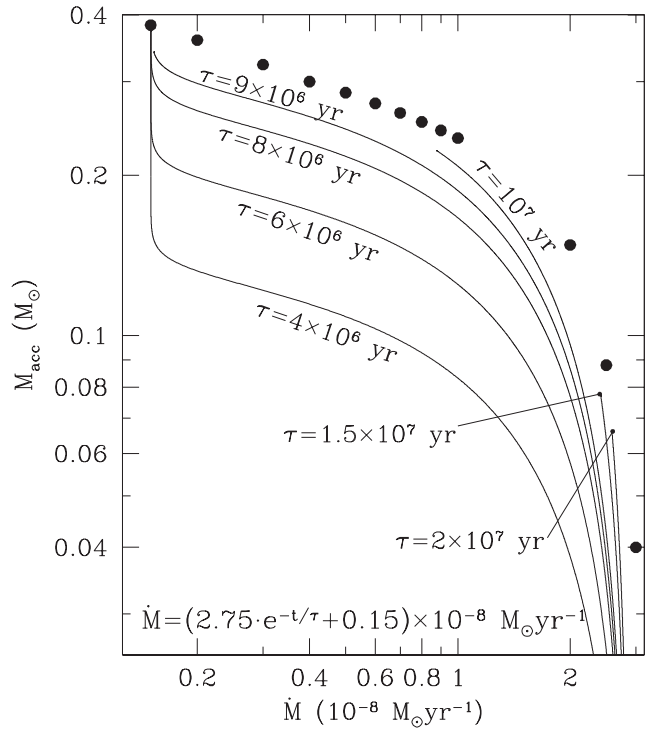}
  \caption{The same as in Fig. \ref{f:detmas}, but for the M092 CO WD 
           accreting He-rich matter with an exponentially decaying 
           accretion rate, as reported in the Figure. Different 
           models have different decay timescale, as labelled. Heavy 
           dots represent total accreted mass for models with constant 
           accretion rate.}
  \label{f:tau}
\end{figure}

In Fig. \ref{f:tau} we plot the amount of accreted mass as a function 
of the accretion rate for the models listed in Table~\ref{t:tau} and, 
for comparison, we also show the total accreted mass of models with 
constant accretion rate (filled dots). As it can be noticed, the mass 
accreted before the onset of the dynamical He-flash 
depends both on the accretion rate and on the accretion history. In 
particular Fig. \ref{f:tau} reveals that the leading parameter 
determining $\rm M_{acc}$ is the actual value of \mdot at the onset 
of the He-flash, while the previous mass transfer process has a minor 
role, just reducing by less than 10\% the value of the accreted mass. 
In any case, if \mdot decreases very rapidly, the accreting WD loses 
memory of the previous evolution and the evolution to the explosion 
occurs exactly as it would if \mdot was kept from the very beginning 
constant and equal to its final value. 

\subsection{Strong Flashes Regime}\label{s:strongfl}
For slightly larger values of the accretion rate the He-flash does not 
become dynamical, as the nuclear timescale at the base of the He-rich layer  
never approaches $\mathrm{\tau_{dyn}}$. However, the released energy is 
huge, so that an expansion to giant dimensions occurs. 
In order to illustrate the evolution of models experiencing this accretion 
regime we show in Fig. \ref{f:hr_sf} the evolution in the HR diagram of 
the model M102 accreting He-rich matter at $3\times 10^{-7}$ \myr. We 
mark by filled dots and letters some crucial moments during the evolution. 
In our analysis we follow the approach used by \cite{pier2000} to discuss 
the evolution of H-accreting models (see also the references therein). 
The evolution is counterclockwise along the track and we start our 
discussion from point A in Fig.~\ref{f:hr_sf}, the bluest point of the 
whole track. This point represents a bifurcation in the evolution of 
He-accreting models. At this stage the He-burning shell is very close 
to the surface and cool and, in addition, the mass of the He-rich zone  
has been reduced below a critical value at which release of nuclear 
energy exceeds release of the gravothermal one. Hence, the energy 
production via He-burning rapidly decreases and the gravothermal energy 
becomes the main source of energy\footnote{The reasons for this 
are discussed in \cite{iben1982}.}. During the following evolution the 
model approaches the WD radius appropriate to its mass, while the 
He-burning luminosity continues to decrease and after $\rm \Delta T_{AB}
\simeq 1555$ yr it attains its minimum ($\mathrm{L_{He}=2.19\times 
10^{33}\ erg\ s^{-1}}$) at point B. 

During the following evolution which lasts for $\rm \Delta T_{BC}\simeq 
5476$ yr, due to the continuous deposition of matter, the physical 
base of the He-shell begins to heat up and nuclear burning via 
3$\alpha$-reactions gradually resumes. Note that, even if the degeneracy 
of the He-burning shell is not high, the local nuclear timescale is shorter 
than the timescale for the thermal response of the star to a structural 
change. As a consequence, He-burning turns into a flash. Very soon, 
the flash triggers the formation of a convective shell (point C in Fig. 
\ref{f:hr_sf}) which rapidly increases in mass, attaining very soon 
the surface (point D). It is important to remark that also the inner border of 
the convective zone moves inward and this causes that the He-burning 
shell becomes more internal. In the model considered here the 
convective unstable zone attains its maximum mass extention after
$\rm \Delta T_{CD}\simeq 16.33$ yr.
The onset of convection has two 
main effects: on one hand the nuclear energy delivered locally by 
He-burning is transferred outward, thus limiting the thermonuclear 
runaway; on the other hand, convection dredges down fresh helium into 
the burning zone so that the thermonuclear runaway speeds up. When 
the evolutionary timescale becomes very short the feeding of the 
He-burning shell by convection becomes not efficient and the structure 
reacts by evolving toward high luminosity. After $\rm \Delta T_{DE}\simeq 
0.48$ day it attains point E along the HR track, where the He-burning 
luminosity is at a maximum: $\mathrm{L_{He}=3.63\times 10^{42}\ erg\ 
s^{-1}}$. The continuous expansion of the whole He-rich zone occurs at the 
expense of its thermal energy, so that He-burning occurs at a progressively 
lower rate; hence, the flash-driven convection starts to recede very soon 
after $\rm \Delta T_{EF}\simeq 3.39$ day (point F in Fig. \ref{f:hr_sf}) 
and it definitively disappears at point G ($\rm \Delta T_{FG}\simeq 10.82$\,yr). 
\begin{figure}   
 \centering
  \includegraphics[width=\columnwidth]{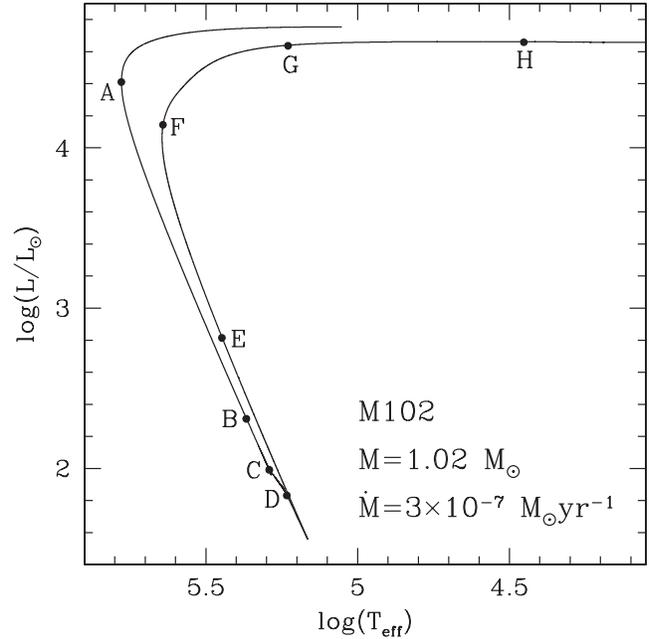}
  \caption{Evolution in the HR diagram of the M102 model accreting He-rich 
           matter at \mdot=$3\times 10^{-7}$\myr. The points along the 
           track represent A: the bluest point; B: minimum He-burning 
           luminosity; C: start of flash-driven convection; D: Maximum 
           extension of the flash-driven convective shell; E: maximum 
           He-burning luminosity; F: beginning of the flash-driven 
           convective shell backtrack; G: die down of flash-driven 
           convection; H: appearance of surface convection. For more 
           details see text.}
  \label{f:hr_sf}
\end{figure}

The He-flash drives the 
transition of the accreting model from a low state, corresponding to 
the cooling of the structure, to a high state, corresponding to the high 
luminosity branch. In fact, the He-flash provides the thermal energy 
needed for a quiescent He-burning to set in, modifying the temperature 
and the density at the base of the He-rich layer. For the Strong Flashes 
regime the amount of nuclear energy delivered during the flash largely 
exceeds the energy required for such a transition. This implies that the 
large thermal energy produced by He-burning is initially locally stored 
and, later on, redistributed along the whole zone above the He-burning 
shell. Under this condition the He-rich layer has a too large thermal content 
which has to be dissipated before quiescent He-burning could set in. 
Hence, the expansion toward the red part of the HR diagram is the natural 
consequence of the strong flash. Obviously, the lower the accretion rate, 
the stronger the resulting He-flash and, hence, the larger the final radius 
attained by the accreting model. During the expansion along the high 
luminosity branch, surface convection sets in (point H) and it penetrates 
inward as the effective temperature decreases. The expansion from point 
G to H occurs in about $\rm \Delta T_{GH}\simeq 13.28$ yr. We forcibly halt 
the computation when the surface temperature becomes smaller than 11300 
K, since the adopted opacity tables are inadequate to describe models 
with lower $\mathrm{T_{eff}}$. Indeed, due to the overlap of the 
flash-driven convective shell and convective envelope, the surface heavy 
elements abundance (mainly $\mathrm{^{12}C}$) increases well above the 
total metal content usually adopted in the computation of low temperature 
opacity tables.

When considering that the accreting WD is a component of an interacting 
binary system, it turns out that during the expansion phase a fraction 
of the matter previously accreted can be lost by the WD, thus limiting 
the growth in mass of the CO core. This problem has been investigated in 
detail by KH04, who derived the amount of mass effectively retained by 
the accreting WD in the framework of the optically thick wind theory 
\citep{kato1994}. According to such an approach, after the He-flash, 
during the expansion to giant dimensions, wind mass loss starts when 
the photospheric temperature decreases to a critical value 
($\mathrm{\log(T_{ph}/K)\simeq 5.45}$). During the redward evolution 
the wind mass loss rate increases as the photospheric temperature 
decreases until the WD attains thermal equilibrium and stops to expand, 
coming back blueward. The continuous loss of the matter reduces the  
layer above the He-burning shell and, 
when the photospheric temperature of the star 
becomes larger than the critical value, it becomes negligible. The following 
evolution is driven by He-burning which reduces the mass of the He-rich layer, 
up to the bluest point in the HR diagram, when the burning dies and the 
structure becomes again supported by gravothermal energy. \cite{kato1994} 
claimed that the strong optically thick wind prevents the onset of a RLOF. 
Moreover, they suggested that, even if a RLOF could occur, a common 
envelope phase is avoided in any case, because the wind velocity is much 
larger than the orbital velocity of the two components in the binary 
system, so that the heating and the consequent acceleration of the lost 
matter is practically inefficient. However, as discussed in KH04, for CO 
WDs less massive than 0.8 \msun the optically thick wind does not occur 
and all the matter is effectively deposited onto the WD, ''{\it if the 
binary separation is large enough for the expanded envelope to reside 
in the Roche lobe}''. Moreover, KH04 analyse the behaviour of CO WDs only 
for very large values of the accretion rate, not exploring the whole 
parameter space reported in our Fig.~\ref{f:regime}.
\begin{table*}  
\caption{For each model listed in Table~\ref{t:initial} experiencing 
         Strong Flashes, we report as a function of the accretion rate 
         \mdot in $10^{-8}$\myr, the value of the total mass at the 
         bluest point along the HR-diagram loop $\rm M_{BP}$ in \msune, 
         the mass transferred up to the onset and after the RLOF episode, 
         $\rm \Delta M^1_{tr}$ and $\rm \Delta M^2_{tr}$, respectively, 
         in $10^{-3}$\msune, the mass lost during the RLOF episode 
         $\rm \Delta M_{L}$ in $10^{-3}$\msune, the accumulation efficiency \acefe, 
         the mass coordinate of the He-burning shell at the epoch of the maximum extention 
         of the flash-driven convective shell $\rm M^{CM}_{He}$ in \msune, the 
         maximum extention of the flash-driven convective shell $\rm \Delta M^{CM}_{He}$
         in $10^{-3}$\msune, the mass coordinate of the He-burning shell at 
         the end of the RLOF $\rm M^{R}_{He}$ in \msune, and the average chemical 
         composition by mass fraction of the matter ejected during the RLOF.}
\centering
\label{t:sf_single}
\begin{tabular}{c c c c c c c c c c c c c}
\hline\hline
\mdot & $\rm M_{BP}$ & $\rm \Delta M^1_{tr}$ & $\rm \Delta M^2_{tr}$ & $\rm \Delta M_{L}$ & \acef & 
$\rm M^{CM}_{He}$ & $\rm \Delta M^{CM}_{He}$ &  $\rm M^{R}_{He}$ & 
$\rm \Delta M^{R}_{He}$ & $\rm {^{4}He}$ & $\rm {^{12}C}$ & $\rm {^{16}O}$\\
\hline
\multicolumn{13} c {M060}\\                 
 2.5 & 0.59743 & 44.780 & 0.328 & 37.45 &  0.170 & 0.58686 & 55.354 & 0.58757 & 17.193 & 0.803 & 0.170 & 0.003\\
 3.0 & 0.59757 & 34.180 & 0.407 & 26.61 &  0.231 & 0.58691 & 44.840 & 0.58762 & 17.525 & 0.801 & 0.172 & 0.004\\
 4.0 & 0.59787 & 25.970 & 0.590 & 17.67 &  0.335 & 0.58726 & 36.577 & 0.58797 & 18.195 & 0.802 & 0.171 & 0.004\\
 5.0 & 0.59818 & 21.570 & 0.778 & 12.78 &  0.428 & 0.58755 & 32.188 & 0.58832 & 18.645 & 0.803 & 0.170 & 0.004\\
 6.0 & 0.59852 & 18.340 & 1.003 &  9.10 &  0.530 & 0.58798 & 28.864 & 0.58863 & 19.128 & 0.803 & 0.171 & 0.005\\
 7.0 & 0.59888 & 15.750 & 1.248 &  6.04 &  0.645 & 0.58844 & 26.166 & 0.58910 & 19.491 & 0.808 & 0.166 & 0.005\\
 8.0 & 0.59928 & 13.480 & 1.507 &  3.47 &  0.769 & 0.58891 & 23.830 & 0.58958 & 19.706 & 0.810 & 0.164 & 0.005\\
 9.0 & 0.59969 & 11.430 & 1.811 &  0.93 &  0.930 & 0.58950 & 21.602 & 0.59015 & 20.039 & 0.822 & 0.152 & 0.004\\
\multicolumn{13} c {M070}\\
 3.0 & 0.70203 & 31.220 & 0.089 & 26.07 &  0.167 & 0.69779 & 35.457 & 0.69869 &  8.495 & 0.758 & 0.212 & 0.003\\
 4.0 & 0.70209 & 22.090 & 0.125 & 16.77 &  0.245 & 0.69784 & 26.341 & 0.69873 &  8.686 & 0.752 & 0.219 & 0.004\\
 5.0 & 0.70216 & 18.460 & 0.163 & 12.96 &  0.304 & 0.69791 & 22.703 & 0.69876 &  8.892 & 0.752 & 0.220 & 0.004\\
 6.0 & 0.70222 & 16.110 & 0.202 & 10.48 &  0.358 & 0.69798 & 20.353 & 0.69883 &  9.026 & 0.753 & 0.219 & 0.004\\
 7.0 & 0.70229 & 14.320 & 0.240 &  8.65 &  0.406 & 0.69803 & 18.578 & 0.69892 &  9.037 & 0.755 & 0.217 & 0.005\\
 8.0 & 0.70236 & 12.880 & 0.282 &  7.10 &  0.461 & 0.69810 & 17.123 & 0.69892 &  9.215 & 0.756 & 0.216 & 0.005\\
 9.0 & 0.70243 & 11.580 & 0.323 &  5.75 &  0.517 & 0.69818 & 15.899 & 0.69900 &  9.253 & 0.765 & 0.208 & 0.004\\
10.0 & 0.70250 & 10.610 & 0.366 &  4.68 &  0.573 & 0.69829 & 14.801 & 0.69908 &  9.336 & 0.763 & 0.209 & 0.005\\
\multicolumn{13} c {M081}\\
 4.0 & 0.81039 & 24.420 & 0.067 & 20.02 &  0.182 & 0.80984 & 24.976 & 0.80954 & 5.255 & 0.703 & 0.260 & 0.005\\
 5.0 & 0.81041 & 18.700 & 0.084 & 14.70 &  0.217 & 0.80974 & 19.380 & 0.80936 & 5.054 & 0.707 & 0.257 & 0.005\\
 6.0 & 0.81043 & 15.730 & 0.098 & 11.75 &  0.258 & 0.80958 & 16.576 & 0.80928 & 5.134 & 0.708 & 0.257 & 0.005\\
 7.0 & 0.81044 & 13.780 & 0.115 &  9.78 &  0.296 & 0.80959 & 14.626 & 0.80930 & 5.142 & 0.710 & 0.255 & 0.005\\
 8.0 & 0.81046 & 12.340 & 0.132 &  8.45 &  0.322 & 0.80954 & 13.254 & 0.80925 & 5.095 & 0.714 & 0.252 & 0.005\\
 9.0 & 0.81047 & 11.130 & 0.148 &  7.24 &  0.358 & 0.80954 & 12.064 & 0.80926 & 5.109 & 0.714 & 0.251 & 0.005\\
10.0 & 0.81049 & 10.130 & 0.163 &  6.33 &  0.385 & 0.80950 & 11.115 & 0.80923 & 5.064 & 0.716 & 0.250 & 0.005\\
20.0 & 0.81066 &  4.720 & 0.247 &  1.01 &  0.797 & 0.80975 &  5.619 & 0.80957 & 4.814 & 0.731 & 0.236 & 0.005\\
\multicolumn{13} c {M092}\\
 5.0 & 0.91899 & 15.590 & 0.015 & 15.27 &  0.022 & 0.91828 & 16.304 & 0.91742 & 1.885 & 0.586 & 0.352 & 0.019\\
 6.0 & 0.91899 & 12.680 & 0.017 & 12.22 &  0.038 & 0.91827 & 13.402 & 0.91772 & 1.727 & 0.603 & 0.344 & 0.012\\
 7.0 & 0.91899 & 10.910 & 0.020 & 10.48 &  0.041 & 0.91827 & 11.633 & 0.91758 & 1.836 & 0.606 & 0.343 & 0.013\\
 8.0 & 0.91900 &  9.620 & 0.023 &  9.13 &  0.053 & 0.91828 & 10.335 & 0.91764 & 1.845 & 0.607 & 0.342 & 0.013\\
 9.0 & 0.91900 &  8.730 & 0.023 &  7.97 &  0.089 & 0.91828 &  9.449 & 0.91805 & 1.703 & 0.609 & 0.342 & 0.013\\
10.0 & 0.91900 &  7.980 & 0.027 &  7.10 &  0.113 & 0.91829 &  8.689 & 0.91807 & 1.810 & 0.611 & 0.340 & 0.013\\
20.0 & 0.91903 &  4.220 & 0.052 &  2.73 &  0.359 & 0.91832 &  4.915 & 0.91873 & 1.781 & 0.621 & 0.334 & 0.013\\
30.0 & 0.91906 &  2.640 & 0.088 &  1.10 &  0.596 & 0.91838 &  3.311 & 0.91862 & 1.980 & 0.643 & 0.315 & 0.011\\
40.0 & 0.91910 &  1.920 & 0.104 &  0.45 &  0.776 & 0.91844 &  2.559 & 0.91878 & 1.791 & 0.676 & 0.286 & 0.008\\
50.0 & 0.91914 &  1.450 & 0.114 &  0.03 &  0.983 & 0.91856 &  1.996 & 0.91893 & 1.630 & 0.809 & 0.161 & 0.002\\
\multicolumn{13} c {M102}\\
 8.0 & 1.02047 &  7.250 & 0.010 &  7.43 & -0.023 & 1.02019 &  7.540 & 1.01955 & 0.740 & 0.545 & 0.355 & 0.020\\
 9.0 & 1.02047 &  6.460 & 0.000 &  6.49 & -0.005 & 1.02021 &  6.730 & 1.01970 & 0.740 & 0.553 & 0.361 & 0.018\\
10.0 & 1.02047 &  5.850 & 0.010 &  5.75 &  0.019 & 1.02020 &  6.110 & 1.01982 & 0.750 & 0.555 & 0.365 & 0.018\\
20.0 & 1.02048 &  3.120 & 0.010 &  2.22 &  0.291 & 1.02017 &  3.430 & 1.02060 & 0.780 & 0.569 & 0.370 & 0.021\\
30.0 & 1.02049 &  2.120 & 0.020 &  1.18 &  0.449 & 1.02018 &  2.420 & 1.02051 & 0.920 & 0.596 & 0.351 & 0.019\\
40.0 & 1.02049 &  1.620 & 0.030 &  0.67 &  0.594 & 1.02019 &  1.910 & 1.02053 & 0.910 & 0.623 & 0.330 & 0.015\\
50.0 & 1.02050 &  1.300 & 0.030 &  0.38 &  0.714 & 1.02023 &  1.550 & 1.02056 & 0.860 & 0.652 & 0.306 & 0.012\\
60.0 & 1.02051 &  1.060 & 0.030 &  0.17 &  0.844 & 1.02025 &  1.300 & 1.02060 & 0.800 & 0.693 & 0.270 & 0.008\\
\hline
\end{tabular}
\end{table*}

\begin{figure}  
 \centering
  \includegraphics[width=\columnwidth]{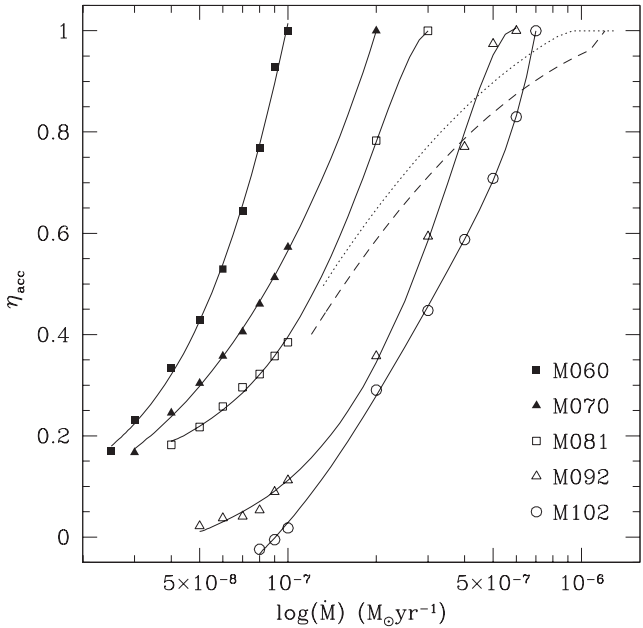}
  \caption{Accumulation efficiency \acef\ as a function of the accretion 
           rate for models of  different initial mass as labelled inside 
           the figure. 
           Solid lines display polynomial fits to the data 
           (see Appendix \ref{app3}).
           The dotted and dashed lines represent the accumulation 
           efficiency determined by KH04 for CO WDs with initial mass 0.9 
           and 1.0 \msune, respectively (their Eqs. 4 and 5).}
  \label{f:accum1}
\end{figure}
At variance with KH04, in the present work we do not assume that the 
optically thick wind operates and we compute the post-flash evolution 
of CO WDs experiencing strong flashes by assuming that a RLOF occurs.
When the surface radius of the accreting WD becomes larger than the 
corresponding $\mathrm{R_{Roche}}$ we subtract mass by requiring that 
the star remains confined inside its lobe. When the WD definitively 
starts to contract, we allow the accretion to resume and 
we follow the evolution up to the bluest point of 
the loop in the HR diagram. Hence, the accumulation efficiency 
along the cycle is computed as:
\begin{equation}
\mathrm{
\eta_{acc}=1-{\frac{\Delta M_{L}}{\Delta M^1_{tr}+\Delta M^2_{tr}}}.
}
\label{e:acef}
\end{equation}
where $\mathrm{\Delta M_{L}}$ is the mass lost during the RLOF, while
$\mathrm{\Delta M^1_{tr}}$ and $\mathrm{\Delta M^2_{tr}}$ are the mass accreted 
onto the WD before and after the RLOF episode, respectively.
In real binaries, the value of $\mathrm{R_{Roche}}$ is determined by 
the parameters of the system itself, i.e. masses of primary and secondary 
components and separation. For this reason, we computed some tests 
by fixing the initial WD mass (namely the M102 model) and varying the 
value of $\rm R_{Roche}$ in the range 1 -- 45 $\rm R_\odot$\footnote{Note 
that the maximum considered value for $\mathrm{R_{Roche}}$ depends on 
the CO WD mass, because as already recalled, we stop the computation 
when the photospheric temperature becomes lower than 11300 K.}. Our 
results show that the uncertainty in the estimated \acef is smaller 
than 7-8 \%. For this reason, we set $\rm R_{Roche}=10\ R_\odot$ in the 
computations of all the models experiencing Strong Flashes. 
The values of \acef computed according to Eq.~(\ref{e:acef}) 
are displayed in Fig. \ref{f:accum1} as a function of the accretion 
rate for each CO WD model. Note that they refer to the first Strong 
He-flash experienced by all the {\sl Heated Models} listed in 
Table~\ref{t:initial}. 
As it can be noticed, decreasing 
the accretion rate, the accumulation efficiency reduces very rapidly. 
Indeed, lowering \mdote, the resulting He-flash is stronger and, hence, 
a larger amount of mass has to be removed in order to dissipate the 
extra-energy content of the whole He-rich layer. For the M102 model, the 
accumulation efficiency for \mdot lower than $10^{-7}$\,\myr, becomes 
negative; this means that during the RLOF episode the thin He-rich layer  
already existing at the bluest point in the HR diagram has been partially 
eroded. Polynomial fits of \acef as a function of the accretion rate
for all the models displayed in Fig. \ref{f:accum1} are provided in  
Appendix \ref{app3}. 

In Table~\ref{t:sf_single} we list, for each initial CO WD model, 
the accretion rate, the total
mass at the bluest point along the loop in the HR diagram, the
amount of mass transferred from the donor before and after the RLOF, 
the mass lost during the RLOF episode and the corresponding accumulation 
efficiency, as well as some other relevant physical quantities. 
By inspecting Table~\ref{t:sf_single} it is evident that for each initial 
CO WD the amount of matter accreted after the RLOF is negligible, even if 
it increases as the accretion rate increases. Moreover, Table~\ref{t:sf_single} 
reveals that, for the computed models, the flash-driven convective shell 
extends all over the accreted layer and the most external zone of the 
He-rich mantle already existing at the epoch of bluest point in the HR 
diagram loop. This implies that at the onset of the RLOF episode the 
zone above the He-burning shell has been completely homogenized and, hence, 
it is depleted in helium and enriched in carbon, the oxygen production 
resulting marginally efficient. In the 
last three columns of Table~\ref{t:sf_single} we report the average chemical 
composition of the ejected matter. As it can be noticed, for a fixed initial 
CO WD, the lower the accretion rate, the stronger the He-flash and, thus, 
the more efficient the He-consumption and the larger the carbon abundance. 
Moreover, the larger the initial CO WD, the lower the matter accreted before 
the RLOF and, hence, the mass of the convective shell, so that, on average, 
the dilution of the He-burning ashes is lower and the resulting $\rm {^{12}C}$ 
abundance in the ejecta becomes larger. 

As shown in Table~\ref{t:sf_single} for a fixed initial CO WD, 
$\mathrm{\Delta M^{R}_{He}}$, the post-RLOF mass of the He-rich 
envelope above the He-burning shell $\rm M^R_{He}$, 
is largely independent of \mdote. This reflects the property that the 
location along the high luminosity branch, i.e. the effective temperature 
and, hence, the surface radius, depend on the mass of the He-layer 
above the burning shell. Notwithstanding, Table~\ref{t:sf_single} reveals 
that $\mathrm{\Delta M^{R}_{He}}$ still slightly depends on the accretion rate: 
in low mass initial CO WDs models it is larger for higher \mdote, while 
in more massive ones it exhibits a maximum for intermediate values of 
\mdote. A further inspection of Table~\ref{t:sf_single} 
reveals that at the end of the RLOF the location of He-burning shell becomes closer 
to the surface as \mdot increases. Both these circumstances suggest that the 
strength of the He-flash, which depends inversely on the accretion rate, 
plays a role in determining the actual value of the retention 
efficiency. 

Encouraged by an anonymous referee to investigate in more detail 
such an issue, we compute several toy models, by fixing the initial CO WD 
(M102 model) and accretion rate (\mdote=$\rm 3\times 10^{-7}$\myr), 
and by activating, after the ignition of He-burning, an extra energy 
source in the layer where helium abundance is larger than 0.01 by 
mass fraction. This allows us to vary the strength of the flash, 
while $\rm \Delta M^1_{tr}$ as well as the ignition point remain 
unaltered. For the sake of simplicity we parametrize the energy 
delivered by this fake source as 
$\rm \varepsilon_{ES}(m)=\beta\cdot\varepsilon_{nuc}(m)$, where 
$\rm \varepsilon_{nuc}(m)$ is the nuclear energy produced at the 
mass coordinate $m$ and $\beta$ a free parameter. Negative value of $\beta$ 
means that energy is subtracted from the He-rich layer. When the 
He-flash quenches and the luminosity of the He-burning shell becomes 
lower than 100 times the surface luminosity, we deactivated the extra 
energy source. Our results are summarized in Table~\ref{t:sf_test} 
(lines 1---4): the larger the energy injected into 
the He-rich layer, the larger the flash-driven convective shell, the larger 
the mass loss during the RLOF, the more internal the position of the 
He-burning shell after the RLOF episode and, hence, the lower the 
accumulation efficiency. 

We perform also another test, by putting $\beta=0$ and preventing the mixing 
in the flash driven convective shell. In this way the He-burning shell 
is not re-fuelled so that the resulting He-flash is definitively 
less strong ($\rm L^{Max}_{He}$ is almost an order of magnitude 
lower than in the standard case) and the corresponding 
accumulation efficiency approaches almost unity (line 5 in 
Table~\ref{t:sf_test}). In order to further test the sensitivity 
of \acef on the amount of helium dredged down by convective mixing, 
we computed an additional model. In this case, 
during the phase when the outer border of the flash-driven 
convective shell coincides with the stellar 
surface, we artificially alter the chemical composition in the outermost 
$10^{-6}$ \msun zone of the WD after each time step by restoring 
the local abundances of all elements as in the accreted matter (see 
\S \ref{s:code}). In this way, the reservoir of helium-rich matter 
which could feed the He-flash is increased by about 50\%.
As shown in Table~\ref{t:sf_test} (line 6), the resulting He-flash 
is stronger and, correspondingly, the retention almost halves.

The results of all the toy models listed 
in Table~\ref{t:sf_test} clearly suggest that the strength of 
the He-flash plays a pivotal role in determining the accumulation 
efficiency. In fact, it determines the energy content of the He-rich 
layer in accreting WDs at the onset of the RLOF and, hence, 
the amount of mass that has to be lost in order to attain the 
physical condition suitable for the accreting WD to start its 
blueward evolution. Moreover, the strength of the He-flash determines 
the maximum inward shift of the He-burning shell during the flash-driven 
convective episode as well as the duration of the expansion phase 
up to the RLOF, affecting the exact value of $\rm M^{R}_{He}$.
Note that the energy delivered by the He-flash is determined mainly by 
the value of \mdot for a fixed initial CO WD, even if our results also 
demonstrate that the convective mixing acts as a propelling mechanism 
of the He-flash itself, thus affecting the final value of the 
accumulation efficiency.

\begin{table} 
\caption{Selected physical properties of the test models computed by 
artificially varying the thermal content of the He-rich layer (see text 
for more details). 
For comparison we report also the standard case $\beta=0.0$. 
$\rm L^{Max}_{He}$ represents the maximum luminosity of the He-burning 
shell and is expressed in $\mathrm{10^{42}\ erg\ s^{-1}}$ unit.
The other quantities are the same as in Table~\ref{t:sf_single} and 
have the same unit.}
\centering
\label{t:sf_test}
\begin{tabular}{r c c c c c c}
\hline\hline
$\beta$ & $\rm L^{Max}_{He}$ & $\rm \Delta M_{L}$ & \acef & 
$\rm \Delta M^{CM}_{He}$ & $\rm M^{R}_{He}$ & $\rm \Delta M^{R}_{He}$ \\
\hline
\multicolumn{7} c {$\rm \varepsilon_{ES}(m)=\beta\cdot\varepsilon_{nuc}(m)$}\\
 0.0 & 4.223 & 0.117 & 0.453 & 0.242 & 1.02053 & 0.91\\
-0.5 & 2.559 & 0.075 & 0.648 & 0.237 & 1.02091 & 0.96\\
 0.2 & 4.609 & 0.124 & 0.423 & 0.244 & 1.02037 & 1.00\\
 0.5 & 5.022 & 0.130 & 0.393 & 0.245 & 1.02029 & 1.02\\
\multicolumn{7} c {No Convective Mixing}\\
0.0  & 0.628 & 0.019 & 0.913 & 0.253 & 1.02160 & 0.83\\
\multicolumn{7} c {Altered Chem. Composition}\\
 -   & 9.282 & 0.158 & 0.251 & 0.242 & 1.02027 & 0.76\\ 
\hline
\end{tabular}
\end{table}

For the sake of comparison, in Fig.~\ref{f:accum1} we also show the values 
of \acef obtained by KH04 in the framework of the optically thick wind 
scenario for a 0.9 (dotted line) and 1.0 \msun (dashed line) CO WDs. The 
differences in the estimated retention efficiency reflect the different 
assumptions concerning the mass loss episode. To make more clear this 
issue, let us recall that the huge energy released during the He-flash 
is stored in the He-shell as thermal energy, since the nuclear timescale 
is shorter than the radiative diffusion timescale due to partial 
degeneracy of the matter. Hence, the flash-driven convective episode 
redistributes the thermal energy excess, so that the thermal content 
of the layer above the He-burning shell increases while its physical 
dimensions (both in 
radius and mass) remain practically unaltered. In this way the 
thermal energy of the He-rich layer becomes too large for its very 
compact configuration and, hence, it has to be dissipated. Such a situation 
is similar to what occurs to an ideal gas evolving at constant volume 
(and, hence, constant density): if the temperature is increased then the 
pressure has to increase. In He-flashing structure the overpressure 
determined by the increase of the thermal content triggers the expansion 
of the whole He-rich layer, making work against gravity. In this way 
the volume increases, the density decreases as well as the specific heat; 
thus, the thermal content of the He-rich zone decreases. 
The Roche geometry defines a finite volume in the space so that 
matter residing inside the corresponding lobe represents the 
expanding WD while the mass passed through it is lost from the 
binary system. Hence, due to the continuos 
expansion, the portion of the star remaining inside the lobe 
has a lower density and, hence, its specific heat decreases.
\begin{table}  
\caption{The same as in Table~\ref{t:sf_single}, but for model M092 
         accreting He-rich matter at \mdot$=5\times10^{-7}$\myr\ and 
         with different values of the Roche lobe radius, as listed 
         (in solar units). }
\centering
\label{t:etadd}
\begin{tabular}{r c c c c c c}
\hline\hline
${\rm R_{Roche}}$ & $\rm \Delta M^1_{tr}$ & $\rm \Delta M^2_{tr}$ & $\rm \Delta M_{L}$ & \acef & 
$\rm M^{R}_{He}$ & $\rm \Delta M^{R}_{He}$ \\
\hline
 0.1 & 1.426 & 0.086 & 0.60 & 0.594 & 0.91863 & 1.335\\  
 1.0 & 1.444 & 0.112 & 0.16 & 0.895 & 0.91886 & 1.566\\
10.0 & 1.450 & 0.114 & 0.03 & 0.983 & 0.91893 & 1.630\\  
\hline
\end{tabular}
\end{table}

\begin{figure}  
 \centering
  \includegraphics[scale=0.45]{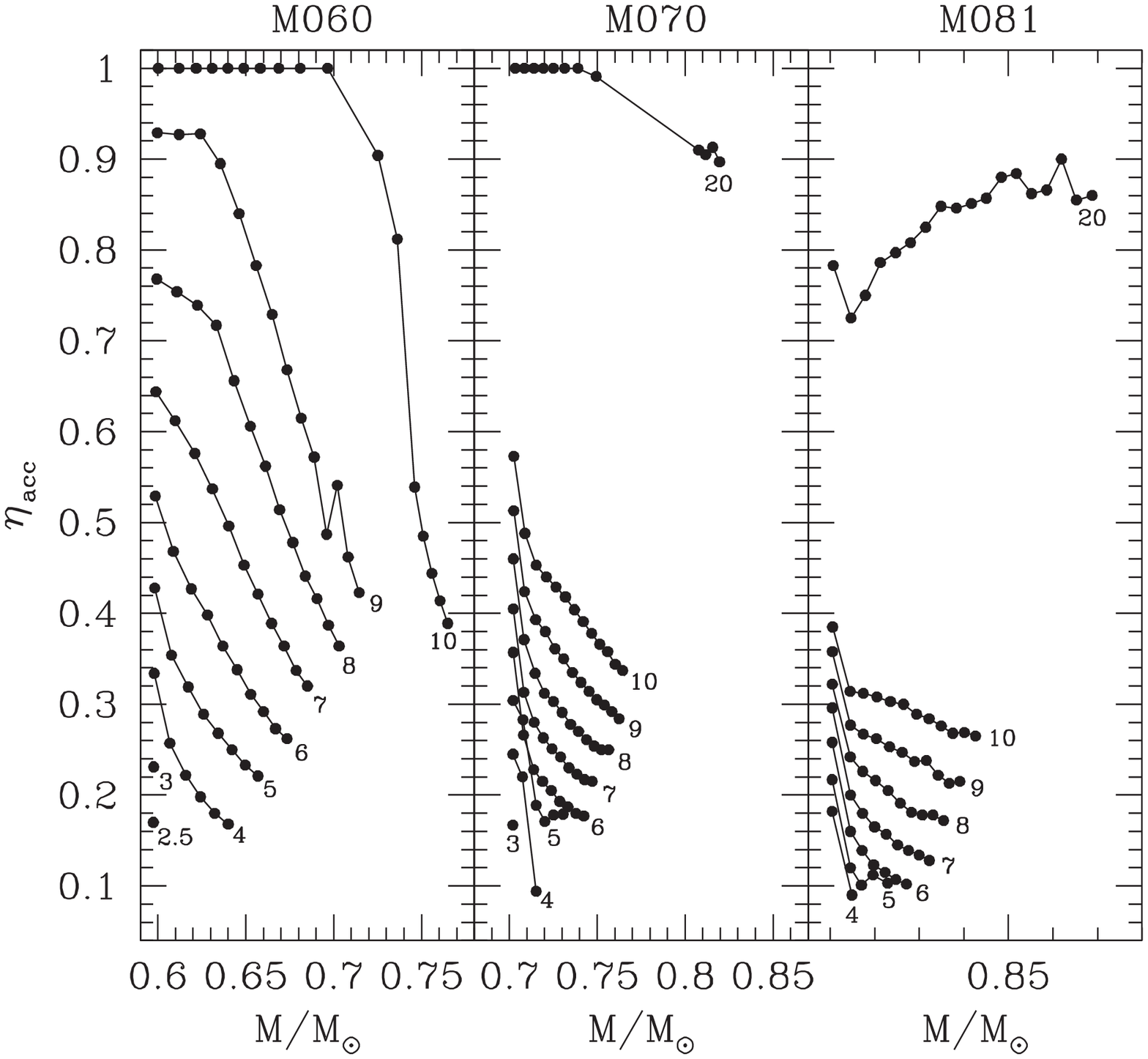}
\vskip 0.1cm
  \includegraphics[scale=0.55]{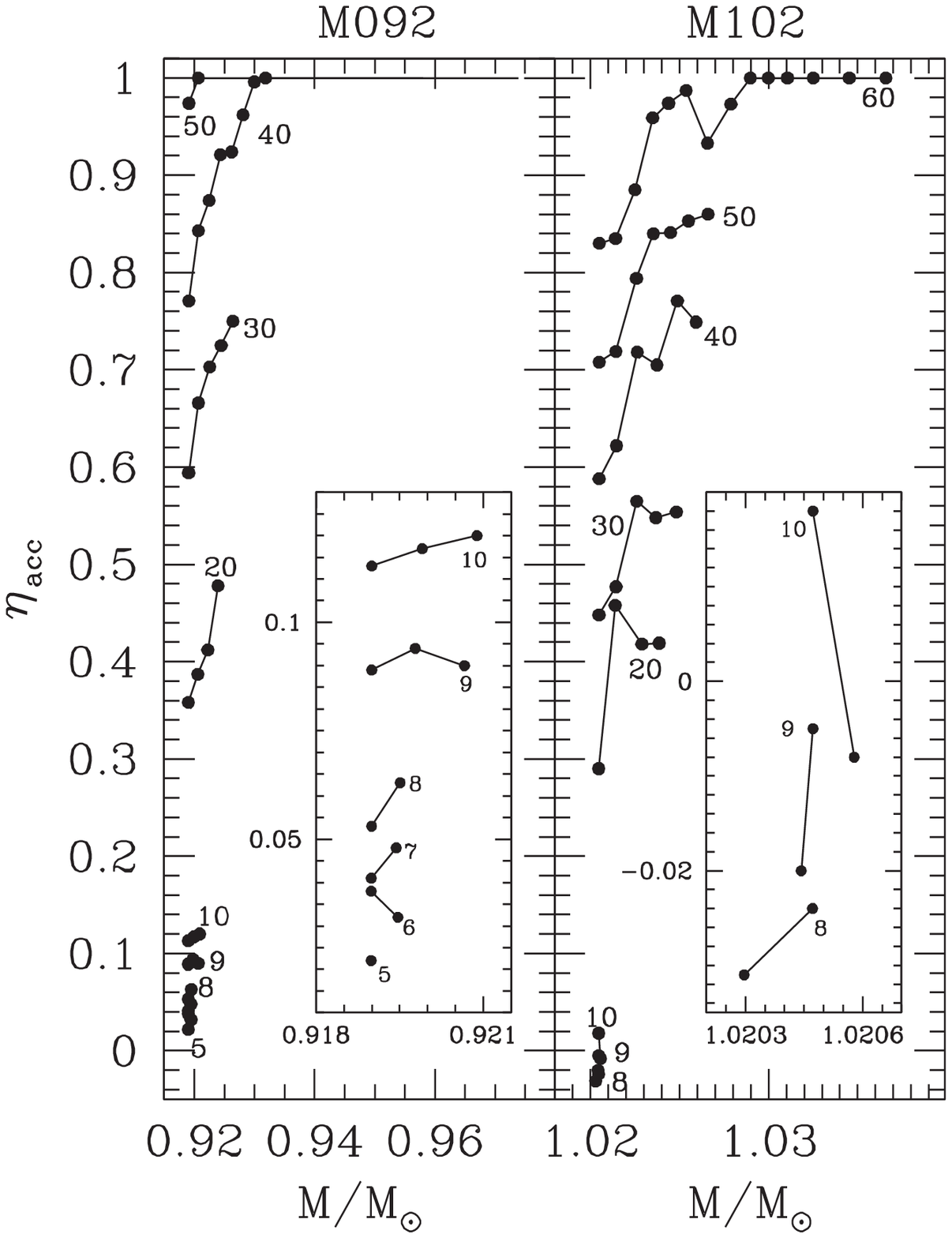}
  \caption{Accumulation efficiency \acef as a function of the WD total 
           mass at the bluest point along the loop in the HR diagram, 
           for all the models listed in Table~\ref{t:initial}. Different 
           curves refer to different accretion rates, as labelled (in 
           $10^{-8}$\myr).}
  \label{f:accum2}
\end{figure}

In the RLOF model expansion determines the mass loss, while  
in the KH04 model wind mass-loss represents an additional mechanism 
favouring the reduction of the thermal content of the expanding 
He-rich layer. As a consequence, 
for low values of the accretion rate, i.e. for very powerful non-dynamical 
He-flashes, in the RLOF case a larger amount of mass 
has to be lost and, consequently, the accumulation efficiency is lower 
than in the case of the KH04 models. For high values of \mdote, i.e. for 
less energetic non-dynamical He-flashes, in the 
KH04 computation mass loss starts very soon, when the accretor 
is still very compact (the surface radius is smaller than 0.1 $R_\odot$) 
so that in this case the expansion does not play a significant role in 
reducing the thermal energy excess. On the other hand, in the Roche lobe 
scenario, the WD continues to expand to larger radii, so that the 
thermal energy excess in the He-envelope is already largely reduced
when the mass loss episode induced by the presence of the Roche lobe 
occurs. 
As a consequence, in the latter case \acef is larger than 
in the KH04 models. Such a conclusion is confirmed by the values of \acef 
we obtain for an additional set of models, 
where we compute accretion of He-rich 
matter at \mdot$=5\times 10^{-7}$\myr onto the M092 ``{\sl Heated Model}'' but varying 
the size of the Roche lobe. As it is shown in Table~\ref{t:etadd}, the 
lower $\mathrm{R_{Roche}}$, the larger the amount of matter lost from 
the accretor to remove the thermal energy excess determined by the 
He-flash. It is worth noticing that, if it is assumed that the strong 
wind by KH04 is at work, the retention efficiency for all these models 
should be \acef=0.858. Table~\ref{t:etadd} also reveals that ${\rm \Delta M^2_{tr}}$
decreases as the Roche lobe radius is reduced. This because, after the RLOF episode, 
during the blueward excursion along the high luminosity branch of the 
loop in the HR diagram the evolutionary timescale becomes longer and a 
sizable amount of matter can be accreted (see also Table~\ref{t:sf_single}). 
As a consequence, the lower 
the Roche lobe radius the shorter the time spent in steady burning 
condition, the smaller the value of $\rm \Delta M^2_{tr}$. 
Moreover we found that also $\rm \Delta M^1_{tr}$ 
depends mildly on the exact value of ${\rm R_{lobe}}$, as the redward evolution
after the end of the flash-driven convective episode is slower with respect 
to the time-span of the He-flash itself (see the discussion at 
the beginning of this section). 
Our results suggest 
that in relatively wide systems the accumulation efficiency weakly 
depends on the radius of the critical lobe, but it may reduce by 
up to a factor 2 in extremely tight systems. 

In order to investigate the asymptotic behaviour of CO WDs experiencing 
Strong Flashes, we computed a sequence of flashes for each WD mass and 
\mdot combination. For low values of \mdot the computation of each flash 
episode is very time consuming, due to the large number of models in the 
sequence because of very small time steps and certain problems in 
determining the physical structure of the accretor (the models are close 
to becoming dynamical). For this reason, in some cases we computed just 
two flash episodes. For each sequence we determine the retention efficiency 
according to Eq.~(\ref{e:acef}) and in Fig. \ref{f:accum2} we plot the 
results. A table listing the accumulation efficiency plotted in Fig. 
\ref{f:accum2} is available online. In the M060 and M070 cases we 
plot also the values of \acef for models accreting at \mdot=$10^{-7}$\myr\,and 
\mdot=$2\times 10^{-7}$\myr, respectively, which start their evolution in 
the Mild flash Regime and, then, experience strong He-flashes (for more 
detail see \S \ref{s:mildfl}).

The thermal content of accreting models, as determined by the accretion 
history before entering the Strong Flashes regime, has an important role 
in determining the actual values of \acefe, as, for example, in AM 
CVn type systems in which \mdot in $\simeq 10^6$\,yr after beginning of 
mass-transfer declines from $\rm \sim (10^{-5}-10^{-6})$\,\myr\ to 
$\rm \sim 10^{-8}$\,\myr\,(see Fig.~\ref{f:amtracks}). In fact, during 
the previous evolution, occurring in the mild flashes regime, an extended 
hot layer is piled-up on the initial CO WD. This represents a ``boundary 
condition'' completely different to the small helium layer present as a 
consequence of the pre-heating procedure. Hence, for fixed \mdote, in 
the models which did not have stages of steady burning or ``mild flashes'' 
accretion, He-flash will be stronger, leading to a larger mass loss and, 
hence, to a lower \acefe. For example, for \mdot$=10^{-7}$\myr, model 
M060 has \acefe=1 when its total mass at the bluest point is equal to 
0.7251 \ms, while, for the same accretion rate, model M070 has \acefe=0.429 
when $\mathrm{M_{tot}=}$0.7265 \ms. The same is also true at \mdot$\rm =2\times 
10^{-7}$\myr\, for the model M070, having \acefe=0.905 at $\rm M_{tot}=$
0.8118 \ms, and M081, having \acefe=0.783 at $\rm M_{tot}=$0.8107 \ms. 
However, pulse by pulse, the thermal content of the He-rich zone is modified, 
as thermal energy is diffused inward on a timescale depending mainly on 
the CO core mass. Therefore, the mean temperature level at the base of 
the accreted He-rich layer becomes a function only of the actual mass of the 
CO core and of the mass of the helium layer (which is determined 
by \mdot for each value of $\mathrm{M_{CO}}$). Therefore, the 
differences in the accreting models with different accretion history are 
smeared off, so that their accumulation efficiency attains an asymptotic 
value. This is evident when comparing M060 and M070 models accreting at 
\mdot=$\rm 10^{-7}$\myr, for which we obtain \acefe=0.389 at 
$\rm M_{tot}=$0.7649 and \acefe=0.337 at \mwd$=$0.7645, respectively. 
For all these models the general trend is that, for fixed \mdote, \acef 
decreases increasing the total mass of the accreting WDs. 

These considerations are still valid for more massive initial CO WDs, 
as it can be derived by an inspection of Fig. \ref{f:accum2}. Indeed, 
for high values of the accretion rate, the accumulation efficiency 
increases in the M092 and M102 models, since the energy delivered by 
both the deposition of matter and the He-flash is employed to heat up
the most external layers of the He-deprived core, thus modifying the degeneracy 
level of the physical base of the He-shell. Therefore, pulse by pulse 
He-flashes become less strong so that the fraction of mass effectively 
retained during the episode increases. On the other hand, for low 
\mdote, the compressional heating timescale becomes longer than 
the inward thermal diffusion timescale. Therefore, He-flashes are 
ignited in more degenerate matter and the corresponding \acef 
decreases as $\mathrm{M_{tot}}$ increases.
\begin{center}
\begin{table*}  
 \caption{Selected properties of models with the same accretion rate 
          and different initial masses. From left to right we list 
          the initial model, the total WD mass at the bluest point 
          in the HR diagram, the corresponding effective temperature 
          and luminosity, the mass coordinate of the He-burning shell, 
          and the corresponding values of density and temperature. 
          In the last column we report the accumulation efficiency. 
          See text for more details.} 
\label{t:mf2sf}
 \begin{tabular}{c c c c c c c c}
  \hline\hline 
  Model & $\rm M_{BP}$ &  $\rm \log(T_{eff})$ & $\rm \log(L/L_\odot)$ & 
          $\rm M_{He}$ & $\rm \log(\rho_{He})$ & $\rm \log(T_{He})$ & \acef \\
  \hline
  \multicolumn{8}{c}{\mdote$\rm =10^{-7}$\ms} \\
  \hline
M060 & 0.736278 & 5.47712 & 3.85658 & 0.73113 & 3.6606 & 8.3296 & 0.81164\\
M070 & 0.737064 & 5.48664 & 3.85790 & 0.73411 & 3.4683 & 8.2744 & 0.40375\\
  \hline
  \multicolumn{8}{c}{\mdote$\rm =3\times 10^{-7} M_\odot$} \\
  \hline
M060 & 0.920639 & 5.65280 & 4.21510 & 0.91970 & 3.3369 & 8.3352 & 0.96985\\
M070 & 0.919911 & 5.66284 & 4.23922 & 0.91864 & 3.5007 & 8.3761 & 0.95435\\
M081 & 0.919113 & 5.65428 & 4.21838 & 0.91816 & 3.3541 & 8.3391 & 0.94949\\
M092 & 0.919064 & 5.68452 & 4.24309 & 0.91832 & 3.3638 & 8.3507 & 0.59358\\
  \hline
 \end{tabular}
\end{table*}
\end{center}

\subsection{Mild Flashes Regime}\label{s:mildfl}

For slightly larger values of the accretion rate, CO WDs accreting He-rich 
matter experience Mild Flashes. In this case the evolution is similar 
to models experiencing Strong Flashes: a He-rich layer is piled-up via 
accretion, determining the compressional heating of the He-shell up to 
the moment when the physical conditions suitable for He-ignition are 
attained. However, as \mdot is larger than in the previously discussed 
case, energy losses can not counterbalance compressional heating, so 
that the He-shell does not become degenerate at all. As a consequence, the 
resulting He-flash is very mild and it delivers just the amount of energy 
to determine the transition of the model from the low to the high state. 
As a result, the radii of CO WDs experiencing this accretion regime 
typically remain smaller than $\rm \sim 0.5$\rsun, so, excluding the 
most compact AM CVn stars, no interaction with the companion could occur 
and all the matter transferred during each cycle is effectively deposited 
onto the accretors. For AM CVn stars with separations $\sim 0.1$\rsun 
formation of a short-living common envelope may be envisioned and then 
\acef should decrease.
\begin{figure}   
 \centering
  \includegraphics[width=\columnwidth]{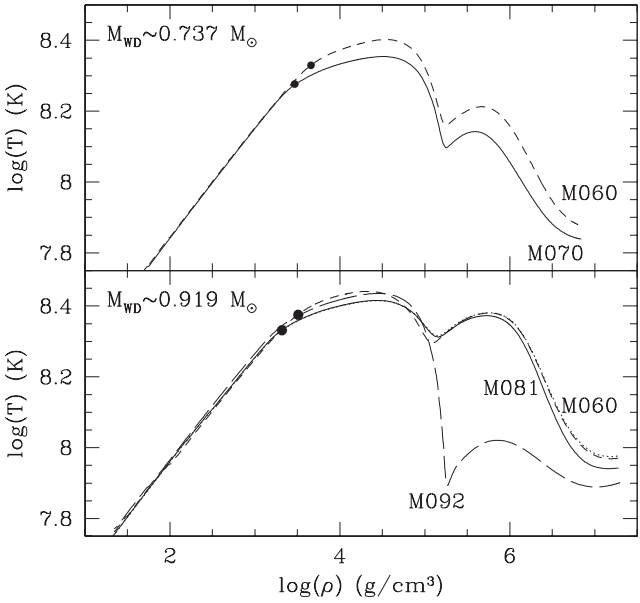}
   \caption{Profiles in the $\rho -T$ for models with the same total 
            mass at the bluest point, but with different initial mass 
            and equal accretion rate (\mdot=$10^{-7}$\myr\ and 
            \mdot=$3\times 10^{-7}$\myr\ in the upper and lower panel, 
            respectively). Filled dots mark the position of the 
            He-burning shell. In the upper panel  the dotted lines 
            represent the M060 {\sl Heated Model}.}
  \label{f:mf2sf}
\end{figure}

It is well known, and we find it as a result of our computations too, 
that for a given value of the accretion rate, He-flashes become stronger, 
as the total mass of the CO WD increases. As a consequence the models 
may transit from mild flashes regime into the regime of the strong ones. 
However, the possibility of such a transition depends not only on the 
actual mass of the accreting WD, but also on the previous accretion 
history which, in turn, fixes the thermal content of the CO core. Such 
an occurrence appears evident when comparing the long term evolution of 
models with the same \mdot and different initial mass (see Table~\ref{t:mf2sf}). 
For example, model M070 accreting at $\rm 10^{-7}$\myr\,experiences strong 
He-flashes from the very beginning (\mwd$=$0.70185\msune), while 
model M060 accreting at the same \mdot enters the Strong Flashes regime 
when its total mass has increased to \mwd$=$0.72514 \msune. In 
the latter model the deposition of matter has deeply modified the 
temperature profile in the CO core underlying the He-rich layer with respect 
to the M070 model, as it is clearly seen in the upper panel of Fig. 
\ref{f:mf2sf}, where we plot the temperature profile as a function of 
density for the two models when their total mass is $\rm \sim 0.737$\msune. 
The same considerations are still valid when considering the evolution 
of a model accreting He-rich matter at \mdote$\rm =3\times 10^{-7}$\myr, 
as it is shown in the lower panel of the same Figure. Note that in this 
case the temperature profiles in the inner zones of the CO cores of 
models M060 (dotted line) and M070 (dashed line) became practically 
coincident. It is worth noting that the difference in the temperature 
profiles of the models with the same total CO core mass and accretion 
rate, but with different accretion histories, leads to different 
accumulation efficiency, as it comes out by an inspection of the last 
column in Table~\ref{t:mf2sf}. Some differences between models M060 and 
M070 are due to the fact that the nuclear energy delivered during the 
last $\approx1$\,Myr of evolution corresponding to the mild flashes 
regime has been transferred inward very efficiently, so that the whole 
structure of the former model became hotter and less dense with respect 
to the M070. As well, the structures do not have exactly the same 
CO-core mass.

\subsection{Steady Accretion Regime}\label{s:steasta}

In the Steady Accretion regime, by definition, the rate at which He is 
converted via nuclear burning into a CO-rich mixture is very close to 
the rate at which He-rich matter is transferred from the donor. As a 
consequence, models experiencing this regime evolve in the HR diagram 
along the high luminosity branch of the typical loop. The long-term 
evolution of these models is determined by the interplay of two 
different factors:\\
\indent 
(i) With increase of the mass of the CO core the luminosity level of 
the models becomes larger. In order to counterbalance larger radiative 
energy losses, the shell has to burn helium at a higher rate. This 
determines a progressive reduction of the mass of the He-rich mantle;\\
\indent
(ii) The external layers of the CO core are hot and expanded, since 
they have been piled-up via nuclear burning. Contraction of this zone 
delivers thermal energy which represents an additional source to balance 
the radiative loses from the surface. Hence, He-rich matter has to be 
burnt at a lower rate. In any case, for a model with a constant or a 
decreasing accretion rate, the mass of the He-rich envelope progressively 
reduces and when it becomes smaller than a critical value the accretor 
enters the Mild Flashes regime. 

In Fig. \ref{f:ste2mf} we show accretion rate at which the transition 
from the Steady to the Mild Flashes regime occurs as a function of the 
WD total mass. For comparison we also plot the transition line as derived 
for Fig. \ref{f:regime} (open circles with thick solid line). This 
figure reveals that the thermal content of the most external zones of 
the CO core just below the He-burning shell affects the value of the WD 
total mass at which the transition occurs. Such a conclusion is reinforced 
when considering that different initial models converge to the same value 
when about (0.05--0.1)\,\msun of He-rich matter has been accreted. 
\begin{figure}  
 \centering
  \includegraphics[width=\columnwidth]{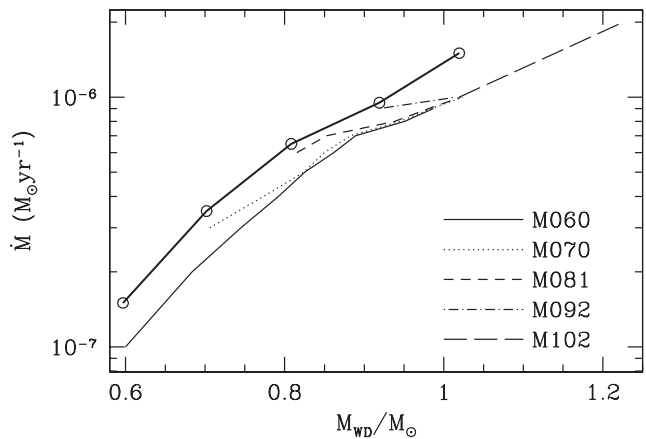}
   \caption{Evolution of the accretion rate at which the transition from 
            the Steady Accretion to the Mild Flashes regimes occurs as a 
            function of the WD total mass . Different lines refer to 
            different initial models. Heavy solid line and the open dots 
            represent the lower limit of the Steady Accretion zone shown 
            in Fig.~\ref{f:regime}.}
  \label{f:ste2mf}           
\end{figure}

\subsection{RG Regime}
\label{s:rgregime}

If the accretion rate is definitively larger than the rate at which 
helium is nuclearly processed into a CO-rich mixture, a massive He-rich mantle 
is piled-up. As a consequence, accreting WD resembles a post-AGB star: 
the more massive the helium envelope, the lower the effective temperature and, hence, 
more expanded the structure. Models in the RG accretion regime evolve 
redward in the HR diagram, developing very soon a surface convective 
layer which penetrates inward as the structure expands. When considering 
that accreting WDs are components of interacting binary systems, it turns 
out that, depending on the geometry of the systems, they could overfill 
their Roche lobe, so that a part (if not all) of the matter transferred 
from the donor has to be ejected from the binary system.
\begin{table*} 
\caption{Physical properties of some selected models experiencing C-ignition. 
         For more details see the text.}
\label{t:coigni}      
\centering          
 \begin{tabular}{c c c c c | c c c c }
  \hline\hline       
  \mdot               & $\rm M_{fin}$ & $\rm M_{ig}$ & $\rm \log(\rho_{ig})$ & 
  $\rm \log(T_{ig})$  & $\rm M_{fin}$ & $\rm M_{ig}$ & $\rm \log(\rho_{ig})$ & $\rm \log(T_{ig})$ \\
  ($10^{-6}$ \myr) & (\msune)  & (\msune)   & ($\rm g\cdot cm^{-3}$) & (K)   & 
  (\msune)  & (\msune)   & ($\rm g\cdot cm^{-3}$) & (K) \\
  \hline
  {\sl }  & \multicolumn{4}{|c|}{\sl He-accreting} & \multicolumn{4}{c}{\sl CO-accreting} \\
  \hline                    
  \multicolumn{9}{c}{M102} \\
  \hline                  
  2 &  1.3746 & 0.0000 & 9.3230 & 8.4321 & 1.3733 & 0.0000 & 9.3230 & 8.4321 \\
  3 &  1.3347 & 1.3204 & 6.3845 & 8.7761 & 1.3559 & 1.3445 & 6.4328 & 8.7743 \\
  \hline                  
  \multicolumn{9}{c}{M092} \\
  \hline
  2 &  1.3747 & 0.0000 & 9.3230 & 8.4321 & 1.3740 & 0.0000 & 9.3195 & 8.4358 \\
  \hline
 \end{tabular}
\end{table*}
 
In order to define the lower limit in the $\mathrm{M_{WD}-\dot{M}}$ 
plane for the RG regime we adopt the same procedure as for models in 
the Strong Flashes regime. In particular, for each model listed in 
Table~\ref{t:initial}, we determine the values of the \mdot for which 
the accreting WD expands, thus attaining an effective temperature of 
11300 K. Hence, we assume that the Roche lobe radius of accretor is 
equal to 10 $\mathrm{R_\odot}$ and we force that $\mathrm{R_{WD}\le 
R_{Roche}}$ by subtracting mass from the WD. We compute the minimum 
rate for the RG regime as:
\begin{equation}
\mathrm{
\dot{M}_{RG}=\dot{M}_{acc}-{\frac{\Delta M_{lost}}{\Delta t}},
}\label{e:mrg}
\end{equation}
where $\mathrm{\dot{M}_{acc}}$ is the rate at which matter is transferred 
from the donor, $\mathrm{\Delta M_{lost}}$ is the amount of mass lost 
via RLOF and $\mathrm{\Delta t}$ -- the time step. Our results suggest 
that for a given initial model the larger is the accretion rate, the 
more rapid is the expansion, so that the transition to the RG regime 
occurs at smaller WD total mass (see Fig. \ref{f:expa}). However, after 
a short transition phase, all the models with the same initial mass 
converge to the same limiting value, clearly indicating that 
\mdote$\mathrm{_{RG}}$ depends on the thermal content of the CO 
core as determined both by the compression and the thermal energy flowing 
inward from the He-burning shell. When comparing models with different 
initial masses the same conclusion is still valid.
\begin{figure}   
 \centering
  \includegraphics[width=\columnwidth]{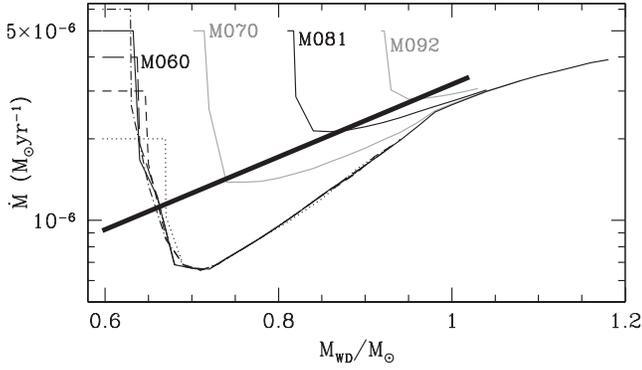}
   \caption{Evolution of \mdote$\rm _{RG}$ as a function of the WD total 
   mass. Solid lines refer to models with different initial mass and the 
   same accretion rate \mdot$\mathrm{=5\times 10^{-6}}$\myr. 
   Dotted, dashed, long-dashed and dot-dashed lines refer to the M060 
   model accreting He-rich matter at $\mathrm{2,\ 3,\ 4\ and\ 6\times 
   10^{-6}}$ \myr, respectively. The heavy solid line represents the upper 
   limit of the Steady Accretion zone, as  in Fig.~\ref{f:regime}.}
 \label{f:expa}           
\end{figure}

If an accreting WD is in the RG regime, it is quite reasonable to assume 
that the mass excess with respect to the maximum value defined by 
Eq.~(\ref{e:mrg}) is lost by the system. However, the further evolution 
of this kind of binary systems strongly depends on the amount of angular 
momentum carried away by the lost matter as it determines the evolution 
of the separation and, hence, the rate at which the donor continues to 
transfer mass. As a matter of fact, if the mass transfer occurs at a rate 
typical of the RG regime both the components of the binary system should 
become immersed in a common envelope. 
\begin{figure}   
 \centering
  \includegraphics[width=\columnwidth]{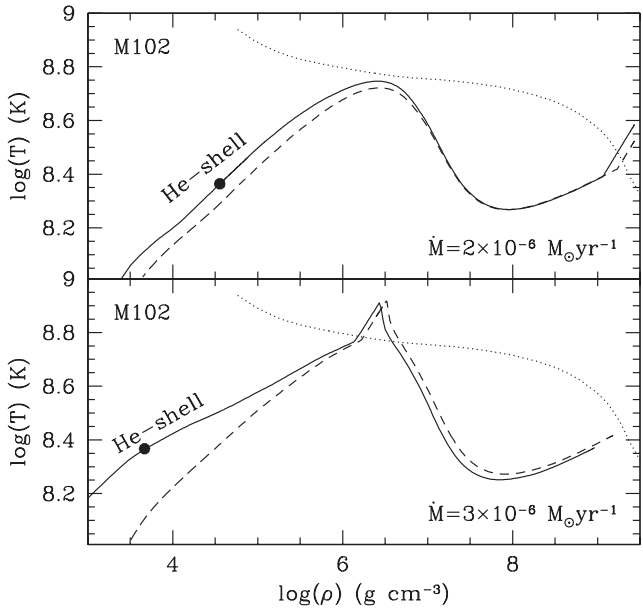}
   \caption{Profiles in the $\rho -T$ plane for some selected models 
            attaining the physical conditions suitable for C-ignition. 
            Solid and dashed lines refer to He-accreting and CO-accreting 
            WDs, respectively. The location of the He-burning shell in
            He-accreting models is marked by a filled dot. The dotted line 
            represents the ignition line.}
\label{f:coigni}           
\end{figure}

\section{Evolution up to the C-ignition}
\label{s:c_igni}

In principle, if the He-donor is massive enough, accreting WDs could 
attain the physical conditions suitable for C-ignition. In this regard, 
it is important to remark that in He-accreting WDs a part of the nuclear 
energy released via nuclear burning is transferred inward though it 
does not provide a significant heating of the underlying CO core. In 
fact, the thermal evolution of the CO core is driven by the deposition 
of CO-rich material, the ashes of the He-burning shell.
As a consequence, it can be argued that for values of the accretion rate 
lower than $\rm (1-2)\times 10^{-7}$ \myr C-burning will be never ignited 
in He-accreting CO WDs. If CO-rich matter is directly accreted onto 
CO WDs, for such values of \mdot the structure 
can grow in mass up to the Chandrasekhar mass limit, when the strong 
homologous compression determines the physical conditions suitable for 
C-ignition at the center (for example, see \citealt{pier2003a} and 
references therein). However, according to the results discussed in 
\S~\ref{s:strongfl}, for \mdot$\rm \le 2\times 10^{-7}$\myr, 
He-accreting WDs enter the Strong Flashes regime, independently of their 
initial mass and the amount of matter effectively accreted decreases 
very rapidly to zero as \mwd\ increases. For this reason, in the 
considered range of accretion rates, the final outcome is a very massive 
CO WD. 

For larger \mdote, the extant results for WD accreting CO-rich matter 
suggest that central C-ignition occurs for \mdot$\rm \le 10^{-6}$\myr, 
while for larger values C-burning is ignited off-center (see 
\citealt{pier2003a} and references therein). In order to define the 
maximum \mdote$\mathrm{_{He}}$ for which a C-deflagration Supernova could 
occur, we computed the long-term evolution of He-accreting WDs with 
accretion rates (1 -- 3)$\times 10^{-6}$\myr. The results are summarised 
in Table~\ref{t:coigni}, where we report the mass coordinate where C-burning 
is ignited, the values of density and temperature at that point and the total 
mass at the ignition time\footnote{We define the ignition point as the mass 
coordinate where the nuclear energy production rate via C-burning is equal 
to the neutrino energy losses rate. The epoch at which such a condition 
is fulfilled defines the ignition time and, hence, the value of the WD 
total mass reported in Table~\ref{t:coigni}.}. In Fig. \ref{f:coigni} we 
plot the profile of the last computed structure in the $\rho -T$ plane 
for the model M102 accreting at \mdot$\rm =2\times 10^{-6}$\myr\ and 
\mdot$\rm =3\times 10^{-6}$\myr\ (upper and lower panels, respectively). 
Solid lines refer to He-accreting WDs, while dashed lines to CO-accreting 
WDs. As it can be noticed, models igniting Carbon at the center have the 
same physical properties, independently of the chemical composition of 
the accreted matter. In this case, the evolution up to the ignition is driven 
by the contraction of the whole accreting WD as it approaches \mch . At 
variance, the temperature and density profiles in models igniting Carbon 
off center depend on the presence or not of the He-burning shell. In fact, 
since C-burning occurs very close to the surface, thermal energy flowing 
from the He-burning shell keeps hotter the underlying CO layer. 

According to the previous considerations, central C-burning in highly 
degenerate physical 
conditions can occur only if $\mathrm{8\times 10^{-7}\le\dot{M}_{He}\le 
2\times 10^{-6}}$\myr . All the He-rich matter accreted in the Steady 
and Mild Flashes regimes is retained by the WD and determines the growth 
in mass of the CO core. The matter accreted in the Strong Flashes regime 
is only partially retained (when \mdot is very low it is not retained 
at all). When the WD enters the Dynamical Flashes regime, it could explode if 
the necessary amount of He-rich matter appropriate to its current mass 
is accreted. 

\section{Retention efficiency of H-accreting WDs.}\label{s:h_eff}

In SD systems, where the donor has a H-rich envelope, a He-rich layer can 
be piled up onto the CO~WD if the H-accretion rate is larger than the 
upper limiting rate for the strong nova-like H-flashes \citep[e.g. 
see][and references therein]{cas98}. If H-accretion occurs steadily, 
the rate of H-burning and, hence, of He-accumulation are by definition 
equal to the accretion rate. 
On the other hand, if the H-accreting WD 
experiences mild H-flashes\footnote{\citet{cas98} assumed that 
mass loss from H-accreting model can be triggered by the dynamical 
acceleration of matter during a novalike flash or induced by a 
RLOF in the case of a non dynamical flash. In the latter case they 
addressed as ``mild H-flasher'' those WDs that do not expand 
to giant dimensions after 
the flash episode, so that all the H-rich matter previously accreted is 
nuclearly processed piling-up a He-rich layer. 
In more recent works \citep[e.g.][and references therein]{wolf2013,idan13} 
this classification has been abandoned because the authors take into account 
the possibility that wind mass loss triggered by the Super-Eddington luminosity 
could occur as a consequence of the H-flash.}, the matter accreted before the H-flash is 
nuclearly processed during the high luminosity phase after the H-flash 
so that {\sl averaging} along the loop the rate of He-deposition is 
almost equal to the rate of H-accretion. 
\begin{table*}  
 \caption{Selected physical properties of models M060, M070, M081 and 
          M092 accreting H- and He-rich matter at a rate given by 
          Eq.~(\ref{e:h_steady}). The listed values of the He-shell 
          position ($\mathrm{M_{He}}$), its density ($\mathrm{\rho_{He}}$) 
          and temperature ($\mathrm{T_{He}}$), the mass of the He-rich 
          ($\mathrm{\Delta M_{He}}$) and H-rich ($\mathrm{\Delta M_{H}}$) 
          layers refer to the epoch of He-flash ignition. 
          $\mathrm{L_{He}^{max}}$ represents the maximum luminosity 
          produced via He-burning during the flash episode. In the last 
          row we report the retention efficiency for the computed models.}
 \label{t:h_steady}    
 \centering          
  \begin{tabular}{l c c  | c c  | c c | c c }
   \hline\hline
   \multicolumn{1}{c}{\it } &
   \multicolumn{2}{c}{M060} & \multicolumn{2}{c}{M070} &
   \multicolumn{2}{c}{M081} & \multicolumn{2}{c}{M092} \\
   \hline
   \multicolumn{1}{c}{\it }                   & He-Acc.  &    H-acc.   & He-Acc.  &  H-acc.   & He-Acc.   &    H-acc.   & He-Acc.  &    H-acc.   \\
   \hline
   $\mathrm{M_{He}}$                             & 0.5871   &    0.5866   &  0.6983  &  0.6981   &  0.8092   &    0.8090   &  0.9188  &   0.9186    \\
   $\mathrm{\rho_{He}\ (10^4 g cm^{-3})}$        & 4.9218   &    3.8365   &  4.6535  &  3.2002   &  4.2944   &    2.8316   &  5.6805  &   3.4903    \\
   $\mathrm{T_{He}\ (10^8 K)}$                   & 1.3902   &    1.7216   &  1.4458  &  1.8366   &  1.5054   &    1.9540   &  1.4653  &   1.9311    \\
   $\mathrm{\Delta M_{tran}\  (10^{-2} M_\odot)}$& 1.3985   &    1.4413   &  0.9268  &  0.8033   &  0.5281   &    0.4460   &  0.4156  &   0.3076    \\
   $\mathrm{\Delta M_{He}\ (10^{-2} M_\odot)}$   & 9.2495   &    9.2780   &  4.9761  &  4.5877   &  1.4159   &    1.3212   &  1.1457  &   1.0369    \\
   $\mathrm{L_{He}^{max}\ (10^{42} erg s^{-1})}$ & 3.1191   &    3.0973   &  7.3273  &  4.1301   &  7.8490   &    5.2372   & 32.5150  &   7.4220    \\
   $\mathrm{\Delta M_{H} \ (10^{-4} M_\odot)}$   &  ---     &    2.2862   &   ---    &  0.9673   &   ---     &    0.5615   &   ---    &   0.1640    \\
   \acef                                         & 0.659    &    0.649    &  0.5209  &  0.5797   &  0.6561   &    0.5928   &  0.3236  &   0.4518    \\
   \hline                  
  \end{tabular}
\end{table*}

Usually the long term evolution of H-accreting CO WDs is determined by 
assuming that the behaviour of the He-rich layer is equal to that in a model 
accreting directly He-rich matter at the same rate. According to the 
range of \mdot values derived by \citet{cas98} and \citet{pier1999} 
for the steady H-burning and mild H-flashes in H-accreting WDs, the 
helium layer piled-up via nuclear burning can undergo either a dynamical 
or a strong non-dynamical flash. However, on a general grounds, the 
H-burning shell can modify the thermal content of the He-rich zone, so 
that H-accreting WDs could have an evolution different from those 
accreting He-rich matter at the same \mdote. \cite{pier2000} demonstrated 
that, for low values of $\mathrm{\dot{M}_H}$, corresponding to the 
Strong He-Flashes regime, the mass of the He-rich layer becomes so 
large that the 
physical base of the He-shell, where the He-flash will be ignited, is 
insulated from the overlaying H-burning shell. In this case the mass 
of the He-rich zone as well as the temperature and the density at the 
He-ignition are practically the same in models accreting H- or He-rich 
matter at the same rate. On the other hand, \cite{cas98} showed that 
for WDs accreting H-rich matter in the Steady regime thermal energy 
flows from the H-burning shell inward, keeping the whole He-rich zone hotter with 
respect to models accreting directly He-rich matter at the same rate. As 
a consequence, in H-accreting models He-burning is ignited when the mass 
of the He-rich zone is smaller. This might have important consequences in 
determining the total matter retention efficiency of H-accreting WDs. 
\begin{figure}  
 \centering
  \includegraphics[width=\columnwidth]{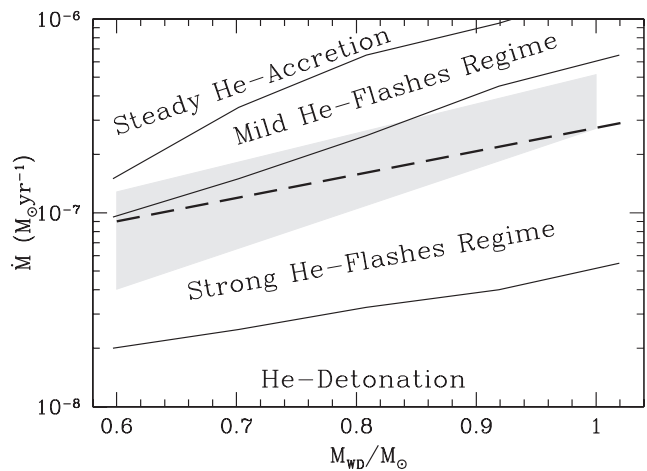}
  \caption{The parameter space suitable for steady accretion of H-rich 
           matter is displayed (grey area) overimposed to the accretion 
           regimes derived in the present work for He-accreting WDs. 
           Heavy dashed line corresponds to Eq.~(\ref{e:h_steady}).}
  \label{map_hacc}           
\end{figure}

To better illustrate this issue we computed some additional models, by 
accreting H-rich matter onto the M060, M070, M081 and M092 {\sl Heated Models}. 
The chemical composition of the accreted matter has been set to X=0.7, 
Y=0.28, Z=0.02, elements heavier than helium having a scaled-solar distribution. 
Moreover, we assumed that the CO WD is accreting in the Steady H-burning 
regime at the fixed accretion rate 
\begin{equation}
\mathrm{
\dot{M}_H=\left(4.543\cdot \frac{M_{WD}}{M_\odot}-1.817\right)\times 10^{-7} M_\odot yr^{-1}.
}
\label{e:h_steady}
\end{equation}

In Fig. \ref{map_hacc} we show the steady accretion regime for H-accreting 
models (grey area) as derived by \citet{pier1999}, overimposed to the possible 
accretion regimes of He-accreting WDs as derived in \S~\ref{s:regimes}. 
The long dashed line represents the value of $\mathrm{\dot{M}_H}$ provided 
by Eq.~(\ref{e:h_steady}). 

For comparison, we compute also models with the same accretion rate, but 
with chemical composition of the accreted matter X=0, Y=0.98, Z=0.02 as 
in \S~\ref{s:regimes}. In Table~\ref{t:h_steady} we report some relevant 
quantities of the computed models. 

In the M070, M081 and M092 models the total amount of mass transferred 
to the WD $\mathrm{\Delta M_{tran}}$ as well as the mass of the He-rich layer  
at the onset of the He-flash $\mathrm{\Delta M_{He}}$ are lower in the 
H-accreting case. Moreover, due to the flow of thermal energy from the 
H-burning shell inward, the bottom of the He-shell, where the He-flash 
is ignited, is hotter and less dense with respect to the same model 
accreting directly He-rich matter. As a consequence, in the H-accreting 
case the He-flash is less strong, as suggested by the lower value of 
$\mathrm{L_{He}^{max}}$. Similar considerations are valid also for the 
M060 case, even if the differences between the H- and He-accreting cases 
are less marked, because the accreted matter represents a small fraction 
($\sim$ 15\%) of the He-rich zone at the onset of the He-flash. Due to the 
He-flash, all the models expand to giant dimension, so we assume 
that the accreting WD fills its Roche lobe when its surface radius 
becomes larger than 10 $\mathrm{R_\odot}$. At the onset of the RLOF we 
stop the accretion and compute the following evolution up to when the 
WD definitively contracts. The obtained retention efficiency is listed 
in the last row of Table~\ref{t:h_steady}. 

Our results are in good agreement with the findings by \citet{cas98}, 
while they contradict the recent work by \citet{newsham13}. A direct 
comparison of the computations in the latetr work and our results is 
possible only for the M070 model accreting at $\rm \sim 1.9\times 
10^{-7}$ \myr corresponding to the model with \mwd=0.7 \ms, 
accreting H-rich matter at \mdot=1.6$\rm \times 10^{-7}$ \myr. In this 
case we find that the strong He-flash is ignited when a mass of $\sim 
0.008$\msun has been accreted onto the WD, while \cite{newsham13} halted 
their computation after the total accreted mass is $\sim 0.00175$\ms, 
i.e. well before the physical conditions for igniting He-burning in the 
He-rich layer could be attained. 
    
Recently, \citet{idan13} (hereinafter ISS13) analyzed the very-long term evolution of 
H-accreting massive WDs. They considered high value of \mdote, 
corresponding to steady H-burning in a ''red-giant-like-star'' and they 
find that all the computed models experience a very powerful He-flash 
driving to the expulsion of the whole accreted He-layer. Their models 
can not be compared directly with the results of our computations because 
we consider initial WDs definitely less massive. Moreover, 
ISS13 included in their computation the effects of thick wind 
which reduces the average value of the accretion rate. In the case of 
their model with \mwd=1.00 \msun accreting H-rich matter at \mdot=$\rm 10^{-6}$
\myr the strong He-flash occurs when the mass of the He-rich layers attains $\rm \Delta 
M_{He}=9.9\times 10^{-3}$\ms, after about 33000 yr (4153 H-flashes with 
a period of 8 yr). This corresponds to an average growth rate of the 
He-rich layer of about $\mathrm{\dot{M}^*_{He}\sim 3\times 10^{-7}}$\myr. 
This model can be compared with our M102 model accreting He-rich matter 
directly at the same accretion rate, even if some caveats has to be borne in mind. 
The initial model in ISS13 is a bare CO core 
with a temperature of $6\times 10^7$ K at the center and of $\sim 10^7$ 
K at the border of the He-deprived core, while in our 
M102 ``{\sl Heated Model}'' the CO core is capped by a He-rich mantle 
and the temperature in the zone below the He-burning shell is larger (see 
Table~\ref{t:initial}). 
Moreover, model M102 has already experienced one very powerfull He-flash 
(pre-heating procedure) while ISS13 focus their attention on the very 
first He-flash. Note that in the latter model the recurrent H-flashes experienced 
by the accreting WD deliver an amount of nuclearly energy definitively 
lower than the pre-heating He-flash in our M102 model. This implies 
that the thermal content of the layer below the He-burning shell in the initial 
model adopted by ISS13 is definitively lower than in our M102 ``Heated Model''.
At the end, as discussed at the beginning of this section, the mass of the layer 
above the He-burning shell at the epoch of the He-flash could be lower 
in WD accreting helium as a by product of H-burning 
because thermal energy is diffused inward from the 
burning shell, thus allowing He to be ignited when a smaller amount of He-rich 
matter has been piled-up. 
In our model M102 accreting He-rich matter directly at $\mathrm{\dot{M}^*_{He}}$
a strong He-flash is ignited after the deposition of 
$\rm \sim 2.12\times 10^{-3}$\msun of matter; the He-flash is ignited at the mass 
coordinate 1.02050\msun and the mass coordinate of the He-burning shell 
moves inward during the flash-driven convective episode down to the
mass coordinate 1.02018\msune. Hence, in our model the mass of the 
He-rich layer above the burning shell 
is at maximum $\rm \sim 2.42\times 10^{-3}$\msun (see Table~\ref{t:sf_single}),  
a factor $\sim$ 4 smaller than 
in the ISS13 computation. In 
our computation the maximum temperature attained during the He-flash 
is $\rm T_{He}=4.87\times 10^8$~K, while the maximum luminosity 
delivered via He-burning is as high as $\mathrm{1.10\times 10^{9} L_\odot}$
(see Table~\ref{t:sf_test}).

Note that the M102 ``{\it Cool Model}'' accreting He-rich matter 
at \mdote$=10^{-7}$\myr has negative retention efficiency (see the 
discussion in \S \ref{s:code}). At variance, the M102 ``{\it Heated Model}''
accreting He-rich matter at the same rate has a small but positive accumulation 
efficiency (\acefe$\le 2$\%). Such an occurence shows that the strength of the 
He-flash depends on the physical conditions (mainly the temperature) below 
the He-burning shell. In order to verify if the discrepancy between 
our results and those in ISS13 depends on the thermal content 
of the adopted initial CO WD, we evolve our M102 ``{\sl Cool Model}'' 
up to the instant when its 
temperature at the center decreases to $4.36\times 10^7$ K and 
its central density increases to $3.98\times 10^7$ ${\rm g\,cm^{-3}}$.
At this epoch the mass coordinate where the He-abundance is larger than 
0.01 by mass fraction has a temperature of $\sim 3\times 10^7$ K
This structure is more similar to the one adopted as starting model 
by ISS13, even if the temperature at the physical 
base of the He-rich layer is a factor of 3 larger. Hence, we accrete 
directly He-rich matter at 
$\mathrm{\dot{M}^*_{He}}$ and we find that a very strong He-flash 
is ignited at the mass coordinate 1.02230 \ms when the accreted 
mass is $\rm \sim 7.12\times 10^{-3}$\ms. During the flash-driven 
convective episode the He-shell moves inward to the mass 
coordinate 1.02017\msune, so that the mass of the He-rich layer 
above the He-burning shell is $\rm \sim 7.55\times 10^{-3}$\msune.
In this case  
the maximum temperature attained during the He-flash and the maximum luminosity 
related to He-burning are $\rm T_{He}=5.31\times 10^8$ K and
$\mathrm{4.73\times 10^{11} L_\odot}$, respectively. 
By comparing this last model with the previous one, it comes out 
that, for a fixed value of \mdote, the amount of matter to be 
accreted to trigger a He-flash and, hence, the strenght of the He-flash itself 
do depend on the thermal content of core underlying the point 
where He-burning is ignited. We do not follow the RLOF episode of this 
additional model, but, according to the discussion in \S \ref{s:strongfl}, 
we can assume that the corresponding value of 
the accumulation efficiency is similar to the 
one obtained for the M102 ``{\sl Heated Model}'' accreting 
He-rich matter at $8\times 10^{-8}$\myr, i.e. \acefe$\le 0$. 
According to the data in Table~\ref{t:sf_single}, this latter model 
accretes $\sim 7.25\times 10^{-3}$ \msun during the evolution prior to 
the RLOF episode and the maximum mass of the He-rich layer above the 
He-burning shell is $7.54\times 10^{-3}$\msun. 
Moreover, during the He-flash the maximum luminosity related to He-burning is 
as high as $7.50\times 10^{11} L_\odot$. 
This result confirms that the discrepancy between our findings and 
those by ISS13 depends only on the thermal content of the initial model. 
In particular, our ``Heated'' M102 model has already experienced a strong 
He-flash, which modified the thermal content below the He-burning shell, while
the initial CO core adopted by ISS13 is very cold, as they do not adopt any 
pre-heating procedure. 

\section{Some Applications}\label{s:applic}

We discuss below several types of binaries with accretion of helium 
onto CO WDs.  Helium WDs and helium stars in close binaries (CB) form in 
the so-called ``case B'' of mass-exchange, when the stars overflow Roche 
lobes in the hydrogen-shell burning stage \citep{kip_weig67}. 
During the subsequent 
evolution, some of them can stably transfer mass onto companion. For the purpose of 
this paper we are interested only in helium WDs and helium-stars with
the lowest rates of mass-transfer, which do not result in formation 
of extended envelopes of WD ($\mathrm{M_{He}\aplt 1.5}$\,\ms).

\subsection{CO WDs in semidetached systems with helium WD companions} 

Helium WD companions to CO WDs have precursors with main-sequence mass 
$\aplt(2.0-2.5)$\,\ms. If the time scale of angular momentum loss 
by a detached pair of He and CO WDs via 
radiation of gravitational waves is shorter than the Hubble time, 
He WD which has larger radius than its companion may overfill its Roche lobe,
forming an ``interacting double-degenerate'' system.
Observationally, these systems are identified with AM CVn stars 
\citep{pac67a}. \citet{nyp01a} nicknamed this variety of AM CVn's 
``WD-family''. Their precursors may be hidden, e.g., among so-called 
``extremely low mass'' (ELM) detached binary white dwarfs with the mass 
of visible component $\aplt0.2$\,\ms, a significant fraction of which 
is expected to lose mass stably after the contact \citep[see][and references 
therein]{brown2013}. Evolutionary considerations and conditions for 
stable mass exchange limit initial masses of the donors and accretors in 
the WD family of AM CVn stars by (0.1 -- 0.3)\,\ms\ and (0.5 -- 1.0)\,\ms, 
respectively \citep{nyp01a,marsh2004,solh2010}. 
As it was shown by \citet{tutu1996} and \citet{nyp01a}, time-delay between formation 
of a detached pair of He- and CO-WD and the onset of mass-transfer
may last from  several Myr to several Gyr. Therefore,  finite entropy 
of Roche-lobe filling WD should be taken into account in evolutionary computations
\citep[see][and references therein]{deloye07}. Non-zero entropy of the 
donors is reflected in the degree of their degeneracy, which may be 
characterized by central ``degeneracy parameter'' 
$\psi=E_{F,c}/kT_c \approx \rho_c/(1.2\times10^{-8}T_c^{3/2}\mathrm{ K^{3/2})\,cm^3\,g^{-1}}$, 
where $\rho_c$ , $T_c$ , $E_{F,c}$ are the
central density, temperature, and electron Fermi energy, respectively. 
``Mass of the donor -- mass loss rate'' relations for two 
systems: (${\mathrm M_{d,0}/\ms,\,M_{a,0}/\ms},\,\psi)=(0.3,\,0.525,\,1.1)$ and 
(0.3,\,1.025,\,3.0) are plotted in  Fig.~\ref{f:amtracks}. 
The former system represents an example of a binary with
a ``hot'' low-mass donor in which RLOF occured very soon after formation,
while the latter system is an extreme 
example of the system with a ``cold'' donor and massive accretor. 
Typical evolutionary tracks for AM~CVn stars should be located 
between these two curves.
The tracks, computed by full-scale evolutionary code with realistic EOS 
and opacities under the assumption of completely conservative mass exchange 
were kindly provided by  C.~Deloye \citep[see][]{deloye07}.
Actually, if completely nonconservative evolution 
is assumed or mass of the donor is varied to also typical value of accretor
mass 0.2\,\ms,  the tracks  in the $\mathrm M_{WD}$ - \mdot 
plane practically do not differ.
Heavy dots overplotted on the lower line in  Fig.~\ref{f:amtracks} represent 
the lower limits of \mdot\ for steady burning, mild flashes and strong 
flashes regimes, respectively, as derived in the present study. 
In the strong flashes regime WD experiences about 10 flashes.
White Dwarf accumulates $\aplt 0.1$\msun before entering the Dynamical Flashes regime and, 
since, for $\simeq 0.7$\msun WD it is necessary to accrete at least about 
0.2\msun for a dynamical flash (see Fig.~\ref{f:detmas}), the latter never happens. 
Thus, our results confirm that the strongest ``last'' flash really should exist, 
possibly producing a ``faint thermonuclear supernova''
(SN~.Ia, \citet{bild2007}.
But note, 
none of the observed explosive events suggested to be SN~.Ia was confirmed 
so far \citep{Drout_SN.Ia_13}.
\begin{figure}  
 \centering
  \includegraphics[width=\columnwidth,angle=-90]{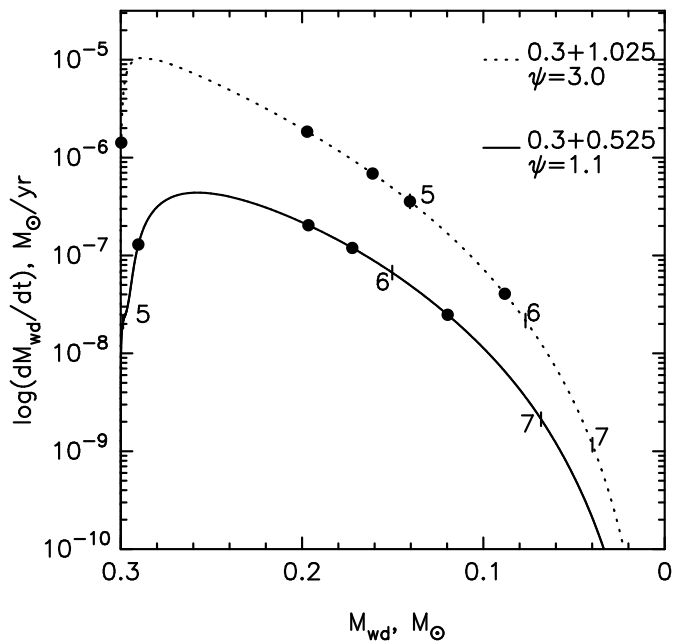}
   \caption{Mass loss rate by finite-entropy He WDs vs. their mass 
            in the systems with initial masses of donor and accretor 
            ($\mathrm{M_{d,0},\, M_{a,0})=(0.3,\,0.525)}$\,\ms\ 
            (solid line) and $\mathrm{(0.3,\,1.025)}$\,\ms\ (dotted line). 
            $\psi$ is the degeneracy parameter (see text).  
            In the former system $\psi=1.1$, while in the latter system $\psi=3.0$.
            Heavy dots at the upper line mark the \mdot limits 
            for RG, steady burning, mild and strong flashes regimes 
            (left to right). 
            At the lower line, the same limits for the latter three 
            regimes are marked, since \mdot in this case never is high 
            enough for the formation of an extended envelope. Ticks on both 
            lines mark time elapsed from the beginning of RLOF -- 0.1, 
            1 and 10 Myr, respectively.}
  \label{f:amtracks}           
\end{figure}

At the beginning of mass transfer in the system with more degenerate donor
and more massive accretor, the donor loses mass for $\simeq15\,000$\,yr at 
a rate exceeding the upper limit for the steady
burning of He by the WD. The mass lost by the donor in this regime is
$\simeq0.087$\,\ms, while the accretor may burn only $0.045$\,\ms. 
It is reasonably to assume that the resulting small amount of ejected matter
cannot lead to the formation of a common envelope.
In the steady accretion, mild and strong flashes regimes the CO WD may accrete 
additionally $\aplt 0.1$\,\ms. Extrapolation of the 
data presented in Table~\ref{t:detona} 
suggests that the accretor will experience a dynamical flash. Even if it
will evolve into a detonation, it is still under debate,
whether detonation of the He-shell may result in a double detonation 
and destruction of the binary 
\citep[see, e.g.,][]{moll_woosley_dd13,shen2014}.
Note, in this case the mass of the accumulated 
helium may be too high to allow the existing theoretical models to reproduce 
correctly observations of SNe \citep{kromer2010}. But such events may 
be hidden among other transients. As well, they hardly contribute 
significantly to the total rate of \sne, since AM CVn stars are rare 
themselves (birthrate $\nu\approx1\times10^{-3}$\,\pyr, \citeauthor{nyp04}
\citeyear{nyp04}) and typically have low-mass accretors.

It is interesting to note that all the possible explosive events in AM~CVn stars
happen during the first several Myr of their lifetime. This means that 
currently in the Galaxy exist only several $10^3$ of these binaries  which
may still be ``nuclearly active'', while the rest of them may show only
accretion-related activity.

However, it is worth to note that, since the delay-time between the formation 
of CO WD + He WD pair and the beginning of RLOF may be up to several Gyr 
and the evolution to a dynamical flash also proceeds in 
Gyr-long time scale, explosions may occur in galaxies of any 
morphological type.          
\begin{figure}
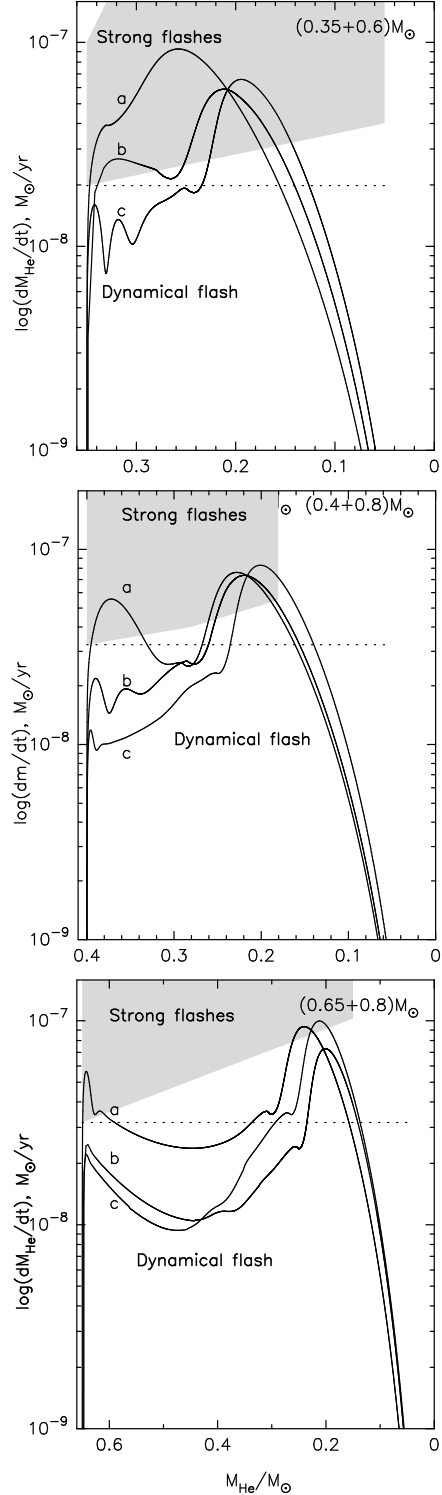
  
 \centering
  \includegraphics[scale=0.32]{fig14a.eps}
\vskip -0.4cm
  \includegraphics[scale=0.32]{fig14b.eps}
\vskip -0.4cm
  \includegraphics[scale=0.32]{fig14c.eps}
   \caption{Mass loss rate by nondegenerate He-donors in 
            $(M_{\mathrm{He}\star},\,M_{\mathrm{WD}})$ binaries
            vs. mass of the He-star.
            Upper panel --- (0.35,\,0.6)\,\msun pair; 
            lines a, b, c show evolution of the systems with   
            (initial period $P_0$ , He-abundance in the core) = 
            (20 min.,\,0.98), (100 min.,\,0.64), and (144 min,\,0.118), respectively.
            Middle panel --- the same for (0.4,\,0.8) pair, with 
            (20 min.\,0.98), (100 min,\,0.51), and (140 min,\,0.09).
            Lower panel --- the same for (0.65,\,0.8) pair, with 
            (35 min.\,0.98), (80 min,\,0.4), and (85 min,\,0.29).
            In the shaded regions of the plots He burns in the Strong 
            Flashes regime (the lower limits of them correspond to 
            a completely conservative 
            evolution, while dotted lines mark the same limit 
            for completely nonconservative evolution).}
   \label{f:tracks}           
\end{figure}

\subsection{CO WD in semidetached systems with low-mass helium-star companions}
\label{s:wdhestar}

For solar chemical composition, the precursors of nondegenerate He-star 
components in close binaries (with minimum mass ${\rm M_{He}}\approx0.33$\,
\msune) have main-sequence mass $\apgt(2.0-2.5)$\,\msune.  
The He-star mass is related to the mass of its MS precursor as 
$\rm M_{ He} \approx 0.08M_{\rm MS}^{1.56}$ (in solar units).
If ${\rm M_{He}} \la 0.8$\,\ms, the stars 
do not expand during core helium burning stage which lasts up to (500 -- 
700)\,Myr. If during this time-span angular momentum loss via 
gravitational waves radiation brings the two components into contact, helium 
star may fill its Roche lobe and, if the conditions for stable mass loss 
are fulfilled, mass transfer onto the companion starts. 
In systems with a CO WD 
these stars first evolve to shorter orbital periods $\simeq10$\,min.
At the epoch of the period minimum, the He-stars masses decrease to (0.20 -- 0.25)\,\ms\
and they start to lose matter that was nuclearly processed in their convective
cores prior to contact (RLOF quenches very fast the nuclear burning --- 
\citealt{skh86}). 
  
Semidetached low-mass He star + CO WD binaries also are suggested to be a variety 
of AM CVn type stars \citep{skh86,iben1991,ty96}, nicknamed ``He-star family'' 
\citep{nyp01a}. Typical mass transfer rates in these systems before the period minimum 
are $\aplt 10^{-7}$\,\myr\ \citep[][see also Fig.~\ref{f:tracks}]{yung2008}. 

In Fig.~\ref{f:tracks} we show three examples of evolutionary tracks of 
He-stars in semidetached systems with WD companions from \citet{yung2008} 
in ``mass of the donor\,-\,mass-loss rate'' plane. The tracks 
shown in Fig.~\ref{f:tracks} were computed in conservative approximation, 
but this does not affect their behaviour compared to the completely 
nonconservative case \citep{yoon_langer04,yung2008}. Initial combination 
of masses $(M_{\mathrm{He}\star},\,M_{\mathrm{WD}})$=(0.35,\,0.6) is 
deemed to be typical for precursors of He-star AM~CVn systems, while 
(0.4,\,0.8)\,\ms\ and (0.65,\,0.8)\,\ms\ may be more rare 
\citep[see Fig.~3 in][]{nyp01a}. 
For each combinations of He-stars and CO WDs mass in Fig. \ref{f:tracks} 
we show three characteristic tracks corresponding to the RLOF by a completely 
unevolved star (line a), a star with He in the core consumed by about 40\% to
60\% (line b) and a star with significantly He-depleted core ($Y_c\simeq0.1$, 
line c); for lower 
$\rm Y_c$ overall contraction begins and the stars can not fill their
critical lobes. Shaded regions in 
the plots show the domain of strong He-flashes (after Fig.~\ref{f:regime} 
above). The lower border of this region is drawn under the assumption 
that He-accretion occurs conservatively, i.e., the mass of WD is the sum 
of its initial mass and mass lost by He-star. For stars experiencing 
strong flashes the border between dynamical and strong flashes regime 
is between shaded regions and dotted lines, representing the transition 
for completely non-conservative evolution.

In the (0.35+0.6)\,\ms\ set of systems, the accreted He never 
experiences a dynamical flash since such systems predominantly 
evolve into the Strong Flashes regime of He-burning 
immediately after the RLOF. Moreover, the donor is not massive enough to 
provide enough mass for a dynamical flash 
after entering this accretion regime
(see Table B2 in the Additional material in 
the online version).

In the (0.4+0.8)\,\ms\ set of systems, the accreting WDs stay in the range 
of mass-exchange rates corresponding to the Dynamical Flashes regime 
for a part of the pre-period-minimum time, but hardly accumulate enough 
He to give rise to a dynamical event (Fig.~\ref{f:detmas}). 
In the Strong Flashes regime they experience $\simeq10$ outbursts and, 
even if the corresponding retention efficiency is small, 
the WD mass should increase. 
For this case we may safely assume that by reentering the Dynamical Flashes 
regime the total mass of WDs is close to 1.0\,\ms. 
Then, WDs of this set should experience 
a dynamical flash shortly after the period 
minimum, when several 0.01\,\ms has been accreted. According to 
\citet{fink2010}, the detonation of He-shell in this case may initiate 
the detonation of carbon close to the center of the CO WD.

Finally, in the most extreme case of (0.65+0.8)\,\ms\ systems 
no strong flashes happen, but dynamical flashes of He onto WD may
occur and double detonations may be expected. 

Like semidetached WD+WD systems, low-mass He-stars evolve to the
period minimum in $\sim(10^6\,-\,10^7)$\,yr only. This 
may mean that most of the AM~CVn stars of He-star family existing in the Galaxy 
already experienced their ``last flash'' (SN~.Ia in the case of detonation)
and continue to evolve without any expected thermonuclear events.
It is possible that some would-be AM CVn's 
ceased their existence due to double-detonations shortly after their birth. 
The estimation of the rate of the latter events should be addressed by means of 
population synthesis calculations taking into account relations between critical 
masses for explosive events, retention efficiency and \mdote, which was 
never made before. The same, in fact, is true for the total population of 
AM~CVn's, since in the existing studies its formation rate was 
restricted by rather \textit{ad hoc} assumptions, while the effects of unstable 
He-burning were not taken into account. The relevance of such a new study 
is also emphasized by the fact that the existing ``theoretical'' models 
predict significantly larger Galactic population of AM~CVn stars than 
observed and it has been suggested that the ``theoretical'' models overestimate 
their number \citep[see, e.g.,][and references therein]{carter2014}.

\subsection{Helium-star channels to SN~Ia}
\label{sec:he_snia}

Above, we discussed the evolution of CO WDs accreting He from low-mass stellar companions
($\aplt 0.8$\,\ms) which do not expand after exhausting He in their cores. 
More massive He-stars may overflow Roche lobe both in the core 
He-burning and in the He-shell burning phases \citep{pac_he71}. 
In the latter case the expansion (up to several 100\,\rsun), which occurs in the 
thermal time scale is limited only by the existence of companion.  
It was shown, e.g., by \citet{it85}, \citet{yl03}, \citet{wang09} that 
upon RLOF both core and shell He-burning stars may lose 
several 0.1\,\ms\ at a rate $\sim10^{-(6\pm1)}$\,\myr, which corresponds for 
the accreting CO WDs to burn helium in steady or flashes regimes, 
depending on the orbital period and 
combination of masses of the components. If RG formation is avoided, initially 
sufficiently massive CO~WD components may accumulate \mch.

As examples of possible evolution we show in 
Fig.~\ref{fig:heevol} \mwd -- \mdot dependence for a system 
harbouring initially 1.23\,\ms\ He-star and  a 0.84\,\msun WD.

We consider 2 cases: 
(a) -- initial period of the system $P_{\rm orb,0}$=0.035 day, 
the donor overflows the Roche lobe practically unevolved ($\mathrm{Y_c\approx}$
0.979); 
(b) -- $P_{\rm orb,0}$=0.2 day, $Y_c=0$, mass of the He-depleted core 0.692\,\ms.
\begin{figure}  
 \centering
  \includegraphics[width=0.75\columnwidth,angle=-90]{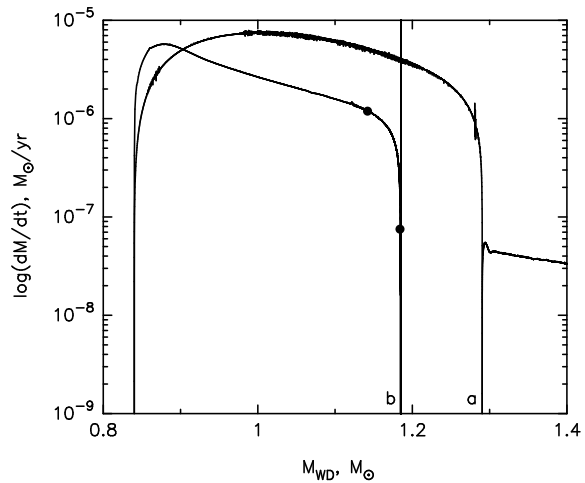}
  \caption{Mass loss rate vs. mass of nondegenerate He-donors in 
           semidetached systems with initial masses 
           $M_{{\mathrm He\star,0}}=1.23$\,\ms, $M_{\mathrm WD,0}
           =0.84$\,\ms\ and initial periods 0.035 day (line a) and   
           0.2 day (line b). Heavy dots overplotted on line b mark 
           the limits of the Strong Flashes regime. A spike of \mdot 
           in line b shows the initial stage of mass transfer in the system 
           formed by the WD-remnant of the He-star and the CO WD.}
  \label{fig:heevol}           
\end{figure}

In the case of the unevolved donor, a short ($\Delta T\approx$0.155\,Myr) 
episode of mass transfer occurs. Then, the accretion rate corresponds subsequently 
to RG, Steady Burning, Mild and Strong Flashes regimes. The amount of  matter 
transferred in the RG regime is 
close to 0.2\,\ms\ and the question whether a common envelope may form,
if optically thick winds are not considered, remains open.
The mass accreted by the WD is not sufficient to attain \mch.
High mass transfer results in a sharp decrease of the nuclear burning rate, 
the donor contracts and detaches from the Roche lobe. Since the system is 
very tight, angular momentum loss via gravitational waves radiation continues to
shrink the orbit. At a certain moment the Roche lobe radius becomes smaller than the 
radius of the star and mass loss resumes. The rate of accretion 
corresponds now to the regime of Dynamical Flashes. Accreting WD is massive 
enough to experience a flash 
virtually immediately after resumption of mass transfer (the mass-loss curve for
the second stage of mass-transfer is, in fact, formal).  

In the case of the evolved system, mass-loss stage reduces to a rapid 
($\Delta\,T\approx$0.18\,Myr) loss of 
most of the He-layer overlying the CO core. Mass-loss does not result in 
quenching of shell He-burning, the core continues to grow and the remnant of 
the donor has a 0.831\,\ms\ He-depleted core, while its total mass is 
0.886\,\ms. It is important to remark that in the Strong Flashes regime 
the donor loses about 0.015\,\ms, while the critical mass of the He-layer
for ignition, as suggested by extrapolation of the data in Fig. \ref{f:accum2}
(see also Table B7),
is about an order of magnitude lower. Thus, WD may experience $\sim 10$ strong flashes 
which may be associated with Helium novae (see \S~\ref{s:henovae}). The mass accumulated 
by the WD may be enough for a dynamical flash when the accretion rate drops into 
the appropriate range.

Double-detonation has been considered as a possible explosion mechanism 
of \sne\ for more than
three decades (see \S~\ref{s:intro}). In particular, \citet{wang13} suggested that 
double-detonations onto sub-Chandrasekhar WD in He-star+CO~WD systems 
produce subluminous supernovae of SN~Iax subtype \citep{foley_iax13}. 
On an observational ground, SN~Iax are suggested to constitute about 1/3 
of all \sne. Note, these events should not currently occur in early-type 
galaxies and, really, none is detected in the elliptical ones. 
Indeed, the evolutionary scenario suggested by \citeauthor{wang13} is similar to 
the one discussed in \S~\ref{s:wdhestar}, but involves also more massive 
donors (up to 1.25\,\ms); an exemplary evolutionary track for such kind of systems
is presented by curve b in Fig.~\ref{fig:heevol}. However, 
the estimated contributions of pertinent \sna\ to the total rate of 
\sne\ in BPS studies, as usually, depend on assumptions in the BPS codes. 
In particular, \citeauthor{wang13} assumed that the only condition for the 
occurrence of a double-detonation is accumulation 
of 0.1\ms\ of He at a rate between $1\times10^{-9}$ and $4\times 10^{-8}$\,\myr on 
a (0.8-1.2)\,\ms\ WD, irrespective of how robustly the resulting dynamical He-flash 
at these conditions leads to SN~Ia. Possibility of strong flashes was neglected.
Note that, also in this case, reproducing the observed features of SN~Iax, 
apart from being faint, is still questionable \citep{kromer2010}.
Moreover, the current Galactic star formation rate of 3.5\,\myr\ assumed 
in the Wang et al. computations is higher than the modern estimates, close to
2\,\myr\ \citep{kenni2012}.
Finally, as mentioned before, CO WDs hardly form with mass exceeding 1.1\,\ms.  
\citeauthor{wang13} 
claim Galactic rates of double-detonations in the considered systems
$\sim 1.5 \times 10^{-3}$\,\pyr, while a more realistic estimate based 
on the data in the quoted study and in the present work, hardly exceeds 
$10^{-4}$\,\pyr.

The possibility of accumulation of \mch\ by He-accreting WD was studied most 
recently by \citet{wang09} by means of BPS, using He-retention efficiencies from
\citet{kato2004}. 
While \citeauthor{wang09}  
claim that the He-donor channel may enable $\sim$30\% of 
the observed \sne\ rate, we should note again that this result depends 
very strongly on the assumption that optically thick stellar wind from the 
accreting WD operates in the RG regime, thus preventing the formation of 
common envelopes, 
and on the limits set for the different accretion regimes. 
In the quoted study mass of the donors in systems with WD accumulating \mch\
may be as high as 2.5\,\ms, while 
our test calculations, performed by relaxing the assumption of existence of 
the optically thick wind, 
show that the maximum mass of the donors is about 1.25\,\ms. For larger masses the 
initial \mdot\ correspond to the RG accretion regime and the formation of common 
envelopes is expected. We note also 
that in the study by \citeauthor{wang09} a significant fraction of \sne\ is produced by
WDs with initial masses between 1 and 1.2 \ms\ and the problem of the origin of
very massive CO~WDs is not considered.
Moreover, the birthrates 
of binaries which contribute to this channel, i.e., have ``proper'' 
$M_{\rm He}$, $M_{\rm WD}$, \porb, heavily rely on the assumed parameters 
in a specific BPS code, especially, on the common envelopes ejection efficiency, 
the estimates of binding energy of the donors, the treatment of the mass transfer 
process in BPS.
For instance, \citet{bly+95} find that the He-star 
channel cannot contribute more than several per cent to the current 
rate of Chandrasekhar-mass \sne\ in the Galaxy. These systems could be 
more important in the early stages of the Galactic evolution when the 
star formation rate was much higher than the current one, since their 
delay time is limited by the sum of lifetimes of the least massive 
He-donors and their precursors ($\approx$ 150\,Myr). 
\begin{figure}  
 \centering
  \includegraphics[width=\columnwidth]{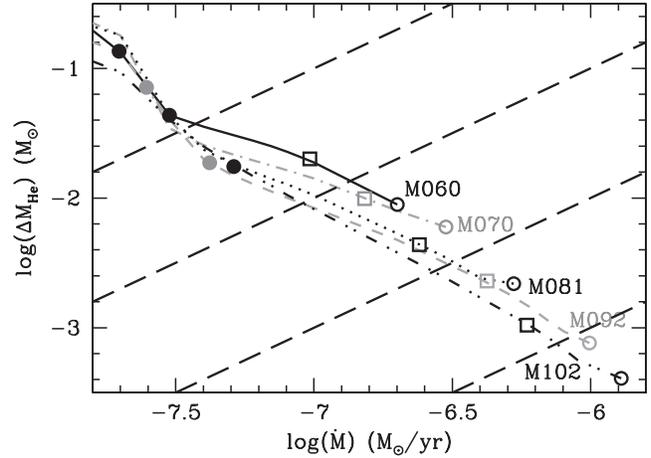}
  \caption{Dependence on the accretion rate of mass of the He-rich layer above 
           the He-burning shell at the onset 
           of the first He-flash for all the models computed in the present work. 
           Different lines are used for different initial WD models, as labeled. 
           Transitions from one 
           accretion regime to another are marked with different symbols: 
           open circles --- From Steady to Mild Flashes regime; open squares 
           --- from Mild to Strong Flashes regime; filled circles ---  
           from Strong to Dynamical Flashes regime. Slanting long dashed lines show 
           recurrence periods from $10^3$ to $10^6$\,yr (bottom to top).}
  \label{f:dmhe}           
\end{figure}

In this scenario the amount of mass which may be transferred by a He-star onto the 
CO-companion and retained by the latter is crucially 
important for the ``helium-ignited violent merger'' scenario for \sne\, 
\citep{pakmor_violent_13}, since, as noted in the Introduction, 
exploding WD accretors should be massive \citep[see discussion in][]
{ruiter_13_violent}. We remind, however, that the evolution of He-star+\cwd\ 
systems is very poorly explored, some of combinations of He stars and 
WD evolve differently than assumed in population synthesis studies, as 
demonstrated by Fig.~\ref{fig:heevol}. Full-scale evolutionary calculations 
of stellar models and their parametrization in population synthesis codes 
produce both different \mdot and amounts of mass lost by He-star 
$\rm \Delta{M_{He}}$. For instance, compare full-fledged evolutionary 
computations for 6.95\,\ms\ star by \citet{it85} and parametrized 
calculations for 6.67\,\ms\ star in \citet{ruiter_13_violent}: in the 
latter $\rm \Delta M_{He}$ is more than twice larger than in the former: 
0.46\,\ms\ \emph{vs.} 0.21\,\ms, thus, much more favourable for 
``He-ignited violent mergers'' scenario. Another issue is whether the amount 
of mass in the He-skin(s) of merging WDs is sufficient for ignition of 
initial He-detonation. 
In evolutionary computations, nascent CO WDs entering the cooling sequence 
retain only traces of He at the surface 
\citep[$<0.001$\,\ms,][]{it85}. \citet{pakmor_violent_13} postulate 
presence of 0.01\,\ms\ of He atop CO-cores of merging WDs. 
\begin{figure*}
 \centering
  \includegraphics[width=0.95\textwidth]{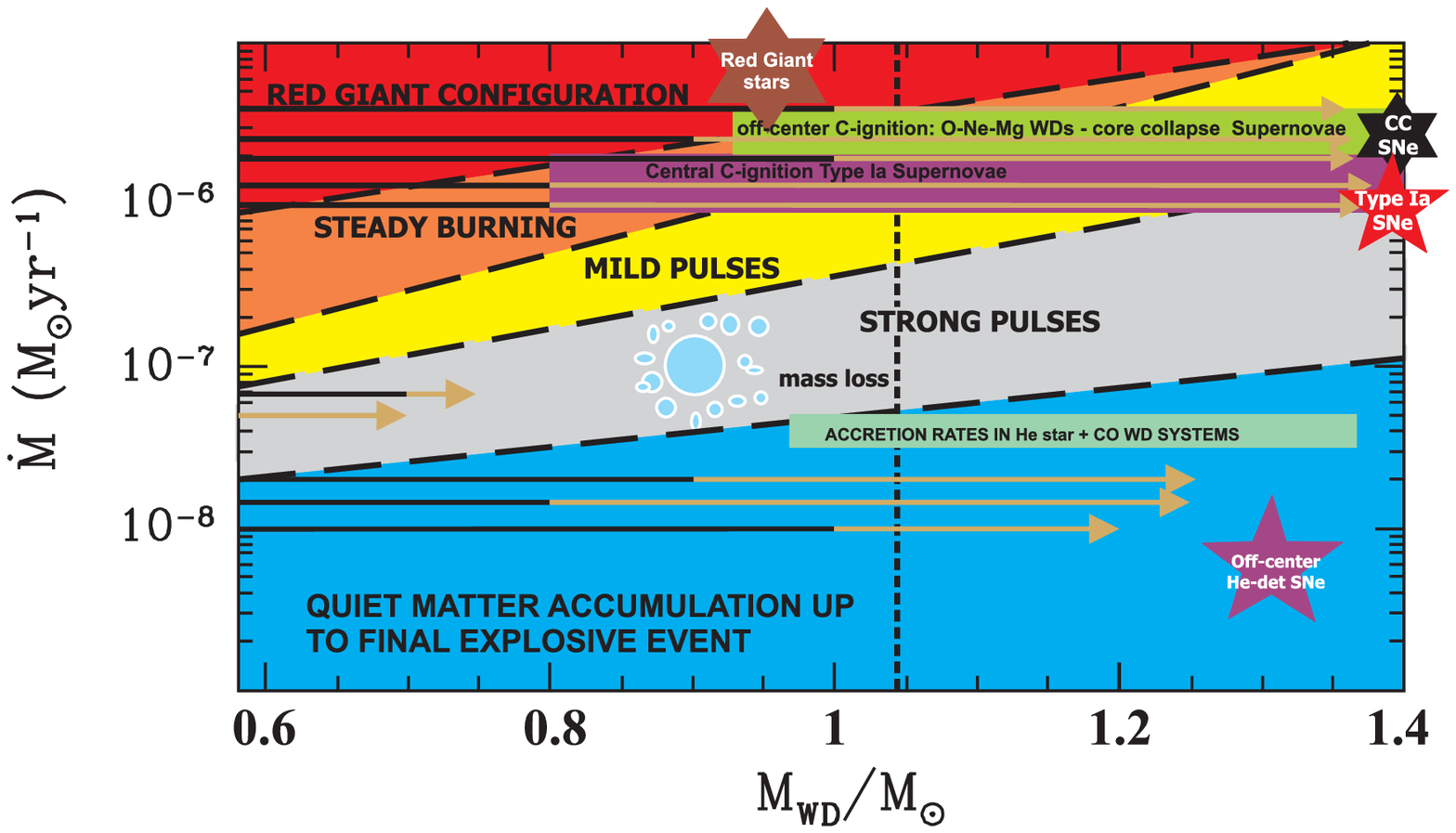}
   \caption{Accretion regimes for He-accreting WDs as a function of the 
            WD total mass (\mwd) and the accretion rate (\mdote). 
            The limits of the different accretion regime 
            on the left of the dotted vertical line has been extrapolated 
            according the results of our computations.
            The possible final outcomes of the accretion process are also 
            displayed. Black horizontal lines represent the initial WD 
            masses while dark-yellow arrows trace the accretion process.}
   \label{f:HE+}           
\end{figure*}

\subsection{He-novae}
\label{s:henovae}

In the Strong Flashes regime, He-accreting WDs probably may manifest 
themselves as He-novae -- an analogue of novae associated with 
thermonuclear runaways onto WD accreting hydrogen \citep{ksh89}. 
In Fig.~\ref{f:dmhe} 
we show dependence of He-ignition masses (masses above the coordinate 
of maximum $\mathrm{\epsilon_{nuc}}$) on mass accretion rate. $\rm 
\Delta M_{He}$ shown in the Figure are lower than the ones computed 
by \citet{kato2008} for similar combinations of \mdot and masses and, 
respectively, recurrence time of flashes is shorter. 

Helium novae are represented to day by only one still not well studied 
object --- V445~Pup \citep{ashok03b,woudt_steeghs05a,ijima2008,woudt2009} 
and several candidate objects which show similar light curves and 
overabundance of He \citep{rosenbush_he_nov08}. V445~Pup apparently, does 
not belong to the ``He-star'' family of AM CVn stars. \citet{woudt2009} found 
that the pre-outburst luminosity of the object was $\log(L/\ls)=4.34\pm0.36$. 
Such a luminosity is compatible with the donor being a {${\mathbf 1.2 - 1.3}$}\,\ms\ 
star burning helium in a shell. If the variability of the system with period 
$\approx 0.65$~day \citep{goran2010} reflects the orbital motion in the system, 
it favours such an interpretation. \citet{kato2008} inferred that 
WD-component of V445~Pup is very massive ($\apgt 1.3\,\ms$), based on the 
fitting of the observed Nova light curve in the framework of the optically 
thick wind theory. 
Based on the test calculations presented in Fig.~\ref{fig:heevol},
we may suggest that the initial mass of the WD in V445~Pup could be close to 0.8\,\ms,
while now it is about 1\,\ms. 

Apparent extreme rarity of He-novae may be due 
to low birthrate of systems with initially massive enough WD and short 
duration of He-shell burning stage. Regretfully, the real orbital period of 
the system is unknown, thus hampering the identification of a possible 
companion to the WD. 

\section{Final remarks}\label{s:final}

We explored systematically the thermal response of non-rotating WDs to 
direct accretion of helium. The initial WD masses adopted in our analysis 
cover the whole range of CO WDs expected to form in close binaries due 
to mass loss by AGB components --- from 0.6 to 1.02 \ms. As well, the 
range of studied accretion rates encompasses evolutionary significant 
range $(1\times10^{-9} - 2\times10^{-6})$\,\myr. Special attention was 
paid to the processes involved in energy exchange between the underlying 
CO core and accreted He-layers. Unlike previous studies, in accretion 
regimes driving to the expansion of the WD, due to either thermonuclear 
flashes at the base of accreted layer or the WD inability to consume 
nuclearly all accreted matter, we considered the possibility of mass 
and angular momentum loss from the system as determined by the 
interaction of the WD envelope with the companion\footnote{To our 
knowledge, this mode of mass loss was studied by \citet{it96symb} 
for a single example of 1\,\ms\ accretor only.}. For convenience of 
discussion, in Fig.~\ref{f:HE+} we reproduce different evolutionary 
models involving direct He-accretion at constant \mdot overimposed to the plot showing 
different burning regimes for He-accreting WDs. Note, abscissa has been 
extended up to 1.4\,\ms\ in order to show different pathways to \sne. 
The limits of the various accretion regimes for WD total 
masses larger than 1.05 \msun (the zone in the plot on the 
right of the dashed vertical line) have been extrapolated on 
the base of the results presented in \S 3 and \S 4.

In agreement with previous studies, we found that, at {\sl constant} 
low accretion rates (from $\aplt 5\times 10^{-8}$\,\myr\ for 1.02\,\msun 
WD to $\aplt 2\times 10^{-8}$\,\myr\ for 0.6\,\msun WD), the accretion 
time scale is much longer than the inward thermal diffusion time scale 
so that conditions at the base of He-layer are defined by the interplay 
between neutrino cooling and energy release by contraction of accreted 
matter. Helium is deposited quietly at the surface of the CO dwarf and 
piled up for a long time without increase of surface temperature and 
luminosity, while temperature at the base of the He-shell increases. 
He-ignition occurs in highly degenerate matter and it becomes dynamical  
by its nature. We found that the minimum masses of He accumulated prior 
to the dynamical flash are 0.02\,\ms\ to 0.102\,\ms\ for 1.02\ms\ to 0.8\ms\ WD, 
respectively (Table~\ref{t:detona}). These masses of accreted He are 
rather close to the minimum values of $\rm \Delta M_{He}$ which can 
robustly trigger double detonations in WDs of these masses \citep{fink2010}. 
Such events may happen in AM~CVn type systems with He-star donors. In 
Fig.~\ref{f:HE+} we show 3 examples of evolutionary ``tracks'' for WDs 
of different initial mass with \mdot in the range $1-3\times 
10^{-8}$\,\myr\ accumulating critical layers for a dynamical event. 

The estimates of the contribution of double detonations to the total 
\sne\ rate are controversial. On one hand, BPS for precursors of AM~CVn 
stars by \citet{nyp01a} with parameters reproducing the population of 
WD binaries in the Galaxy shows that double detonation \sne\ should be 
a very rare outcome, since binaries with massive enough accretors and 
He-star or He WD donors enabling appropriate accretion rate are hardly 
formed. On the other hand, \citet{ruiter2011} find that, for certain 
combination of BPS parameters, double detonation of sub-Chandrasekhar 
WDs may be a dominating mechanism for \sne\ with delay times shorter 
than about 5 Gyr; they also found two distinct groups of precursors: 
short-living ones, harbouring He-star donors, and long-living ones, 
with He WD donors. Similarly, massive accretors and high rate of 
double-detonations in He-star AM CVn systems was found by \citet{wang13}. 
The reliability of certain set of results could be demonstrated if, for 
the same sets of BPS input parameters used to model \sne, should be 
possible to reproduce well observed and numerous enough groups of stars 
preceding in evolutionary paths \sne, like local samples of WD binaries.
It is worth noting again that the extant calculations of double-detonating WD 
still fail to reproduce all \sne\ observables \citep{kromer2010}. In 
Appendix \ref{app2} we provide polynomial fits of $\rm \Delta M_{He}$ 
necessary for a dynamical flash as a function of the accretion rate. ``Failed 
double-detonations'', i.e. single detonations of He which do not trigger 
explosion of the underlying core may be hidden among weak \sna\ or other 
transients with magnitudes between those of Novae and Supernovae 
\citep{kasliwal_bridge12}. 

For higher \mdot we find the limits of He-burning in ``strong flashes 
regime''. For semidetached systems composed by low-mass He stars and 
WD donors with evolution driven by gravitational  waves radiation, 
almost constant accretion rate 
may be a fair approximation in the stage before they reach the minimum 
of the periods \citep{yung2008}. For systems with He WD donors and CO 
WD accretors decaying \mdot is typical. For decaying mass transfer rate 
we show that the amount of matter accumulated prior to the strong flash 
is a function of accretion history, with \mdot being a deciding factor. 
This zone may be associated with He-Novae, of which one object is confirmed 
to day (V445~Pup - \citealt{ashok03b}). 
For stars in this zone strong mass loss is typical. If \mdot is 
close to the transition line between Strong and Dynamical Flashes regimes, 
the retention efficiency may be negative. This happens because, during 
the previous evolution, He-burning did not attain the surface of the 
accreted WD, so that a small He-rich zone remained on the top of the CO 
core. Hence, negative \acef implies that the He-flash is ignited in this 
pre-existing helium layer and, moreover, that almost all the matter above 
the ignition point is ejected. 

The values of retention efficiency we obtained are, for WDs of the same 
mass, lower than the ones by KH04. We explain such an occurrence as the
consequence of the assumption that mass loss from the system is the 
result of the interaction of the WD bloated envelope with the companion, 
while KH04 account for mass loss from the accretor according to the 
optically thick wind theory. Polynomial fits to the obtained values of 
retention efficiencies \acef as a function of the accretion rate are 
provided in Appendix \ref{app3}. 

In SD-systems with H-rich donor, if the accretion proceeds in the regime 
of steady H-burning, accumulation of helium via nuclear burning occurs 
at a rate typical for the Strong He-flashes regime. In this case, the 
comparison between models accreting He directly or as a consequence of 
H-burning at the same \mdot shows that in the latter case the mass of the 
He-rich zone is lower at the onset of the He-flash, due to the presence 
of the H-burning 
shell which keeps hotter the underlying He-layer. In any case, the 
resulting accumulation efficiencies are very similar. It is worth noting 
that, according to BPS, the H-rich SD scenario can not account for the 
observed rate of \sne\ unless unrealistic unlimited accumulation of 
hydrogen onto CO WDs is assumed \citep[see the discussion in][]
{claeys_sn_14}.
 
The next two regimes of He-burning are those of mild flashes and steady 
burning. Mild flashes occur if ignition starts in a nondegenerate matter. 
The increase of the WD radius is not large and accumulation efficiency is 
close to 1. In the steady burning regime He is converted into C+O mixture 
at a rate equal to \mdote . A caveat should be entered that steady burning 
is attained only after a strong initial pulse has been experienced by the 
accreting structure in order to settle the external matter in a thermal 
equilibrium appropriate to accrete and burn deposited matter. 
Steady burning and mild flashes regimes open windows in the space of 
parameters suitable for accreting WDs to evolve up to an explosive final. 
For instance, accreting dwarf may experience steady accretion during the 
first stage of the evolution and afterward enter the mild flashes regime, 
thus accumulating mass up to the Chandrasekhar limit. Five detailed 
evolutionary tracks have been followed in this area which end by carbon 
ignition. The models with initial WD mass from 
0.8\,\ms\ to 1.02\,\ms\ with a lower accretion rate succeed to ignite 
carbon in a deeply degenerate environment and an explosive outcome is 
attained. For slightly higher values of \mdote, the maximum of temperature 
is off-center so that Carbon is ignited in an external layer. 
In this case it can be argued that C-burning propagates inward via heat conduction, 
transforming the CO core into an O-Ne-Mg one \citep{saio85,saio98}}. 
Such a structure, while accretion continues, will experience 
an accretion-induced collapse into a neutron star. Those windows of outcomes 
are very narrow in the accretion rate range of parameters when constant 
\mdot is assumed. The most important requirement to get C-ignition as a 
final outcome is that the strong He-flashes are avoided. The windows of 
opportunities can get larger if accretion rates can change during accretion. 
Results of BPS by \citet{wang09} suggest that this channel may be responsible 
for about 30\% of \sne, given that there exists an appropriate observed 
population of sufficiently massive He-stars ($\rm M_{He}\ga0.8$\,\ms) 
with massive WD companions (\mwd$\ga1.0$\,\ms) and that the formation of 
common envelopes at accretion rates exceeding $\rm \dot{M}_{RG}$ may be 
avoided (see \S~\ref{sec:he_snia}).

Finally, for accretion rates from $10^{-6}$\,\myr\ for 0.6\,\ms WD to 
$3.5\times 10^{-6}$ for 1.02\,\ms\ the mass deposition onto the WD 
delivers a huge amount of gravothermal energy.
The consequence is that the transferred matter forms an extended envelope 
acquiring the shape of a red giant. The whole structure becomes embedded 
in the common envelope.

\section*{Acknowledgements}
L.P. acknowledges support from the Italian
Ministry of Education, University and Research under the FIRB2008 
program (RBFR08549F-002) and from the PRIN-INAF 2011 project 
``Multiple populations in Globular Clusters: their role in the
Galaxy assembly''. 
LRY acknowledges support by RFBR grant 14-02-00604 and Presidium 
of RAS program P-21. 
The authors acknowledge C. Deloye for providing detailed evolutionary tracks
for finite entropy WDs.
The authors thank an anonymous referee for helpful comments and suggestions. \\
This research has made use of NASA's Astrophysics Data System.

\bibliographystyle{mn2e}	
\bibliography{pty}

\begin{thebibliography}{120}
\expandafter\ifx\csname natexlab\endcsname\relax\def\natexlab#1{#1}\fi

\bibitem[{{Ashok} \& {Banerjee}(2003)}]{ashok03b}
{Ashok} N.~M., {Banerjee} D.~P.~K., 2003, \aap, 409, 1007

\bibitem[{{Bildsten} {et~al}\mbox{.}(2007){Bildsten}, {Shen}, {Weinberg}, \&
  {Nelemans}}]{bild2007}
{Bildsten} L., {Shen} K.~J., {Weinberg} N.~N., {Nelemans} G., 2007, \apjl, 662,
  L95

\bibitem[{{Bildsten} {et~al}\mbox{.}(2006){Bildsten}, {Townsley}, {Deloye}, \&
  {Nelemans}}]{bilds2006}
{Bildsten} L., {Townsley} D.~M., {Deloye} C.~J., {Nelemans} G., 2006, \apj,
  640, 466

\bibitem[{{Bours}, {Toonen} \& {Nelemans}(2013){Bours}, {Toonen}, \&
  {Nelemans}}]{bours_retention_13}
{Bours} M.~C.~P., {Toonen} S., {Nelemans} G., 2013, \aap, 552, A24

\bibitem[{{Branch} {et~al}\mbox{.}(1995){Branch}, {Livio}, {Yungelson},
  {Boffi}, \& {Baron}}]{bly+95}
{Branch} D., {Livio} M., {Yungelson} L.~R., {Boffi} F.~R., {Baron} E., 1995,
  \pasp, 107, 1019

\bibitem[{{Brown} {et~al}\mbox{.}(2013){Brown}, {Kilic}, {Allende Prieto},
  {Gianninas}, \& {Kenyon}}]{brown2013}
{Brown} W.~R., {Kilic} M., {Allende Prieto} C., {Gianninas} A., {Kenyon} S.~J.,
  2013, \apj, 769, 66

\bibitem[{{Carter} {et~al}\mbox{.}(2014){Carter}, {G{\"a}nsicke}, {Steeghs},
  {Marsh}, {Breedt}, {Kupfer}, {Gentile Fusillo}, {Groot}, \&
  {Nelemans}}]{carter2014}
{Carter} P.~J. {et~al.}, 2014, \mnras, 439, 2848

\bibitem[{{Cassisi}, {Iben} \& {Tornambe}(1998){Cassisi}, {Iben}, \&
  {Tornambe}}]{cas98}
{Cassisi} S., {Iben} I.~J., {Tornambe} A., 1998, \apj, 496, 376

\bibitem[{{Chieffi} \& {Straniero}(1989)}]{chieffi1989}
{Chieffi} A., {Straniero} O., 1989, \apjs, 71, 47

\bibitem[{{Claeys} {et~al}\mbox{.}(2014){Claeys}, {Pols}, {Izzard}, {Vink}, \&
  {Verbunt}}]{claeys_sn_14}
{Claeys} J.~S.~W., {Pols} O.~R., {Izzard} R.~G., {Vink} J., {Verbunt} F.~W.~M.,
  2014, \aap, 563, A83

\bibitem[{{Dan} {et~al}\mbox{.}(2013){Dan}, {Rosswog}, {Br{\"u}ggen}, \&
  {Podsiadlowski}}]{dan2013}
{Dan} M., {Rosswog} S., {Br{\"u}ggen} M., {Podsiadlowski} P., 2013, \mnras

\bibitem[{{Dan} {et~al}\mbox{.}(2012){Dan}, {Rosswog}, {Guillochon}, \&
  {Ramirez-Ruiz}}]{dan2012}
{Dan} M., {Rosswog} S., {Guillochon} J., {Ramirez-Ruiz} E., 2012, \mnras, 422,
  2417

\bibitem[{{Deloye} {et~al}\mbox{.}(2007){Deloye}, {Taam}, {Winisdoerffer}, \&
  {Chabrier}}]{deloye07}
{Deloye} C.~J., {Taam} R.~E., {Winisdoerffer} C., {Chabrier} G., 2007, \mnras,
  381, 525

\bibitem[{{Drout} {et~al}\mbox{.}(2013){Drout}, {Soderberg}, {Mazzali},
  {Parrent}, {Margutti}, {Milisavljevic}, {Sanders}, {Chornock}, {Foley},
  {Kirshner}, {Filippenko}, {Li}, {Brown}, {Cenko}, {Chakraborti}, {Challis},
  {Friedman}, {Ganeshalingam}, {Hicken}, {Jensen}, {Modjaz}, {Perets},
  {Silverman}, \& {Wong}}]{Drout_SN.Ia_13}
{Drout} M.~R. {et~al.}, 2013, \apj, 774, 58

\bibitem[{{Fink}, {Hillebrandt} \& {R{\"o}pke}(2007){Fink}, {Hillebrandt}, \&
  {R{\"o}pke}}]{fink07}
{Fink} M., {Hillebrandt} W., {R{\"o}pke} F.~K., 2007, \aap, 476, 1133

\bibitem[{{Fink} {et~al}\mbox{.}(2010){Fink}, {R{\"o}pke}, {Hillebrandt},
  {Seitenzahl}, {Sim}, \& {Kromer}}]{fink2010}
{Fink} M., {R{\"o}pke} F.~K., {Hillebrandt} W., {Seitenzahl} I.~R., {Sim}
  S.~A., {Kromer} M., 2010, \aap, 514, A53

\bibitem[{{Foley} {et~al}\mbox{.}(2013){Foley}, {Challis}, {Chornock},
  {Ganeshalingam}, {Li}, {Marion}, {Morrell}, {Pignata}, {Stritzinger},
  {Silverman}, {Wang}, {Anderson}, {Filippenko}, {Freedman}, {Hamuy}, {Jha},
  {Kirshner}, {McCully}, {Persson}, {Phillips}, {Reichart}, \&
  {Soderberg}}]{foley_iax13}
{Foley} R.~J. {et~al.}, 2013, \apj, 767, 57

\bibitem[{{Fujimoto} \& {Sugimoto}(1982)}]{fu_su1982}
{Fujimoto} M.~Y., {Sugimoto} D., 1982, \apj, 257, 291

\bibitem[{{Garc{\'{\i}}a-Senz}, {Bravo} \& {Woosley}(1999){Garc{\'{\i}}a-Senz},
  {Bravo}, \& {Woosley}}]{garciasenz99}
{Garc{\'{\i}}a-Senz} D., {Bravo} E., {Woosley} S.~E., 1999, \aap, 349, 177

\bibitem[{{Goranskij} {et~al}\mbox{.}(2010){Goranskij}, {Shugarov}, {Zharova},
  {Kroll}, \& {Barsukova}}]{goran2010}
{Goranskij} V., {Shugarov} S., {Zharova} A., {Kroll} P., {Barsukova} E.~A.,
  2010, Peremennye Zvezdy, 30, 4

\bibitem[{{Guillochon} {et~al}\mbox{.}(2010){Guillochon}, {Dan},
  {Ramirez-Ruiz}, \& {Rosswog}}]{Guillochon_10}
{Guillochon} J., {Dan} M., {Ramirez-Ruiz} E., {Rosswog} S., 2010, \apjl, 709,
  L64

\bibitem[{{Hillebrandt} {et~al}\mbox{.}(2013){Hillebrandt}, {Kromer},
  {R{\"o}pke}, \& {Ruiter}}]{hillen13}
{Hillebrandt} W., {Kromer} M., {R{\"o}pke} F.~K., {Ruiter} A.~J., 2013,
  Frontiers of Physics, 8, 116

\bibitem[{H\"{o}flich {et~al}\mbox{.}(2013)H\"{o}flich, Dragulin, Mitchell,
  Penney, Sadler, Diamond, \& Gerardy}]{hoeflich_snia_2013}
H\"{o}flich P., Dragulin P., Mitchell J., Penney B., Sadler B., Diamond T.,
  Gerardy C., 2013, Frontiers of Physics, 8, 144

\bibitem[{{Huebner} {et~al}\mbox{.}(1977){Huebner}, {Merts}, {Magee}, \&
  {Argo}}]{hueb1977}
{Huebner} W.~F., {Merts} A.~L., {Magee} N.~H., {Argo} M.~F., 1977, Los Alamos
  Scientific Library Report, LA-6760-M, 1

\bibitem[{{Iben} \& {Tutukov}(1985)}]{it85}
{Iben} I., {Tutukov} A.~V., 1985, \apjs, 58, 661

\bibitem[{{Iben}(1982)}]{iben1982}
{Iben}, Jr. I., 1982, \apj, 259, 244

\bibitem[{{Iben} \& {Tutukov}(1984)}]{iben1984}
{Iben}, Jr. I., {Tutukov} A.~V., 1984, \apjs, 54, 335

\bibitem[{{Iben} \& {Tutukov}(1989)}]{iben1989}
{Iben}, Jr. I., {Tutukov} A.~V., 1989, \apj, 342, 430

\bibitem[{{Iben} \& {Tutukov}(1991)}]{iben1991}
{Iben}, Jr. I., {Tutukov} A.~V., 1991, \apj, 370, 615

\bibitem[{{Iben} \& {Tutukov}(1996)}]{it96symb}
{Iben} I.~J., {Tutukov} A.~V., 1996, \apjs, 105, 145

\bibitem[{{Idan}, {Shaviv} \& {Shaviv}(2013){Idan}, {Shaviv}, \&
  {Shaviv}}]{idan13}
{Idan} I., {Shaviv} N.~J., {Shaviv} G., 2013, \mnras, 433, 2884

\bibitem[{{Iglesias} \& {Rogers}(1996)}]{igles1996}
{Iglesias} C.~A., {Rogers} F.~J., 1996, \apj, 464, 943

\bibitem[{{Iijima} \& {Nakanishi}(2008)}]{ijima2008}
{Iijima} T., {Nakanishi} H., 2008, \aap, 482, 865

\bibitem[{{Kasliwal}(2012)}]{kasliwal_bridge12}
{Kasliwal} M.~M., 2012, \pasa, 29, 482

\bibitem[{{Kato} \& {Hachisu}(1994)}]{kato1994}
{Kato} M., {Hachisu} I., 1994, \apj, 437, 802

\bibitem[{{Kato} \& {Hachisu}(2004)}]{kato2004}
{Kato} M., {Hachisu} I., 2004, \apjl, 613, L129

\bibitem[{{Kato} {et~al}\mbox{.}(2008){Kato}, {Hachisu}, {Kiyota}, \&
  {Saio}}]{kato2008}
{Kato} M., {Hachisu} I., {Kiyota} S., {Saio} H., 2008, \apj, 684, 1366

\bibitem[{{Kato}, {Saio} \& {Hachisu}(1989){Kato}, {Saio}, \&
  {Hachisu}}]{ksh89}
{Kato} M., {Saio} H., {Hachisu} I., 1989, \apj, 340, 509

\bibitem[{{Kennicutt} \& {Evans}(2012)}]{kenni2012}
{Kennicutt} R.~C., {Evans} N.~J., 2012, \araa, 50, 531

\bibitem[{{Kippenhahn} \& {Weigert}(1967)}]{kip_weig67}
{Kippenhahn} R., {Weigert} A., 1967, \zap, 65, 251

\bibitem[{{Kromer} {et~al}\mbox{.}(2013){Kromer}, {Pakmor}, {Taubenberger},
  {Pignata}, {Fink}, {R{\"o}pke}, {Seitenzahl}, {Sim}, \&
  {Hillebrandt}}]{kromer_violent13}
{Kromer} M. {et~al.}, 2013, \apjl, 778, L18

\bibitem[{{Kromer} {et~al}\mbox{.}(2010){Kromer}, {Sim}, {Fink}, {R{\"o}pke},
  {Seitenzahl}, \& {Hillebrandt}}]{kromer2010}
{Kromer} M., {Sim} S.~A., {Fink} M., {R{\"o}pke} F.~K., {Seitenzahl} I.~R.,
  {Hillebrandt} W., 2010, \apj, 719, 1067

\bibitem[{Limongi \& Tornamb\`{e}(1991)}]{lt91}
Limongi M., Tornamb\`{e} A., 1991, \apj, 371, 317

\bibitem[{{Livne}(1990)}]{livne1990}
{Livne} E., 1990, \apjl, 354, L53

\bibitem[{{Livne} \& {Arnett}(1995)}]{la95}
{Livne} E., {Arnett} D., 1995, \apj, 452, 62

\bibitem[{Livne \& Glasner(1991)}]{lg91}
Livne E., Glasner A., 1991, {\apj}, 370, 272

\bibitem[{{Maoz}, {Mannucci} \& {Nelemans}(2013){Maoz}, {Mannucci}, \&
  {Nelemans}}]{mmn_snia_rev13}
{Maoz} D., {Mannucci} F., {Nelemans} G., 2013, ArXiv e-prints

\bibitem[{{Marsh}, {Nelemans} \& {Steeghs}(2004){Marsh}, {Nelemans}, \&
  {Steeghs}}]{marsh2004}
{Marsh} T.~R., {Nelemans} G., {Steeghs} D., 2004, \mnras, 350, 113

\bibitem[{{Mennekens} {et~al}\mbox{.}(2010){Mennekens}, {Vanbeveren}, {De
  Greve}, \& {De Donder}}]{mennekens10}
{Mennekens} N., {Vanbeveren} D., {De Greve} J.~P., {De Donder} E., 2010, \aap,
  515, A89+

\bibitem[{{Moll} {et~al}\mbox{.}(2014){Moll}, {Raskin}, {Kasen}, \&
  {Woosley}}]{moll_prompt13}
{Moll} R., {Raskin} C., {Kasen} D., {Woosley} S.~E., 2014, \apj, 785, 105

\bibitem[{{Moll} \& {Woosley}(2013)}]{moll_woosley_dd13}
{Moll} R., {Woosley} S.~E., 2013, \apj, 774, 137

\bibitem[{{Moore}, {Townsley} \& {Bildsten}(2013){Moore}, {Townsley}, \&
  {Bildsten}}]{moore_dd13}
{Moore} K., {Townsley} D.~M., {Bildsten} L., 2013, \apj, 776, 97

\bibitem[{{Nariai}, {Nomoto} \& {Sugimoto}(1980){Nariai}, {Nomoto}, \&
  {Sugimoto}}]{nns1980}
{Nariai} K., {Nomoto} K., {Sugimoto} D., 1980, \pasj, 32, 473

\bibitem[{{Nelemans} {et~al}\mbox{.}(2001){Nelemans}, {Portegies Zwart},
  {Verbunt}, \& {Yungelson}}]{nyp01a}
{Nelemans} G., {Portegies Zwart} S.~F., {Verbunt} F., {Yungelson} L.~R., 2001,
  \aap, 368, 939

\bibitem[{{Nelemans}, {Toonen} \& {Bours}(2013){Nelemans}, {Toonen}, \&
  {Bours}}]{nele2012}
{Nelemans} G., {Toonen} S., {Bours} M., 2013, in IAU Symposium, Vol. 281, IAU
  Symposium, {Di Stefano} R., {Orio} M., {Moe} M., eds., pp. 225--231

\bibitem[{{Nelemans}, {Yungelson} \& {Portegies Zwart}(2004){Nelemans},
  {Yungelson}, \& {Portegies Zwart}}]{nyp04}
{Nelemans} G., {Yungelson} L.~R., {Portegies Zwart} S.~F., 2004, \mnras, 349,
  181

\bibitem[{{Newsham}, {Starrfield} \& {Timmes}(2013){Newsham}, {Starrfield}, \&
  {Timmes}}]{newsham13}
{Newsham} G., {Starrfield} S., {Timmes} F., 2013, ArXiv e-prints

\bibitem[{{Nomoto}(1980)}]{nomoto80}
{Nomoto} K., 1980, \ssr, 27, 563

\bibitem[{{Nomoto}(1982{\natexlab{a}})}]{nom82b}
{Nomoto} K., 1982{\natexlab{a}}, {\apj}, 257, 780

\bibitem[{{Nomoto}(1982{\natexlab{b}})}]{nomoto1982a}
{Nomoto} K., 1982{\natexlab{b}}, \apj, 253, 798

\bibitem[{{Nomoto} \& {Hashimoto}(1987)}]{nh1987}
{Nomoto} K., {Hashimoto} M., 1987, \apss, 131, 395

\bibitem[{{Nomoto}, {Nariai} \& {Sugimoto}(1979){Nomoto}, {Nariai}, \&
  {Sugimoto}}]{nks1979}
{Nomoto} K., {Nariai} K., {Sugimoto} D., 1979, \pasj, 31, 287

\bibitem[{{Paczy{\'n}ski}(1967)}]{pac67a}
{Paczy{\'n}ski} B., 1967, Acta Astron., 17, 287

\bibitem[{{Paczy{\'n}ski}(1971)}]{pac_he71}
{Paczy{\'n}ski} B., 1971, \actaa, 21, 1

\bibitem[{{Pakmor} {et~al}\mbox{.}(2011){Pakmor}, {Hachinger}, {R{\"o}pke}, \&
  {Hillebrandt}}]{pakmor2011}
{Pakmor} R., {Hachinger} S., {R{\"o}pke} F.~K., {Hillebrandt} W., 2011, \aap,
  528, A117

\bibitem[{{Pakmor} {et~al}\mbox{.}(2010){Pakmor}, {Kromer}, {R{\"o}pke}, {Sim},
  {Ruiter}, \& {Hillebrandt}}]{pakmor2010}
{Pakmor} R., {Kromer} M., {R{\"o}pke} F.~K., {Sim} S.~A., {Ruiter} A.~J.,
  {Hillebrandt} W., 2010, \nat, 463, 61

\bibitem[{{Pakmor} {et~al}\mbox{.}(2012){Pakmor}, {Kromer}, {Taubenberger},
  {Sim}, {R{\"o}pke}, \& {Hillebrandt}}]{pakmor_violent12}
{Pakmor} R., {Kromer} M., {Taubenberger} S., {Sim} S.~A., {R{\"o}pke} F.~K.,
  {Hillebrandt} W., 2012, \apjl, 747, L10

\bibitem[{{Pakmor} {et~al}\mbox{.}(2013){Pakmor}, {Kromer}, {Taubenberger}, \&
  {Springel}}]{pakmor_violent_13}
{Pakmor} R., {Kromer} M., {Taubenberger} S., {Springel} V., 2013, \apjl, 770,
  L8

\bibitem[{{Piersanti} {et~al}\mbox{.}(1999){Piersanti}, {Cassisi}, {Iben}, \&
  {Tornamb{\'e}}}]{pier1999}
{Piersanti} L., {Cassisi} S., {Iben}, Jr. I., {Tornamb{\'e}} A., 1999, \apjl,
  521, L59

\bibitem[{{Piersanti} {et~al}\mbox{.}(2000){Piersanti}, {Cassisi}, {Iben}, \&
  {Tornamb{\'e}}}]{pier2000}
{Piersanti} L., {Cassisi} S., {Iben}, Jr. I., {Tornamb{\'e}} A., 2000, \apj,
  535, 932

\bibitem[{{Piersanti}, {Cassisi} \& {Tornamb{\'e}}(2001){Piersanti}, {Cassisi},
  \& {Tornamb{\'e}}}]{pier2001}
{Piersanti} L., {Cassisi} S., {Tornamb{\'e}} A., 2001, \apj, 558, 916

\bibitem[{{Piersanti} {et~al}\mbox{.}(2003){Piersanti}, {Gagliardi}, {Iben}, \&
  {Tornamb{\'e}}}]{pier2003a}
{Piersanti} L., {Gagliardi} S., {Iben} I.~J., {Tornamb{\'e}} A., 2003, \apj,
  583, 885

\bibitem[{{Piersanti}, {Straniero} \& {Cristallo}(2007){Piersanti},
  {Straniero}, \& {Cristallo}}]{pier2007}
{Piersanti} L., {Straniero} O., {Cristallo} S., 2007, \aap, 462, 1051

\bibitem[{{Piro}, {Thompson} \& {Kochanek}(2014){Piro}, {Thompson}, \&
  {Kochanek}}]{piro2013}
{Piro} A.~L., {Thompson} T.~A., {Kochanek} C.~S., 2014, \mnras, 438, 3456

\bibitem[{{Postnov} \& {Yungelson}(2014)}]{postnov14}
{Postnov} K.~A., {Yungelson} L.~R., 2014, Living Reviews in Relativity, 17, 3

\bibitem[{{Potekhin} {et~al}\mbox{.}(1999){Potekhin}, {Baiko}, {Haensel}, \&
  {Yakovlev}}]{pote1999}
{Potekhin} A.~Y., {Baiko} D.~A., {Haensel} P., {Yakovlev} D.~G., 1999, \aap,
  346, 345

\bibitem[{{Prada Moroni} \& {Straniero}(2002)}]{prada2002}
{Prada Moroni} P.~G., {Straniero} O., 2002, \apj, 581, 585

\bibitem[{{Raskin} {et~al}\mbox{.}(2012){Raskin}, {Scannapieco}, {Fryer},
  {Rockefeller}, \& {Timmes}}]{raskin_remnants12}
{Raskin} C., {Scannapieco} E., {Fryer} C., {Rockefeller} G., {Timmes} F.~X.,
  2012, \apj, 746, 62

\bibitem[{{Rosenbush}(2008)}]{rosenbush_he_nov08}
{Rosenbush} A.~E., 2008, in Astronomical Society of the Pacific Conference
  Series, Vol. 391, Hydrogen-Deficient Stars, {Werner} A., {Rauch} T., eds., p.
  271

\bibitem[{{Ruiter}, {Belczynski} \& {Fryer}(2009){Ruiter}, {Belczynski}, \&
  {Fryer}}]{ruiter09}
{Ruiter} A.~J., {Belczynski} K., {Fryer} C., 2009, \apj, 699, 2026

\bibitem[{{Ruiter} {et~al}\mbox{.}(2011){Ruiter}, {Belczynski}, {Sim},
  {Hillebrandt}, {Fryer}, {Fink}, \& {Kromer}}]{ruiter2011}
{Ruiter} A.~J., {Belczynski} K., {Sim} S.~A., {Hillebrandt} W., {Fryer} C.~L.,
  {Fink} M., {Kromer} M., 2011, \mnras, 417, 408

\bibitem[{{Ruiter} {et~al}\mbox{.}(2013){Ruiter}, {Sim}, {Pakmor}, {Kromer},
  {Seitenzahl}, {Belczynski}, {Fink}, {Herzog}, {Hillebrandt}, {R{\"o}pke}, \&
  {Taubenberger}}]{ruiter_13_violent}
{Ruiter} A.~J. {et~al.}, 2013, \mnras, 429, 1425

\bibitem[{{Saio} \& {Nomoto}(1985)}]{saio85}
{Saio} H., {Nomoto} K., 1985, \aap, 150, L21

\bibitem[{{Saio} \& {Nomoto}(1998)}]{saio98}
{Saio} H., {Nomoto} K., 1998, \apj, 500, 388

\bibitem[{Savonije, de~Kool \& van~den Heuvel(1986)Savonije, de~Kool, \&
  van~den Heuvel}]{skh86}
Savonije G.~J., de~Kool M., van~den Heuvel E. P.~J., 1986, \aap, 155, 51

\bibitem[{{Schwab} {et~al}\mbox{.}(2012){Schwab}, {Shen}, {Quataert}, {Dan}, \&
  {Rosswog}}]{schwab2012}
{Schwab} J., {Shen} K.~J., {Quataert} E., {Dan} M., {Rosswog} S., 2012, \mnras,
  427, 190

\bibitem[{{Shen} \& {Bildsten}(2007)}]{shen2007}
{Shen} K.~J., {Bildsten} L., 2007, \apj, 660, 1444

\bibitem[{{Shen} \& {Bildsten}(2009)}]{shen2009}
{Shen} K.~J., {Bildsten} L., 2009, \apj, 699, 1365

\bibitem[{{Shen} \& {Bildsten}(2014{\natexlab{a}})}]{shen_bildsten_dd13}
{Shen} K.~J., {Bildsten} L., 2014{\natexlab{a}}, \apj, 785, 61

\bibitem[{{Shen} \& {Bildsten}(2014{\natexlab{b}})}]{shen2014}
{Shen} K.~J., {Bildsten} L., 2014{\natexlab{b}}, \apj, 785, 61

\bibitem[{{Shen} {et~al}\mbox{.}(2010){Shen}, {Kasen}, {Weinberg}, {Bildsten},
  \& {Scannapieco}}]{shen_.ia10}
{Shen} K.~J., {Kasen} D., {Weinberg} N.~N., {Bildsten} L., {Scannapieco} E.,
  2010, \apj, 715, 767

\bibitem[{{Sim} {et~al}\mbox{.}(2012){Sim}, {Fink}, {Kromer}, {R{\"o}pke},
  {Ruiter}, \& {Hillebrandt}}]{sim2012}
{Sim} S.~A., {Fink} M., {Kromer} M., {R{\"o}pke} F.~K., {Ruiter} A.~J.,
  {Hillebrandt} W., 2012, \mnras, 420, 3003

\bibitem[{{Sim} {et~al}\mbox{.}(2010){Sim}, {R{\"o}pke}, {Hillebrandt},
  {Kromer}, {Pakmor}, {Fink}, {Ruiter}, \& {Seitenzahl}}]{sim2010}
{Sim} S.~A., {R{\"o}pke} F.~K., {Hillebrandt} W., {Kromer} M., {Pakmor} R.,
  {Fink} M., {Ruiter} A.~J., {Seitenzahl} I.~R., 2010, \apjl, 714, L52

\bibitem[{{Solheim}(2010)}]{solh2010}
{Solheim} J., 2010, \pasp, 122, 1133

\bibitem[{{Solheim} \& {Yungelson}(2005)}]{sy05}
{Solheim} J.-E., {Yungelson} L.~R., 2005, in ASP Conf. Ser. 334: 14th European
  Workshop on White Dwarfs, {Koester} D., {Moehler} S., eds., p. 387

\bibitem[{{Sparks} \& {Endal}(1980)}]{sparks1980}
{Sparks} W.~M., {Endal} A.~S., 1980, \apj, 237, 130

\bibitem[{{Straniero}(1988)}]{stra1988}
{Straniero} O., 1988, \aaps, 76, 157

\bibitem[{{Straniero}, {Cristallo} \& {Piersanti}(2014){Straniero},
  {Cristallo}, \& {Piersanti}}]{stran2014}
{Straniero} O., {Cristallo} S., {Piersanti} L., 2014, \apj, 785, 77

\bibitem[{{Straniero}, {Gallino} \& {Cristallo}(2006){Straniero}, {Gallino}, \&
  {Cristallo}}]{stran2006}
{Straniero} O., {Gallino} R., {Cristallo} S., 2006, Nuclear Physics A, 777, 311

\bibitem[{{Sugimoto} \& {Fujimoto}(1978)}]{sufu1978}
{Sugimoto} D., {Fujimoto} M.~Y., 1978, \pasj, 30, 467

\bibitem[{{Taam}(1980{\natexlab{a}})}]{taam1980a}
{Taam} R.~E., 1980{\natexlab{a}}, \apj, 237, 142

\bibitem[{{Taam}(1980{\natexlab{b}})}]{taam1980b}
{Taam} R.~E., 1980{\natexlab{b}}, \apj, 242, 749

\bibitem[{{Toonen}, {Nelemans} \& {Portegies Zwart}(2012){Toonen}, {Nelemans},
  \& {Portegies Zwart}}]{toonen12}
{Toonen} S., {Nelemans} G., {Portegies Zwart} S., 2012, \aap, 546, A70

\bibitem[{{Townsley}, {Moore} \& {Bildsten}(2012){Townsley}, {Moore}, \&
  {Bildsten}}]{townsley12}
{Townsley} D.~M., {Moore} K., {Bildsten} L., 2012, \apj, 755, 4

\bibitem[{{Tutukov} \& {Yungelson}(1996)}]{tutu1996}
{Tutukov} A., {Yungelson} L., 1996, \mnras, 280, 1035

\bibitem[{Tutukov \& Yungelson(1996)}]{ty96}
Tutukov A.~V., Yungelson L.~R., 1996, {\mnras}, 280, 1035

\bibitem[{{Uus}(1970)}]{uus1970}
{Uus} U., 1970, Nauchnye Informatsii, 17, 25

\bibitem[{{Waldman} {et~al}\mbox{.}(2011){Waldman}, {Sauer}, {Livne}, {Perets},
  {Glasner}, {Mazzali}, {Truran}, \& {Gal-Yam}}]{wald2011}
{Waldman} R., {Sauer} D., {Livne} E., {Perets} H., {Glasner} A., {Mazzali} P.,
  {Truran} J.~W., {Gal-Yam} A., 2011, \apj, 738, 21

\bibitem[{{Wang}, {Justham} \& {Han}(2013){Wang}, {Justham}, \& {Han}}]{wang13}
{Wang} B., {Justham} S., {Han} Z., 2013, \aap, 559, A94

\bibitem[{{Wang} {et~al}\mbox{.}(2009){Wang}, {Meng}, {Chen}, \&
  {Han}}]{wang09}
{Wang} B., {Meng} X., {Chen} X., {Han} Z., 2009, \mnras, 395, 847

\bibitem[{Webbink(1984)}]{web84}
Webbink R.~F., 1984, \apj, 277, 355

\bibitem[{{Wolf} {et~al}\mbox{.}(2013){Wolf}, {Bildsten}, {Brooks}, \&
  {Paxton}}]{wolf2013}
{Wolf} W.~M., {Bildsten} L., {Brooks} J., {Paxton} B., 2013, \apj, 777, 136

\bibitem[{{Woosley} \& {Kasen}(2011)}]{woosley_kasen_dd11}
{Woosley} S.~E., {Kasen} D., 2011, \apj, 734, 38

\bibitem[{{Woosley}, {Taam} \& {Weaver}(1986){Woosley}, {Taam}, \&
  {Weaver}}]{wtw1986}
{Woosley} S.~E., {Taam} R.~E., {Weaver} T.~A., 1986, \apj, 301, 601

\bibitem[{{Woosley} \& {Weaver}(1994)}]{ww94}
{Woosley} S.~E., {Weaver} T.~A., 1994, \apj, 423, 371

\bibitem[{{Woudt} \& {Steeghs}(2005)}]{woudt_steeghs05a}
{Woudt} P.~A., {Steeghs} D., 2005, in Astronomical Society of the Pacific
  Conference Series, p. 451

\bibitem[{{Woudt} {et~al}\mbox{.}(2009){Woudt}, {Steeghs}, {Karovska},
  {Warner}, {Groot}, {Nelemans}, {Roelofs}, {Marsh}, {Nagayama}, {Smits}, \&
  {O'Brien}}]{woudt2009}
{Woudt} P.~A. {et~al.}, 2009, \apj, 706, 738

\bibitem[{{Yoon} \& {Langer}(2003)}]{yl03}
{Yoon} S.-C., {Langer} N., 2003, \aap, 412, L53

\bibitem[{{Yoon} \& {Langer}(2004)}]{yoon_langer04}
{Yoon} S.-C., {Langer} N., 2004, \aap, 419, 645

\bibitem[{{Yungelson}(2008)}]{yung2008}
{Yungelson} L.~R., 2008, Astronomy Letters, 34, 620

\end{thebibliography}
\newpage

\appendix

\section{Interpolation Formulae}\label{app0}
In this Appendix we provide interpolation formulae for the data reported 
in the manuscript which could be directly used in population synthesis 
codes to describe the evolution of WDs  accreting He-rich matter.

\subsection{Accretion Regimes}\label{app1}

The values of \mdot as a function of the 
WD initial mass at which the transition from one accretion regime to 
another occurs (see Fig. \ref{f:regime}) is given by: 
\begin{equation}
\mathrm{
\log(\dot{M})=A+B\cdot M_{WD}, 
}\label{e:a1}
\end{equation}
where \mdot is expressed in \myr\,and \mwd\,in \ms. In Table~\ref{t:a1} 
we report the values of the coefficients A and B for the various 
accretion regimes. These fits are valid for WDs initial masses in the 
range (0.59678 -- 1.01948) \msune.

\begin{table}  
\caption{From left to right we report the values of A and the corresponding 
         standard deviation, the values of B and the corresponding standard 
         deviation in Eq.~(\ref{e:a1}), as well as the correlation 
         coefficient $\mathrm{R^{2}}$. 
         RG/SS represents the transition line between the RG and the 
         Steady  Accretion regimes;
         SS/MF -- transition line between the Steady Accretion and the 
         Mild Flashes regimes;
         MF/SF -- transition line between the Mild Flashes and the 
         Strong Flashes regimes;
         SF/Dt -- transition line between the Strong and Dynamical Flashes 
         regimes.} 
  \label{t:a1}
  \centering
  \begin{tabular}{l r r r r r }
  \hline
      &   A    & $\sigma_A^2$ &    B  &  $\sigma_B^2$ & ${\rm R^2}$ \\
\hline
RG/SS & -6.840 &    0.026     & 1.349 &      0.029    & 0.999 \\
SS/MF & -8.115 &    0.151     & 2.290 &      0.170    & 0.986 \\
MF/SF & -8.233 &    0.029     & 2.022 &      0.042    & 0.999 \\
SF/Dt & -8.313 &    0.026     & 1.018 &      0.037    & 0.997 \\
\hline
   \end{tabular}
\end{table}

\subsection{Dynamical Flashes Regime}\label{app2}

For a fixed value of the initial WD mass, for models experiencing a 
dynamical He-flash as a result of direct accretion of He-rich matter, the mass 
accreted before the onset of the dynamical flash ($\mathrm{\Delta M_{He}}$) 
can be expressed as a function of \mdot by the following relation:
\begin{equation}
\mathrm{
\Delta M_{He}=\sum_{i=0}^4 F_i \cdot \dot{M}^i,
}\label{e:a2}
\end{equation}
where $\mathrm{\Delta M_{He}}$ is  in \ms\ and \mdot in $10^{-8}$\myr, 
respectively. In Table~\ref{t:a2} we report the values of the coefficients 
in Eq.~(\ref{e:a2}) for the 5 initial WDs models considered in the present 
work (see Table~\ref{t:initial}). 

\begin{table}  
\caption{For each initial WD model, we report the values of the 
$\mathrm{F_i}$ coefficients and the corresponding standard deviation 
$\sigma^2_{F_i}$ in Eq.~(\ref{e:a2}). We also report the value of the 
correlation coefficient ${\rm R^2}$, as well as the range of validity of the 
fits ($\mathrm{\dot{M}_{min} - \dot{M}_{max}}$ in $10^{-8}$\ms\pyr ).}
\label{t:a2}
\centering
\begin{tabular}{l | r | r | r }
\hline
& \multicolumn{1}{c}{M060} & \multicolumn{1}{c}{M070} & \multicolumn{1}{c}{M081}\\
\hline
$F_0$ & 0.718	$\pm$ 0,012 &  0.625 $\pm$ 0.012 & 0.528 $\pm$ 0.008 \\
$F_1$ &-0.762	$\pm$ 0.067 & -0.671 $\pm$ 0.062 &-0.623 $\pm$ 0.047 \\ 
$F_2$ & 0.744	$\pm$ 0.101 &  0.598 $\pm$ 0.082 & 0.665 $\pm$ 0.078 \\
$F_3$ &-0.259 $\pm$ 0.042 & -0.183 $\pm$ 0.027 &-0.324 $\pm$ 0.045 \\
$F_4$ &\multicolumn{1}{c}{-}&\multicolumn{1}{c}{-}& 0.052 $\pm$ 0.008 \\
${\rm R^2}$ &\multicolumn{1}{c}{0.996} &\multicolumn{1}{c}{0.994} &\multicolumn{1}{c}{0.998} \\
$\mathrm{\dot{M}_{min} - \dot{M}_{max}}$ &
\multicolumn{1}{c}{0.1 -- 1.5} & \multicolumn{1}{c}{0.1 -- 2.0} & \multicolumn{1}{c}{0.15 -- 2.5} \\
\hline
& \multicolumn{1}{c}{M092} & \multicolumn{1}{c}{M102} & \multicolumn{1}{c}{\it }\\
\hline
$F_0$ & 0.440	$\pm$ 0.007 & 0.319	 $\pm$ 0.009  & \multicolumn{1}{c}{\it }\\
$F_1$ &-0.489	$\pm$ 0.035 &-0.240	 $\pm$ 0.030  & \multicolumn{1}{c}{\it }\\ 
$F_2$ & 0.452	$\pm$ 0.052 & 0.105	 $\pm$ 0.027  & \multicolumn{1}{c}{\it }\\
$F_3$ &-0.198	$\pm$ 0.026 &-0.024	 $\pm$ 0.009  & \multicolumn{1}{c}{\it }\\
$F_4$ & 0.029	$\pm$ 0.004 & 0.0021 $\pm$ 0.0009 & \multicolumn{1}{c}{\it }\\
${\rm R^2}$ &\multicolumn{1}{c}{0.998} &\multicolumn{1}{c}{0.989} &\multicolumn{1}{c}{\it } \\
$\mathrm{\dot{M}_{min} - \dot{M}_{max}}$ &
\multicolumn{1}{c}{0.15 -- 3.0} & \multicolumn{1}{c}{0.15 -- 5.0} & \multicolumn{1}{c}{\it }\\
\hline
\end{tabular}
\end{table}

\subsection{Strong Flashes Regime}\label{app3}
For a fixed value of the mass of accreting WD, the accumulation efficiency 
\acef as a function of \mdot for the models experiencing the first strong 
non-dynamical He-flash is provided by the following relation:
\begin{equation}
\mathrm{
\eta_{acc}=\sum_{i=0}^3 G_i \cdot \dot{M}^i,
}\label{e:a3}
\end{equation}
where \mdot is expressed in $10^{-8}$\ms\pyr. In Table~\ref{t:a3} we 
report the values of the coefficients for the 5 initial WDs models 
considered in the present work (see Table~\ref{t:initial}). \\
Eq.~(\ref{e:a3}) provide the value of \acef for the first He-flash 
experienced by an accreting WD entering in the Strong Flashes regime. 
\begin{table}  
\caption{For each initial WD model, we report the values of the 
         ${\rm G_i}$ coefficients and the corresponding standard 
         deviation $\sigma^2_{G_i}$ in Eq.~(\ref{e:a3}). We also report 
         the value of the correlation coefficient ${\rm R^2}$, as well as the 
         range of validity of the fits ($\mathrm{\dot{M}_{min} - 
         \dot{M}_{max}}$ in $10^{-8}$\ms\pyr ).}
\label{t:a3}
\centering
\begin{tabular}{l | c | c  }
\hline
& \multicolumn{1}{c}{M060} & \multicolumn{1}{c}{M070} \\
\hline
$G_0$ & 0.006 $\pm$ 0.121         &-0.035 $\pm$ 0.030        \\
$G_1$ & 0.051 $\pm$ 0.070         & 0.075 $\pm$ 0.012        \\ 
$G_2$ & 0.0083$\pm$ 0.0121        &-0.0018$\pm$ 0.0014       \\
$G_3$ &(-3.317$\pm$ 6.4)$10^{-4}$ &(3.266$\pm$ 4.2)$10^{-5}$ \\
${\rm R^2}$ & \multicolumn{1}{c}{0.996} &\multicolumn{1}{c}{0.999} \\
$\mathrm{\dot{M}_{min} - \dot{M}_{max}}$ &
\multicolumn{1}{c}{2.5 - 10}& \multicolumn{1}{c}{3 - 20} \\
\hline
& \multicolumn{1}{c}{M081}& \multicolumn{1}{c}{M092} \\
\hline
$G_0$ & 0.093  $\pm$ 0.021         &-0.0759 $\pm$ 0.026           \\
$G_1$ & 0.018  $\pm$ 0.005         & 0.0154 $\pm$ 0.004           \\
$G_2$ & 0.0016 $\pm$ 0.0004        & 0.0004 $\pm$ 0.0002          \\
$G_3$ &(-4.111$\pm$ 0.73)$10^{-5}$ &(-5.905 $\pm$ 1.56) $10^{-6}$ \\
${\rm R^2}$ & \multicolumn{1}{c}{0.999} & \multicolumn{1}{c}{0.998}\\
$\mathrm{\dot{M}_{min} - \dot{M}_{max}}$ &
\multicolumn{1}{c}{4 - 30} & \multicolumn{1}{c}{5 - 60} \\ 
\hline
& \multicolumn{1}{c}{M102}& \multicolumn{1}{c}{\it } \\
\hline
$G_0$ &-0.323 $\pm$ 0.017   & \\           
$G_1$ & 0.041 $\pm$ 0.002   & \\           
$G_2$ &-0.0007$\pm$ 0.00006 & \\            
$G_3$ &(4,733 $\pm$ 0.55) $10^{-6}$ & \\ 
${\rm R^2}$ & \multicolumn{1}{c}{0.999} & \multicolumn{1}{c}{\it }\\
$\mathrm{\dot{M}_{min} - \dot{M}_{max}}$ &             
\multicolumn{1}{c}{8 - 70} & \multicolumn{1}{c}{\it } \\   
\hline
\end{tabular}
\end{table}
\label{lastpage}
\onecolumn
\pagestyle{empty}
\pagenumbering{Roman}
\setcounter{page}{1}
\setcounter{table}{1}
\renewcommand{\thetable}{B\arabic{table}}
\LTcapwidth=\textwidth
\begin{center}
{\Huge\it \ }\\
{\Huge\it \ }\\
{\Huge\it \ }\\
{\Huge\it \ }\\
{\Huge\it \ }\\
{\Huge\it \ }\\
{\Huge\it \ }\\
{\Huge\it \ }\\
{\Huge\it \ }\\
{\Huge\it \ }\\
{\Huge\it \ }\\
{\Huge\bf Additional Material}
\end{center}
\newpage
\pagestyle{plain}
\pagenumbering{Roman}
\setcounter{page}{1}
\setcounter{table}{1}
\renewcommand{\thetable}{B\arabic{table}}
\LTcapwidth=\textwidth
\begin{longtable}{r r r r r c c c }
\caption{This is the complete version of Table 2 in the manuscript. For 
         each computed model we list as a function of the accretion rate 
         \mdot in $10^{-9}$\myr, the final mass $\mathrm{M_{fin}}$ 
         in \msune, the accreted mass $\mathrm{M_{acc}}$ in \msune, the 
         accretion time $\mathrm{T_{acc}}$ in $10^6$ yr, the mass 
         coordinate where He-burning is ignited $\mathrm{M_{ign}}$ in 
         \msune, the temperature $\rm T_{ign}$ in $10^7$ K and density 
         $\rm \rho_{ign}$ in $10^{6}\,\mathrm{g\,cm^{-3}}$ when He-burning 
         is ignited. The last column gives the mass of helium buffer 
         $\mathrm{\Delta M_{He}^{pk}}$ in \msune.}
\label{t:detonabis}\\
\hline\hline
        \mdote & $\rm M_{fin}$ & $\rm M_{acc}$    & $\rm T_{acc}$ & 
 $\rm M_{ign}$ & $\rm T_{ign}$ & $\rm \rho_{ign}$ & $\rm \Delta M_{He}^{pk}$ \\
\hline
\endfirsthead
\multicolumn{8}{c}%
{{\tablename\ \thetable{} -- continued from previous page}} \\
\hline
        \mdote & $\rm M_{fin}$ & $\rm M_{acc}$    & $\rm T_{acc}$ & 
 $\rm M_{ign}$ & $\rm T_{ign}$ & $\rm \rho_{ign}$ & $\rm \Delta M_{He}^{pk}$ \\
\hline
\endhead
\hline \multicolumn{8}{|r|}{{Continued on next page}} \\ \hline
\endfoot

\hline \hline
\endlastfoot

\multicolumn{8} c {M060} \\
  1 & 1.255 & 0.658 & 658.351 & 0.587 & 5.246 & 72.061 & 0.737 \\
  2 & 1.183 & 0.586 & 293.041 & 0.589 & 6.492 & 31.369 & 0.662 \\
  3 & 1.136 & 0.540 & 179.837 & 0.598 & 7.042 & 19.879 & 0.615 \\
  4 & 1.107 & 0.510 & 127.510 & 0.600 & 7.428 & 15.272 & 0.586 \\
  5 & 1.092 & 0.495 &  99.076 & 0.601 & 7.579 & 13.452 & 0.571 \\
  6 & 1.076 & 0.480 &  79.925 & 0.601 & 7.730 & 11.773 & 0.555 \\
  7 & 1.063 & 0.466 &  66.549 & 0.604 & 7.856 & 10.447 & 0.541 \\
  8 & 1.051 & 0.454 &  56.733 & 0.604 & 7.969 &  9.475 & 0.529 \\
  9 & 1.039 & 0.443 &  49.190 & 0.605 & 8.067 &  8.634 & 0.517 \\
 10 & 1.029 & 0.432 &  43.200 & 0.607 & 8.153 &  7.892 & 0.506 \\
 15 & 0.971 & 0.374 &  24.929 & 0.608 & 8.619 &  4.995 & 0.448 \\
\multicolumn{8} c {M070} \\
  1 & 1.277 & 0.575 & 698.028 & 0.698 & 5.124 & 75.664 & 0.616 \\
  2 & 1.208 & 0.506 & 252.920 & 0.705 & 6.516 & 30.655 & 0.542 \\
  3 & 1.163 & 0.461 & 153.800 & 0.699 & 7.045 & 19.582 & 0.498 \\
  4 & 1.137 & 0.435 & 108.790 & 0.714 & 7.488 & 14.697 & 0.471 \\
  5 & 1.123 & 0.421 &  84.156 & 0.715 & 7.648 & 12.802 & 0.457 \\
  6 & 1.107 & 0.405 &  67.575 & 0.716 & 7.809 & 11.093 & 0.442 \\
  7 & 1.094 & 0.393 &  56.089 & 0.716 & 7.939 &  9.909 & 0.429 \\
  8 & 1.083 & 0.381 &  47.670 & 0.716 & 8.054 &  8.974 & 0.422 \\
  9 & 1.073 & 0.371 &  41.231 & 0.715 & 8.151 &  8.218 & 0.412 \\
 10 & 1.063 & 0.361 &  36.144 & 0.715 & 8.244 &  7.553 & 0.402 \\
 20 & 0.913 & 0.211 &  10.568 & 0.735 & 9.442 &  1.839 & 0.252 \\
\multicolumn{8} c {M081} \\
 1.5 & 1.263 & 0.453 & 301.691 & 0.818 & 6.071 & 45.067 & 0.461 \\
 2   & 1.237 & 0.426 & 213.209 & 0.821 & 6.552 & 31.429 & 0.435 \\
 3   & 1.196 & 0.386 & 128.592 & 0.821 & 7.194 & 19.413 & 0.395 \\
 4   & 1.172 & 0.362 &  90.399 & 0.824 & 7.520 & 14.758 & 0.371 \\
 5   & 1.158 & 0.347 &  69.449 & 0.824 & 7.695 & 12.654 & 0.356 \\
 6   & 1.144 & 0.333 &  55.565 & 0.825 & 7.851 & 10.913 & 0.342 \\
 7   & 1.132 & 0.322 &  45.950 & 0.827 & 7.980 &  9.657 & 0.331 \\
 8   & 1.122 & 0.312 &  38.961 & 0.827 & 8.084 &  8.722 & 0.321 \\
 9   & 1.113 & 0.302 &  33.590 & 0.828 & 8.179 &  7.914 & 0.311 \\
10   & 1.104 & 0.293 &  29.341 & 0.829 & 8.268 &  7.235 & 0.302 \\
20   & 0.995 & 0.185 &   9.231 & 0.837 & 9.255 &  2.318 & 0.194 \\
25   & 0.912 & 0.102 &   4.069 & 0.826 & 9.947 &  0.929 & 0.111 \\
\multicolumn{8} c {M092} \\
 1.5 & 1.301 & 0.383 & 255.221 & 0.918 &  5,799 & 59.903 & 0.392 \\
 2   & 1.277 & 0.359 & 179.566 & 0.918 &  6,320 & 40.465 & 0.368 \\
 3   & 1.241 & 0.323 & 107.717 & 0.929 &  6,922 & 23.058 & 0.332 \\
 4   & 1.218 & 0.300 &  75.020 & 0.918 &  7,308 & 17.578 & 0.309 \\
 5   & 1.204 & 0.286 &  57.194 & 0.930 &  7,530 & 14.224 & 0.295 \\
 6   & 1.191 & 0.273 &  45.462 & 0.930 &  7,703 & 12.102 & 0.283 \\
 7   & 1.180 & 0.262 &  37.377 & 0.930 &  7,844 & 10.624 & 0.271 \\
 8   & 1.170 & 0.252 &  31.495 & 0.930 &  7,966 &  9.511 & 0.261 \\ 
 9   & 1.161 & 0.243 &  27.026 & 0.929 &  8,067 &  8.609 & 0.253 \\ 
10   & 1.153 & 0.235 &  23.506 & 0.929 &  8,160 &  7.852 & 0.243 \\ 
20   & 1.066 & 0.148 &   7.406 & 0.933 &  9,066 &  2.788 & 0.156 \\   
25   & 1.006 & 0.088 &   3.525 & 0.930 &  9.778 &  1.235 & 0.096 \\
30   & 0.963 & 0.045 &   1.491 & 0.922 & 10.534 &  0.590 & 0.053 \\
\multicolumn{8} c {M102} \\
 1.5 & 1.318 & 0.299 & 199.647 & 1.019 &  5.883 & 92.234 & 0.305 \\
 2   & 1.298 & 0.279 & 139.650 & 1.019 &  6.388 & 39.497 & 0.285 \\
 3   & 1.267 & 0.248 &  82.597 & 1.029 &  7.011 & 21.999 & 0.253 \\
 4   & 1.247 & 0.228 &  57.000 & 1.029 &  7.428 & 16.198 & 0.233 \\
 5   & 1.234 & 0.215 &  42.958 & 1.029 &  7.641 & 13.298 & 0.198 \\
 6   & 1.222 & 0.203 &  33.858 & 1.029 &  7.816 & 11.259 & 0.187 \\
 7   & 1.212 & 0.193 &  27.634 & 1.029 &  7.961 &  9.853 & 0.177 \\
 8   & 1.204 & 0.185 &  23.138 & 1.028 &  8.075 &  8.791 & 0.190 \\
 9   & 1.196 & 0.177 &  19.720 & 1.028 &  8.175 &  7.931 & 0.183 \\
10   & 1.189 & 0.171 &  17.050 & 1.028 &  8.269 &  7.217 & 0.176 \\
20   & 1.122 & 0.103 &   5.145 & 1.029 &  9.147 &  2.692 & 0.108 \\
30   & 1.065 & 0.046 &   1.545 & 1.025 & 10.132 &  0.956 & 0.051 \\
40   & 1.047 & 0.028 &   0.689 & 1.022 & 10.733 &  0.568 & 0.033 \\
50   & 1.039 & 0.020 &   0.400 & 1.021 & 11.148 &  0.427 & 0.025 \\
\end{longtable}
\par\noindent{\it }
\par\noindent{\it }
\par\noindent{\it }
\par\noindent{\it }
\setcounter{table}{6}
 \begin{longtable}{c c c c c} 
 \caption{For models experiencing Strong He-flash, we report the values 
          of the accretion rate (\mdot in $10^{-8}$ \myr), the WD 
          total mass at the bluest point along the loop in the HR diagram 
          $\mathrm{M_{BP}}$ in \msune, the mass transferred during a 
          complete loop $\mathrm{M_{tran}}$ in $10^{-2}$ \msune), the 
          mass effectively accreted ($\mathrm{M_{accr}}$ in $10^{-2}$\msune) 
          and the accumulation efficiency \acef as a function of the 
          accretion rate \mdot in \myr. The accumulation efficiencies 
          pulse by pulse are plotted as a function of $\mathrm{M_{BP}}$ 
          in Figure 7 of the manuscript.}\\
 \hline\hline
 \mdot & $\rm M_{BP}$ & $\rm M_{tran}$ & $\rm M_{accr}$ & \acef \\
 \hline
\endfirsthead
\multicolumn{5}{c}%
{{\tablename\ \thetable{} -- continued from previous page}} \\
\hline
 \mdot & $\rm M_{BP}$ & $\rm M_{tran}$ & $\rm M_{accr}$ & \acef \\
 \hline
\endhead
\hline \multicolumn{5}{|r|}{{Continued on next page}} \\ \hline
\endfoot

\hline \hline
\endlastfoot
  \multicolumn{5}{c}{M060}\\
 \hline                    
  4 & 0.597871 & 2.657 & 0.888 & 0.33440 \\
    & 0.606756 & 3.483 & 0.896 & 0.25722 \\
    & 0.615716 & 3.810 & 0.845 & 0.22178 \\
    & 0.624166 & 4.082 & 0.808 & 0.19793 \\
    & 0.632245 & 4.303 & 0.775 & 0.18021 \\
    & 0.639999 & 4.464 & 0.750 & 0.16790 \\
  5 & 0.598180 & 2.236 & 0.957 & 0.42791 \\
    & 0.607748 & 2.676 & 0.947 & 0.35382 \\
    & 0.617215 & 2.761 & 0.880 & 0.31880 \\
    & 0.626016 & 2.849 & 0.824 & 0.28918 \\
    & 0.634254 & 2.943 & 0.790 & 0.26843 \\
    & 0.642153 & 3.016 & 0.755 & 0.25029 \\
    & 0.649701 & 3.080 & 0.717 & 0.23268 \\
  6 & 0.656867 & 3.141 & 0.694 & 0.22093 \\
    & 0.598519 & 1.936 & 1.024 & 0.52926 \\
    & 0.608763 & 2.177 & 1.019 & 0.46794 \\
    & 0.618949 & 2.168 & 0.926 & 0.42690 \\
    & 0.628206 & 2.201 & 0.876 & 0.39818 \\
    & 0.636968 & 2.233 & 0.813 & 0.36403 \\
    & 0.645097 & 2.282 & 0.770 & 0.33754 \\
    & 0.652801 & 2.325 & 0.723 & 0.31112 \\
    & 0.660036 & 2.371 & 0.691 & 0.29152 \\
    & 0.666947 & 2.407 & 0.657 & 0.27305 \\
    & 0.673518 & 2.439 & 0.639 & 0.26209 \\
  7 & 0.598876 & 1.701 & 1.096 & 0.64417 \\
    & 0.609834 & 1.818 & 1.113 & 0.61235 \\
    & 0.620968 & 1.745 & 1.005 & 0.57616 \\
    & 0.631022 & 1.732 & 0.930 & 0.53701 \\
    & 0.640324 & 1.748 & 0.866 & 0.49579 \\
    & 0.648988 & 1.775 & 0.803 & 0.45254 \\
    & 0.657021 & 1.812 & 0.763 & 0.42099 \\
    & 0.664650 & 1.847 & 0.718 & 0.38896 \\
    & 0.671833 & 1.883 & 0.686 & 0.36423 \\
    & 0.678692 & 1.911 & 0.644 & 0.33686 \\
    & 0.685131 & 1.942 & 0.621 & 0.31961 \\
  8 & 0.599275 & 1.500 & 1.152 & 0.76798 \\
    & 0.610793 & 1.542 & 1.163 & 0.75406 \\
    & 0.622422 & 1.462 & 1.080 & 0.73909 \\
    & 0.633224 & 1.423 & 1.020 & 0.71659 \\
    & 0.643420 & 1.401 & 0.919 & 0.65584 \\
    & 0.652611 & 1.416 & 0.858 & 0.60614 \\
    & 0.661193 & 1.435 & 0.807 & 0.56234 \\
    & 0.669264 & 1.458 & 0.750 & 0.51415 \\
    & 0.676760 & 1.489 & 0.712 & 0.47832 \\
    & 0.683881 & 1.520 & 0.671 & 0.44132 \\
    & 0.690590 & 1.551 & 0.646 & 0.41648 \\
    & 0.697049 & 1.573 & 0.608 & 0.38661 \\
    & 0.703130 & 1.595 & 0.580 & 0.36358 \\
  9 & 0.599693 & 1.325 & 1.231 & 0.92894 \\
    & 0.612005 & 1.309 & 1.214 & 0.92684 \\
    & 0.624141 & 1.231 & 1.143 & 0.92830 \\
    & 0.635569 & 1.180 & 1.056 & 0.89493 \\
    & 0.646133 & 1.160 & 0.974 & 0.83958 \\
    & 0.655871 & 1.163 & 0.911 & 0.78282 \\
    & 0.664978 & 1.173 & 0.855 & 0.72872 \\
    & 0.673528 & 1.189 & 0.794 & 0.66770 \\
    & 0.681470 & 1.212 & 0.745 & 0.61506 \\
    & 0.688924 & 1.236 & 0.707 & 0.57239 \\
    & 0.695998 & 1.250 & 0.609 & 0.48723 \\
    & 0.702088 & 1.144 & 0.619 & 0.54144 \\
    & 0.708283 & 1.354 & 0.626 & 0.46188 \\
    & 0.714539 & 1.325 & 0.561 & 0.42296 \\
 10 & 0.599693 & 1.325 & 1.231 & 1.00000 \\
    & 0.696504 & 2.863 & 2.863 & 1.00000 \\
    & 0.725135 & 1.233 & 1.114 & 0.90365 \\
    & 0.736278 & 1.215 & 0.986 & 0.81164 \\
    & 0.746138 & 0.899 & 0.484 & 0.53881 \\
    & 0.750982 & 1.004 & 0.487 & 0.48546 \\
    & 0.755856 & 1.046 & 0.464 & 0.44354 \\
    & 0.760496 & 1.077 & 0.446 & 0.41398 \\
    & 0.764953 & 1.098 & 0.427 & 0.38857 \\
 \hline                    
 \multicolumn{5} {c} {M070}\\
 \hline                    
 4 & 0.70209 & 2.222 & 0.545 & 0.245 \\ 
 	 & 0.70754 & 3.493 & 0.767 & 0.220 \\ 
 	 & 0.71521 & 3.619 & 0.339 & 0.094 \\ 
 	 & 0.71860 & 2.459 & 0.487 & 0.198 \\ 
 5 & 0.70216 & 1.863 & 0.566 & 0.304 \\ 
 	 & 0.70782 & 2.625 & 0.743 & 0.283 \\ 
 	 & 0.71525 & 2.576 & 0.487 & 0.189 \\ 
 	 & 0.72012 & 2.893 & 0.495 & 0.171 \\ 
 	 & 0.72506 & 3.074 & 0.548 & 0.178 \\ 
 	 & 0.73054 & 3.072 & 0.550 & 0.179 \\ 
 6 & 0.70222 & 1.633 & 0.584 & 0.357 \\ 
 	 & 0.70806 & 2.175 & 0.577 & 0.266 \\ 
 	 & 0.71384 & 2.207 & 0.504 & 0.228 \\ 
 	 & 0.71887 & 2.307 & 0.496 & 0.215 \\ 
 	 & 0.72383 & 2.371 & 0.486 & 0.205 \\ 
 	 & 0.72869 & 2.414 & 0.465 & 0.193 \\ 
 	 & 0.73334 & 2.441 & 0.456 & 0.187 \\ 
 	 & 0.73790 & 2.455 & 0.443 & 0.180 \\ 
 	 & 0.74233 & 2.466 & 0.437 & 0.177 \\ 
 7 & 0.70229 & 1.457 & 0.591 & 0.405 \\ 
 	 & 0.70820 & 1.886 & 0.590 & 0.313 \\ 
 	 & 0.71410 & 1.864 & 0.522 & 0.280 \\ 
 	 & 0.71932 & 1.898 & 0.499 & 0.263 \\ 
 	 & 0.72431 & 1.937 & 0.486 & 0.251 \\ 
 	 & 0.72916 & 1.961 & 0.475 & 0.242 \\ 
 	 & 0.73391 & 1.967 & 0.453 & 0.230 \\ 
 	 & 0.73844 & 1.978 & 0.441 & 0.223 \\ 
 	 & 0.74285 & 1.978 & 0.430 & 0.217 \\ 
 	 & 0.74715 & 1.984 & 0.426 & 0.215 \\ 
 8 & 0.70236 & 1.317 & 0.606 & 0.460 \\ 
 	 & 0.70842 & 1.662 & 0.617 & 0.371 \\ 
 	 & 0.71459 & 1.608 & 0.537 & 0.334 \\ 
 	 & 0.71996 & 1.628 & 0.508 & 0.312 \\ 
 	 & 0.72504 & 1.645 & 0.498 & 0.303 \\ 
 	 & 0.73002 & 1.648 & 0.480 & 0.291 \\ 
 	 & 0.73481 & 1.654 & 0.460 & 0.278 \\ 
 	 & 0.73941 & 1.657 & 0.448 & 0.270 \\ 
 	 & 0.74389 & 1.652 & 0.431 & 0.261 \\ 
 	 & 0.74820 & 1.651 & 0.419 & 0.254 \\ 
 	 & 0.75239 & 1.642 & 0.410 & 0.250 \\ 
 	 & 0.75649 & 1.635 & 0.409 & 0.250 \\ 
 9 & 0.70243 & 1.200 & 0.615 & 0.513 \\ 
 	 & 0.70858 & 1.487 & 0.630 & 0.424 \\ 
 	 & 0.71488 & 1.422 & 0.559 & 0.393 \\ 
 	 & 0.72047 & 1.420 & 0.539 & 0.380 \\ 
 	 & 0.72587 & 1.417 & 0.511 & 0.361 \\ 
 	 & 0.73098 & 1.421 & 0.497 & 0.350 \\ 
 	 & 0.73594 & 1.419 & 0.476 & 0.335 \\ 
 	 & 0.74070 & 1.419 & 0.460 & 0.324 \\ 
 	 & 0.74530 & 1.418 & 0.445 & 0.314 \\ 
 	 & 0.74975 & 1.415 & 0.432 & 0.305 \\ 
 	 & 0.75406 & 1.412 & 0.422 & 0.299 \\ 
 	 & 0.75829 & 1.407 & 0.410 & 0.292 \\ 
 	 & 0.76239 & 1.395 & 0.396 & 0.284 \\ 
10 & 0.70250 & 1.098 & 0.629 & 0.573 \\
	 & 0.70879 & 1.337 & 0.653 & 0.488 \\
	 & 0.71531 & 1.262 & 0.571 & 0.453 \\
	 & 0.72102 & 1.256 & 0.553 & 0.440 \\
	 & 0.72655 & 1.247 & 0.535 & 0.429 \\
	 & 0.73190 & 1.237 & 0.517 & 0.418 \\
	 & 0.73706 & 1.231 & 0.497 & 0.404 \\
	 & 0.74204 & 1.227 & 0.479 & 0.391 \\
	 & 0.74683 & 1.223 & 0.462 & 0.378 \\
	 & 0.75145 & 1.220 & 0.447 & 0.366 \\
	 & 0.75592 & 1.218 & 0.436 & 0.358 \\
	 & 0.76028 & 1.214 & 0.418 & 0.344 \\
	 & 0.76446 & 1.209 & 0.408 & 0.337 \\
 \hline                    
 \multicolumn{5} {c} {M081}\\
 \hline                    
 4 & 0.81040 & 2.449 & 0.446 & 0.182 \\
   & 0.81486 & 3.223 & 0.291 & 0.090 \\
 5 & 0.81041 & 1.879 & 0.408 & 0.217 \\
   & 0.81449 & 2.082 & 0.249 & 0.120 \\
   & 0.81699 & 2.590 & 0.261 & 0.101 \\
   & 0.81960 & 2.892 & 0.325 & 0.112 \\
   & 0.82284 & 2.911 & 0.299 & 0.103 \\
 6 & 0.81043 & 1.584 & 0.408 & 0.258 \\
   & 0.81451 & 1.627 & 0.260 & 0.160 \\
   & 0.81711 & 1.885 & 0.262 & 0.139 \\
   & 0.81973 & 2.055 & 0.254 & 0.123 \\
   & 0.82227 & 2.143 & 0.246 & 0.115 \\
   & 0.82473 & 2.218 & 0.236 & 0.107 \\
   & 0.82710 & 2.282 & 0.234 & 0.102 \\
 7 & 0.81044 & 1.390 & 0.411 & 0.296 \\
   & 0.81456 & 1.343 & 0.268 & 0.200 \\
   & 0.81724 & 1.504 & 0.271 & 0.180 \\
   & 0.81995 & 1.584 & 0.261 & 0.165 \\
   & 0.82256 & 1.627 & 0.255 & 0.157 \\
   & 0.82511 & 1.684 & 0.245 & 0.145 \\
   & 0.82756 & 1.705 & 0.237 & 0.139 \\
   & 0.82993 & 1.722 & 0.231 & 0.134 \\
   & 0.83223 & 1.733 & 0.222 & 0.128 \\
 8 & 0.81046 & 1.248 & 0.402 & 0.322 \\
   & 0.81448 & 1.159 & 0.281 & 0.242 \\
   & 0.81729 & 1.271 & 0.287 & 0.226 \\
   & 0.82016 & 1.306 & 0.282 & 0.216 \\
   & 0.82298 & 1.326 & 0.272 & 0.205 \\
   & 0.82570 & 1.322 & 0.253 & 0.191 \\
   & 0.82823 & 1.356 & 0.245 & 0.181 \\
   & 0.83068 & 1.334 & 0.237 & 0.178 \\
   & 0.83305 & 1.360 & 0.242 & 0.178 \\
   & 0.83547 & 1.349 & 0.233 & 0.172 \\
 9 & 0.81048 & 1.129 & 0.404 & 0.358 \\
   & 0.81451 & 1.039 & 0.287 & 0.277 \\
   & 0.81739 & 1.113 & 0.297 & 0.267 \\
   & 0.82036 & 1.119 & 0.293 & 0.262 \\
   & 0.82329 & 1.129 & 0.286 & 0.253 \\
   & 0.82615 & 1.124 & 0.278 & 0.247 \\
   & 0.82893 & 1.115 & 0.264 & 0.237 \\
   & 0.83157 & 1.114 & 0.265 & 0.238 \\
   & 0.83422 & 1.121 & 0.248 & 0.222 \\
   & 0.83670 & 1.126 & 0.239 & 0.213 \\
   & 0.83910 & 1.109 & 0.238 & 0.215 \\
10 & 0.81049 & 1.030 & 0.396 & 0.385 \\
   & 0.81446 & 0.941 & 0.295 & 0.314 \\
   & 0.81741 & 0.979 & 0.305 & 0.312 \\
   & 0.82046 & 0.987 & 0.304 & 0.308 \\
   & 0.82350 & 0.975 & 0.296 & 0.303 \\
   & 0.82645 & 0.974 & 0.292 & 0.300 \\
   & 0.82938 & 0.972 & 0.281 & 0.289 \\
   & 0.83219 & 0.956 & 0.272 & 0.284 \\
   & 0.83491 & 0.954 & 0.263 & 0.276 \\
   & 0.83754 & 0.951 & 0.255 & 0.268 \\
   & 0.84009 & 0.936 & 0.252 & 0.269 \\
   & 0.84261 & 0.944 & 0.251 & 0.265 \\
20 & 0.81067 & 0.506 & 0.396 & 0.783 \\
   & 0.81463 & 0.450 & 0.326 & 0.725 \\
   & 0.81789 & 0.444 & 0.333 & 0.750 \\
   & 0.82122 & 0.438 & 0.344 & 0.786 \\
   & 0.82467 & 0.423 & 0.337 & 0.797 \\
   & 0.82804 & 0.417 & 0.337 & 0.808 \\
   & 0.83141 & 0.418 & 0.345 & 0.825 \\
   & 0.83486 & 0.410 & 0.348 & 0.848 \\
   & 0.83833 & 0.396 & 0.335 & 0.846 \\
   & 0.84169 & 0.396 & 0.337 & 0.851 \\
   & 0.84506 & 0.393 & 0.337 & 0.857 \\
   & 0.84843 & 0.384 & 0.338 & 0.880 \\
   & 0.85181 & 0.383 & 0.338 & 0.884 \\
   & 0.85519 & 0.394 & 0.340 & 0.862 \\
   & 0.85859 & 0.384 & 0.332 & 0.866 \\
   & 0.86191 & 0.376 & 0.338 & 0.900 \\
   & 0.86529 & 0.412 & 0.352 & 0.855 \\
   & 0.86882 & 0.391 & 0.336 & 0.860 \\
 \hline                    
 \multicolumn{5} {c} {M092}\\
 \hline                    
 5 & 0.91899 & 1.562 & 0.034 & 0.022 \\
 6 & 0.91899 & 1.271 & 0.048 & 0.038 \\
   & 0.91947 & 1.531 & 0.049 & 0.032 \\
 7 & 0.91899 & 1.094 & 0.045 & 0.041 \\
   & 0.91944 & 1.211 & 0.058 & 0.048 \\
 8 & 0.91899 & 0.966 & 0.051 & 0.053 \\
   & 0.91951 & 1.040 & 0.065 & 0.063 \\
 9 & 0.91900 & 0.877 & 0.078 & 0.089 \\
   & 0.91978 & 0.941 & 0.088 & 0.094 \\
   & 0.92066 & 0.961 & 0.087 & 0.090 \\
10 & 0.91900 & 0.802 & 0.090 & 0.113 \\
   & 0.91990 & 0.840 & 0.098 & 0.117 \\
   & 0.92089 & 0.846 & 0.101 & 0.120 \\
20 & 0.91903 & 0.429 & 0.153 & 0.358 \\
   & 0.92057 & 0.432 & 0.167 & 0.387 \\
   & 0.92224 & 0.415 & 0.171 & 0.412 \\
   & 0.92395 & 0.363 & 0.174 & 0.478 \\
30 & 0.91906 & 0.274 & 0.163 & 0.594 \\
   & 0.92069 & 0.279 & 0.186 & 0.666 \\
   & 0.92255 & 0.275 & 0.193 & 0.703 \\
   & 0.92449 & 0.266 & 0.193 & 0.725 \\
   & 0.92641 & 0.261 & 0.195 & 0.750 \\
40 & 0.91910 & 0.204 & 0.157 & 0.771 \\
   & 0.92067 & 0.214 & 0.181 & 0.843 \\
   & 0.92248 & 0.212 & 0.186 & 0.874 \\
   & 0.92433 & 0.204 & 0.188 & 0.921 \\
   & 0.92621 & 0.206 & 0.190 & 0.924 \\
   & 0.92812 & 0.194 & 0.187 & 0.962 \\
   & 0.92998 & 0.185 & 0.184 & 0.996 \\
   & 0.93182 & 0.188 & 0.188 & 1.000 \\
   & 0.93370 & 0.191 & 0.191 & 1.000 \\
   & 0.93560 & 0.199 & 0.199 & 1.000 \\
   & 0.93760 & 1.374 & 1.374 & 1.000 \\
   & 0.95134 & 0.361 & 0.361 & 1.000 \\
   & 0.95495 & 0.313 & 0.313 & 1.000 \\
   & 0.95808 & 0.323 & 0.323 & 1.000 \\
50 & 0.91914 & 0.158 & 0.154 & 0.974 \\
   & 0.92067 & 0.164 & 0.164 & 1.000 \\
   & 0.92231 & 0.166 & 0.166 & 1.000 \\
   & 0.92397 & 0.186 & 0.186 & 1.000 \\
   & 0.92583 & 0.236 & 0.236 & 1.000 \\
   & 0.92819 & 0.332 & 0.332 & 1.000 \\
   & 0.93151 & 0.457 & 0.457 & 1.000 \\
   & 0.93608 & 0.397 & 0.397 & 1.000 \\
   & 0.94004 & 0.429 & 0.429 & 1.000 \\
   & 0.94434 & 0.309 & 0.309 & 1.000 \\
   & 0.94743 & 0.297 & 0.297 & 1.000 \\
   & 0.95039 & 0.341 & 0.341 & 1.000 \\
   & 0.95380 & 0.271 & 0.271 & 1.000 \\
   & 0.95651 & 0.331 & 0.331 & 1.000 \\
   & 0.95982 & 0.098 & 0.098 & 1.000 \\
   & 0.96080 & 0.120 & 0.120 & 1.000 \\
   & 0.96200 & 0.144 & 0.144 & 1.000 \\
   & 0.96344 & 0.188 & 0.188 & 1.000 \\
   & 0.96532 & 0.259 & 0.259 & 1.000 \\
   & 0.96791 & 0.298 & 0.298 & 1.000 \\
   & 0.97089 & 0.301 & 0.301 & 1.000 \\
   & 0.97390 & 0.296 & 0.296 & 1.000 \\
 \hline                    
 \multicolumn{5} {c} {M102}\\
 \hline                    
 8 & 1.02047 & 0.727 & -0.017 & -0.024 \\
   & 1.02030 & 0.806 & -0.025 & -0.031 \\
 9 & 1.02047 & 0.648 & -0.003 & -0.005 \\
   & 1.02044 & 0.700 & -0.014 & -0.020 \\
10 & 1.02047 & 0.586 &  0.011 &  0.018 \\
   & 1.02058 & 0.617 & -0.005 & -0.008 \\
20 & 1.02048 & 0.315 &  0.091 &  0.290 \\
   & 1.02139 & 0.328 &  0.150 &  0.458 \\
   & 1.02290 & 0.231 &  0.097 &  0.418 \\
   & 1.02386 & 0.275 &  0.115 &  0.419 \\
30 & 1.02049 & 0.215 &  0.096 &  0.448 \\
   & 1.02145 & 0.243 &  0.116 &  0.477 \\
   & 1.02261 & 0.189 &  0.107 &  0.565 \\
   & 1.02368 & 0.210 &  0.115 &  0.548 \\
   & 1.02483 & 0.182 &  0.101 &  0.554 \\
40 & 1.02049 & 0.166 &  0.097 &  0.588 \\
   & 1.02147 & 0.188 &  0.117 &  0.622 \\
   & 1.02264 & 0.155 &  0.111 &  0.718 \\
   & 1.02375 & 0.162 &  0.114 &  0.705 \\
   & 1.02489 & 0.136 &  0.105 &  0.771 \\
   & 1.02594 & 0.149 &  0.112 &  0.749 \\
50 & 1.02050 & 0.134 &  0.095 &  0.708 \\
   & 1.02145 & 0.158 &  0.114 &  0.719 \\
   & 1.02259 & 0.121 &  0.096 &  0.794 \\
   & 1.02355 & 0.113 &  0.095 &  0.840 \\
   & 1.02450 & 0.120 &  0.101 &  0.841 \\
   & 1.02551 & 0.128 &  0.109 &  0.853 \\
   & 1.02661 & 0.121 &  0.104 &  0.860 \\
60 & 1.02051 & 0.110 &  0.092 &  0.830 \\
   & 1.02143 & 0.130 &  0.109 &  0.835 \\
   & 1.02251 & 0.111 &  0.098 &  0.885 \\
   & 1.02350 & 0.093 &  0.090 &  0.959 \\
   & 1.02439 & 0.102 &  0.099 &  0.974 \\
   & 1.02539 & 0.120 &  0.119 &  0.987 \\
   & 1.02657 & 0.142 &  0.132 &  0.933 \\
   & 1.02789 & 0.112 &  0.109 &  0.973 \\
   & 1.02898 & 0.101 &  0.101 &  1.000 \\
   & 1.02999 & 0.107 &  0.107 &  1.000 \\
   & 1.03107 & 0.144 &  0.144 &  1.000 \\
   & 1.03251 & 0.203 &  0.203 &  1.000 \\
   & 1.03453 & 0.205 &  0.205 &  1.000 \\
   & 1.03658 & 0.212 &  0.212 &  1.000 \\
 \hline                    
 \end{longtable}

\end{document}